\documentclass[11pt,a4paper]{article}
\usepackage{amssymb,enumitem,booktabs,xcolor,todonotes,microtype,setspace}
\usepackage[top=1.4in, bottom=1.4in, left=1.35in, right=1.35in]{geometry}
\usepackage{natbib}
\usepackage{url} % not crucial - just used below for the URL
\usepackage[pdftex,colorlinks=true,hypertexnames=false,pagebackref]{hyperref}
\definecolor{darkblue}{rgb}{0,0,.6}
\hypersetup{citecolor=darkblue,linkcolor=darkblue,urlcolor=darkblue}
\usepackage{amsmath}
\usepackage{amsfonts}
\usepackage{graphics}
\usepackage{palatino,eulervm}
\usepackage{graphicx}
\usepackage{lscape}
\usepackage{rotating}
\usepackage[linewidth=1pt]{mdframed}
\usepackage{subcaption}
\allowdisplaybreaks[4]
\DeclareMathOperator*{\argmin}{arg\,min}

\providecommand{\U}[1]{\protect\rule{.1in}{.1in}}

\graphicspath{{plots/}}

\usepackage{amsthm,thmtools}
\usepackage{mathrsfs}
\usepackage{float}

\usepackage[autostyle]{csquotes}

\declaretheorem{theorem}
\declaretheorem{lemma}
\makeatletter
\def\th@newremark{\th@remark\thm@headfont{\bfseries}}
\makeatletter
\theoremstyle{newremark}
\newtheorem{remark}{Remark}
\newtheorem{prop}{Proposition}

\newtheorem{assumption}{Assumption}
\declaretheoremstyle[
  spaceabove=6pt, spacebelow=6pt,
  headfont=\bfseries,
  notefont=\mdseries, notebraces={(}{)},
bodyfont=\normalfont,
  postheadspace=0.5em,
]{mystyle}

\begin{document}

\title{\Large\bf Large-Scale Curve Time Series with Common Stochastic Trends\thanks{D. Li acknowledges research support from the CPG and SRG funds at the University of Macau. Phillips acknowledges research support from the Kelly Fund at the University of Auckland and the KLC Fellowship at Singapore Management University.}}
\author{{\normalsize Degui Li\thanks{Faculty of Business Administration, Asia-Pacific Academy of Economics and Management, and Department of Economics,  University of Macau, China. Email: deguili@um.edu.mo.},\ \ \ Yuning Li\thanks{School of Business and Society, University of York, UK. Email: yuning.li@york.ac.uk. },\ \ \ Peter C.B. Phillips\thanks{Cowles Foundation for Research in Economics,
Yale University, US. Email: peter.phillips@yale.edu.}}\\
{\normalsize\em $^\dag$University of Macau, $^\ddag$University of York, $^\S$Yale University,}\\
{\normalsize\em $^\S$University of Auckland and $^\S$Singapore Management University}}
\date{\normalsize This version: \today}

\maketitle

\centerline{\bf Abstract}

\medskip

This paper studies high-dimensional curve time series with common stochastic trends. A dual functional factor model structure is adopted with a high-dimensional factor model for the observed curve time series and a low-dimensional factor model for the latent curves with common trends. A functional PCA technique is applied to estimate the common stochastic trends and functional factor loadings. Under some regularity conditions we derive the mean square convergence and limit distribution theory for the developed estimates, allowing the dimension and sample size to jointly diverge to infinity. We propose an easy-to-implement criterion to consistently select the number of common stochastic trends and further discuss model estimation when the nonstationary factors are cointegrated. Extensive Monte-Carlo simulations and two empirical applications to large-scale temperature curves in Australia and log-price curves of S\&P $500$ stocks are conducted, showing finite-sample performance and providing practical implementations of the new methodology.

\medskip


\newpage

%%%%%%%%%%%%%%%%%%%

\section{Introduction}\label{sec1}
\renewcommand{\theequation}{1.\arabic{equation}}
\setcounter{equation}{0}

The past few decades have seen notable developments in modeling curve time series, a sequence of random curves or functions often defined within a bounded set. Many estimation, inference and forecasting techniques have been proposed to tackle curve time series \citep[e.g.,][]{B00,RS05,HK12,PJ25a} which arise in a variety of areas such as climatology, transportation, finance, demography and health sciences. One may reduce the infinite dimension of curve time series to a finite dimension through a functional version of principal component analysis (PCA), and subsequently apply classic time series models such as VAR \citep{Lu06} to the finite-dimensional time series which retain much of the dynamic sample information. The existing literature often assumes the curve time series to be stationary, thereby facilitating theory development using standard asymptotics. Stationarity may be too restrictive in many applications and is often rejected when we test practical curve time series data. For example, \cite{CKP16} find evidence of a unit root structure for intra-month distribution curves of S\&P 500 index returns; \cite{ARS18} reject the null hypothesis of stationarity for Australian temperature curves; \cite{LRS23} detect a nonstationary feature in US treasury yield curves; and \cite{PJ25b} find unit root behavior in Engel curve data for leisure, health, and food expenditure among ageing seniors in Singapore.  

\smallskip

There have been some attempts in recent years to relax the stationarity restriction for curve time series. \cite{HKR14} introduce a functional KPSS test \citep{KPSS92} for stationarity of curve time series; \cite{CKP16} study nonstationary time series of state density curves by decomposing an infinite-dimensional Hilbert space into the nonstationary I(1) and stationary subspaces; \cite{BSS17} establish the Granger-Johansen representation theorem for I(1) autoregressive curve processes, which has been further extended by \cite{BS20} and \cite{FP20} to I(2) and more general I($d$) autoregressive curve processes; \cite{LRS23} introduce a nonstationary fractionally integrated curve time series framework, covering the nonstationary I(1) curve as a special case; \cite{NSS23} propose a variance ratio-type test to determine the dimension of the nonstationary subspace of the cointegrated curve time series; and \cite{PJ25b} develop ADF and semiparametric unit root tests for curve time series autoregression. The aforementioned literature limits attention to a single curve time series with nonstationarity. \cite{P25} considers rank selection in vector autoregression with multiple curve time series but in a parametric setting. In practice, we often have to jointly model a large number of curve time series driven by some common stochastic trends which are usually latent. For example, thousands of stock return curve time series in financial markets may be driven by latent market and industry factors; curve time series of temperature and rainfall recorded in hundreds of weather stations may be affected by common weather patterns in the region. Hence, it is imperative for adequate empirical modeling to develop a flexible curve time series framework that accommodates large dimensionality and unobserved factors as well as nonstationarity.  

\smallskip 

The approximate factor model has proven to be an effective tool for analyzing large-scale real-valued panel data \citep[e.g.,][]{CR83, BN02}. \cite{BN04} proposed a so-called PANIC method under the factor model framework to test for unit roots in the idiosyncratic components and determine the number of common stochastic trends that are present among the cointegrated nonstationary factors. That work was extended in \cite{BC09} to accommodate structural breaks. \cite{B04} used classic PCA to estimate common stochastic trends and factor loadings, assuming the idiosyncratic components to be stationary over time; and \cite{BLL21} examined an approximate factor model for nonstationary panels with the primary concern of  impulse-response function estimation with cointegrated factors within a vector error-correction model (VECM) specification. 

\smallskip

The main focus of the present paper centers on the interaction of recent advances in nonstationary curve time series and large-dimensional approximate factor models. The goal is to build a fully functional factor model approach designed for curve time series with common stochastic trends. There has been increasing interest in extending the approximate factor model to curve time series under stationarity conditions. For a single or a small number of curve time series, \cite{HSH12}, \cite{KMZ15} and \cite{KMRT18} consider low-dimensional functional factor models, where either factors or factor loadings take functional values. For large-scale curve time series with the cross-sectional size increasing with the temporal dimension, \cite{GQW21} consider a high-dimensional functional factor model with functional factors and real-valued loadings, whereas \cite{TNH23a,TNH23b} introduce a different functional factor model with functional loadings and real-valued factors, proposing functional PCA to estimate the functional common and idiosyncratic components. In addition, \cite{LLSX24} recently introduced a dual functional factor model for high-dimensional stationary curve time series, providing estimates of the functional covariance structure. As far as we know, there is no literature on high-dimensional factor models for nonstationary curve time series. The present paper employs such a framework, adopting a dual functional factor model structure that admits common stochastic trends in high-dimensional curve time series, thereby allowing practical implementation with many financial market and climatic curve time series that manifest nonstationary behavior.   

\smallskip

With a high-dimensional functional factor structure for large-scale curve time series, we decompose each functional observation into common and idiosyncratic components. In particular, we define the common component via an integral operator and allow both the factors and factor loadings to be functional, giving a more flexible structure than those in \cite{GQW21} and \cite{TNH23a, TNH23b}. For the latent factor curves with common stochastic trends, we impose another functional factor model structure via multivariate series approximation, where the number of real-valued stochastic trends is allowed to diverge slowly to infinity. As in \cite{TNH23b}, functional PCA methods are used to estimate the common stochastic trends and functional factor loadings. Under some technical but justifiable assumptions, we derive the mean square convergence of the estimated common trends, where the convergence rate relies on the dimension, time series length and number of common trends. To facilitate inference we establish limit theory for the estimated common trends and functional factor loadings, extending Theorems 2 and 3 in \cite{B04} to large-scale curve time series with a diverging number of common trends. In particular, super-fast convergence is established for the functional factor loading estimate and its limit distribution is derived as if the common stochastic trends were known.

\smallskip

Practical implementation of functional PCA, like other PCA applications, requires consistent estimation of the number of common stochastic trends. A suitable information-based selection criterion is employed for this purpose, modifying existing criteria that have been extensively studied in the literature \citep[e.g.,][]{BN02}. The proposed criterion is easily implemented and consistency follows straightforwardly. A more general model setting is also considered in which the integrated factors are themselves cointegrated and the idiosyncratic components may be nonstationary. In this setting a functional version of PANIC is proposed for estimating factors (via first-order differences) and functional factor loadings.     

\smallskip

Extensive simulations are conducted to assess numerical performance of the methods in finite samples. The findings reveal that when the factors are full-rank integrated and functional idiosyncratic components are stationary, functional PCA estimates of common stochastic trends and functional factor loadings are more efficient than those obtained via functional PANIC; but when the integrated factors are rank-reduced and functional idiosyncratic components are nonstationary, functional PANIC estimation continues to work well but functional PCA estimation is inconsistent. In the empirical applications, functional PCA is used to analyze temperature curve time series for Australia over the period 1943-2022, and functional PANIC is used to analyze log-price curves of S\&P $500$ stocks from January 2023 to November 2023. The empirical results confirm the existence of common stochastic trends for both these  datasets of large-scale curve time series.

\smallskip

The rest of the paper is organized as follows. Section \ref{sec2} introduces the dual functional factor model framework. Section \ref{sec3} describes functional PCA estimation and provides the relevant theory. Section \ref{sec4} proposes a modified information criterion to estimate the number of common stochastic trends. Section \ref{sec5} discusses model estimation with cointegrated factors. Sections \ref{sec6} and \ref{sec7} present the simulation study and the empirical applications. Section \ref{sec8} concludes. Proofs of the main asymptotic theorems are given in Appendix A. Some useful technical lemmas with proofs are in Appendix B. Throughout the paper, we define a separable Hilbert space ${\mathscr H}$ as a set of real measurable functions $f(\cdot)$ on a compact set ${\mathbb C}$ such that $\int_{\mathbb C} f^2(u)du<\infty$, with the inner product of $f_1$ and $f_2$ as $\langle f_1,f_2\rangle=\int_{\mathbb C}f_1(u)f_2(u)du$, and the norm as $\Vert f\Vert=\langle f,f\rangle^{1/2}$. We further generalize $\langle \cdot,\cdot\rangle$ to handle vectors or matrices of functions: for ${\boldsymbol F}_1=(f_{1,ki})$ and ${\boldsymbol F}_2=(f_{2,kj})$ which are $k_1\times k_2$ and $k_1\times k_3$ matrices of functions, define $\langle \boldsymbol F_1^{^\intercal},\boldsymbol F_2\rangle$ as a $k_2\times k_3$ matrix whose $(i,j)$-th entry is $\sum_{k=1}^{k_1}\int f_{1,ki}(u)f_{2,kj}(u)du$, and for a vector of functions $\boldsymbol F$, write $\Vert \boldsymbol F\Vert=\langle \boldsymbol F^{^\intercal},\boldsymbol F\rangle^{1/2}$. We also use the notation $\Vert\cdot\Vert$ as the Euclidean norm of a vector and the operator norm of a matrix or a continuous linear operator whenever no ambiguity arises and let $\Vert\cdot\Vert_F$ be the Frobenius norm of a matrix. Let $a_{n}\sim b_{n}$, $a_{n}\propto b_{n}$ and $a_{n}\gg b_{n}$ denote that $a_{n}/b_{n}\rightarrow1$, $0<\underline{c}\leq a_{n}/b_{n}\leq\overline{c}<\infty$and $b_{n}/a_{n}\rightarrow0$, respectively. For brevity ``with probability approaching one' is written ``{\em w.p.a.1}".

\section{Dual functional factor model structure}\label{sec2}
\renewcommand{\theequation}{2.\arabic{equation}}
\setcounter{equation}{0}

$N$-vectors of curve time series $Z_t=(Z_{1t},\cdots,Z_{Nt})^{^\intercal}$, where $Z_{it}=\left(Z_{it}(u):\ u\in{\mathbb C}_{i}\right)\in{\mathscr H}_{i}$ for $t=1,\cdots,T$ are observed, with ${\mathscr H}_{i}$ being a separable Hilbert space defined as a set of measurable and square-integrable functions on a bounded set ${\mathbb C}_{i}$. We may decompose $Z_{it}$ into a common component and an idiosyncratic component as follows
\begin{equation}\label{eq2.1}
Z_{it}=\chi_{it}+\varepsilon_{it},\ \ i=1,\cdots,N,\ t=1,\cdots,T,
\end{equation}
where $\chi_{it}=\left(\chi_{it}(u):\ u\in{\mathbb C}_{i}\right)$ whose dynamic patterns are driven by some latent factors, and $\varepsilon_{it}=\left(\varepsilon_{it}(u):\ u\in{\mathbb C}_{i}\right)$ is allowed to be correlated over $i$ and $t$. The functional factor model (\ref{eq2.1}) is similar to the classic approximate factor model studied in \cite{CR83} and \cite{BN02} with the exception that all the components in (\ref{eq2.1}) take functional values. But the formulation of the common component $\chi_{it}$ is non-trivial. Broadly speaking, two different ways have been recommended in the recent literature to define $\chi_{it}$: \cite{GQW21} construct $\chi_{it}$ as a product of real-valued factor loadings and functional factors, whereas \cite{TNH23a} define $\chi_{it}$ as a product of functional factor loadings and real-valued factors. The factor number is assumed fixed in \cite{GQW21} and \cite{TNH23a,TNH23b} to achieve dimension reduction in large-dimensional curve time series modeling. Both approaches involve real-valued components, either as parametric factor loadings or as real-valued factors. As in \cite{LLSX24}, we introduce a more flexible functional factor model, constructing $\chi_{it}$ via an integral operator and allowing both the factors and factor loadings to be functional.

\smallskip

Let $F_t=(F_{1t},\cdots,F_{kt})^{^\intercal}$, where $F_{jt}=\left(F_{jt}(u):\ u\in{\mathbb C}_{j}^\ast\right)\in{\mathscr H}_{j}^\ast$ and ${\mathscr H}_{j}^\ast$ is defined similarly to ${\mathscr H}_{i}$ but with ${\mathbb C}_{i}$ replaced by a possibly different bounded set ${\mathbb C}_{j}^\ast$. For $i=1,\cdots,N$ and $j=1,\cdots,k$, we let ${\cal B}_{ij}$ be a linear (kernel) integral operator defined by
\[
{\cal B}_{ij}f(u)=\int_{{\mathbb C}_{j}^\ast} B_{ij}(u,v)f(v)dv,\ \ f\in{\mathscr H}_{j}^\ast,\ \ u\in{\mathbb C}_{i},
\]
where $B_{ij}=\left(B_{ij}(u,v):\ u\in{\mathbb C}_{i}, v\in{\mathbb C}_{j}^\ast\right)$ denotes the kernel of the linear operator ${\cal B}_{ij}$, and write $\chi_{it}$ as
\begin{equation}\label{eq2.2}
\chi_{it}(u)=\sum_{j=1}^k{\cal B}_{ij}F_{jt}(u)= \sum_{j=1}^{k}\int_{{\mathbb C}_{j}^\ast}B_{ij}(u,v)F_{jt}(v)dv,\ \ u\in{\mathbb C}_{i}.
\end{equation}   
Combining (\ref{eq2.1}) and (\ref{eq2.2}), we obtain a fully functional factor model structure which is more general than those in \cite{GQW21} and \cite{TNH23a,TNH23b}.  Neither the factor loading operator ${\cal B}_{ij}$ nor functional factor $F_t$ is known a priori. As in \cite{HG18}, we allow the curve time series observations $Z_{it}$, $i=1,\cdots,N$, and the latent functional factors $F_{jt}$, $j=1,\cdots,k$, to be defined on different domains, i.e., ${\mathbb C}_{i}$ and ${\mathbb C}_{j}^\ast$ may vary over $i$ and $j$. The number of functional factors is unknown but assumed to be a finite positive integer. 

\smallskip

Although the linear integral operator provides a flexible structure for functional common components, it makes the estimation of functional factors and factor loadings challenging. To address the difficulty we impose a low-dimensional functional factor model representation for $F_t$ via the following series expansion
\begin{equation}\label{eq2.3}
F_{jt}(u)=\Phi_j(u)^{^\intercal}G_t+\eta_{jt}(u),\ \ u\in{\mathbb C}_{j}^\ast,\ \ j=1,\cdots,k,
\end{equation}
where $\Phi_j=\left(\phi_{j1},\cdots,\phi_{jq}\right)^{^\intercal}$ is a $q$-dimensional vector of deterministic basis functions, $G_t$ is a $q$-dimensional vector of nonstationary real-valued factors, $\eta_{jt}$ denotes the series approximation error which can be either stationary or nonstationary, and $q$ is a positive integer which may slowly diverge to infinity. Model (\ref{eq2.3}) extends the low-dimensional functional factor model studied in \cite{KMRT18} and \cite{MGG22} to multivariate curve time series. It is also similar to the multivariate Karhunen-Lo\`eve representation in \cite{HG18} if $\left(\phi_{1l},\cdots,\phi_{kl}\right)^{^\intercal}$ is an orthonormal basis vector of eigenfunctions and $q$ is set as the truncation parameter. In the present paper, the integrated factor $G_t$ is generated by 
\begin{equation}\label{eq2.4}
\Delta G_t=(1-L) G_t= \xi_t,
\end{equation}
where $L$ is the lag operator and $\{\xi_t\}$ is a sequence of stationary $I(0)$ random vectors. Without loss of generality, we assume that the initial value $G_0=(G_{10},\cdots,G_{q0})^{^\intercal}$ satisfies $\max_{1\leq j\leq q}|G_{j0}|=O_P(1)$. 

\smallskip

Writing 
\begin{equation}\label{eq2.5}
\Lambda_i(u)=\sum_{j=1}^{k} {\cal B}_{ij}\Phi_j(u),\ \ \chi_{it}^{\eta}(u)=\sum_{j=1}^{k}{\cal B}_{ij}\eta_{jt}(u),
\end{equation}
with the high-dimensional factor structure (\ref{eq2.1}) and (\ref{eq2.2}) for the observed curve time series and the low-dimensional factor structure (\ref{eq2.3}) for the latent factor curves, we obtain 
\begin{equation}\label{eq2.6}
Z_{it}=\Lambda_{i}^{^\intercal}G_{t}+\chi_{it}^{\eta}+\varepsilon_{it},\ \ i=1,\cdots,N,\ \ t=1,\cdots,T,
\end{equation} 
where $\Lambda_i=\left(\Lambda_{i}(u):\ u\in{\mathbb C}_{i}\right)$ and $\chi_{it}^{\eta}=\left(\chi_{it}^{\eta}(u):\ u\in{\mathbb C}_{i}\right)$. For practical purposes our main interest lies in estimating $\Lambda_i$ and $G_t$, and determining $q$, the number of common stochastic trends. In the context of stationary curve time series, \cite{TNH23a, TNH23b}'s high-dimensional functional factor model can be seen as a special case of (\ref{eq2.6}) with $\chi_{it}^{\eta}\equiv0$ and $q$ being a finite positive integer. Model (\ref{eq2.6}) also extends the nonstationary factor model in \cite{B04}, \cite{BN04} and \cite{BLL21} from real-valued time series to more general curve time series.

%%%%%%%%%%%%%%%%%%%

\section{Functional PCA estimation methodology and theory}\label{sec3}
\renewcommand{\theequation}{3.\arabic{equation}}
\setcounter{equation}{0}

This section introduces functional PCA methodology to estimate $\Lambda_i$ and $G_t$. PCA has been commonly used to estimate factors and factor loadings (subject to appropriate rotation) in the standard factor model for a large panel of real-valued time series \citep[e.g.,][]{BN02, SW02, B04, BN04, BLL21}. We extend the technique to high-dimensional nonstationary curve time series. Here we assume the number of common stochastic trends is known and $G_t$ is a vector of full-rank integrated variables. Section \ref{sec4} introduces an easy-to-implement criterion to estimate $q$ and Section \ref{sec5} considers the more general setting of cointegrated factors. 

\subsection{Functional PCA}\label{sec3.1}

Since $\Lambda_i$ and $G_t$ are not identifiable in the functional factor model (\ref{eq2.6}), identification restrictions are imposed in the functional PCA algorithm using 
\begin{equation}\label{eq3.1}
\frac{1}{T^2}\sum_{t=1}^TG_tG_t^{^\intercal}={\boldsymbol I}_{q}\ \ \ {\rm and}\ \ \ \frac{1}{N}\sum_{i=1}^N \int_{u\in{\mathbb C}_{i}}\Lambda_i(u)\Lambda_i(u)^{^\intercal}du\ \ {\rm is\ diagonal},
\end{equation}
where ${\boldsymbol I}_q$ is a $q\times q$ identity matrix. These identification conditions are comparable to those used by \cite{B04} for the traditional factor model. Since $G_t$ is integrated, the normalization rate in (\ref{eq3.1}) is $T^2$ instead of $T$ (for stationary time series). Eigenanalysis is conducted on the matrix
\begin{equation}\label{eq3.2}
\widetilde{\boldsymbol\Omega}=\left(\widetilde\Omega_{ts}\right)_{T\times T}\ \ {\rm with}\ \ \widetilde\Omega_{ts}=\frac{1}{N}\sum_{i=1}^N \int_{u\in{\mathbb C}_{i}}Z_{it}(u)Z_{is}(u)du,
\end{equation}
with $\widetilde {\boldsymbol G}=(\widetilde G_1,\cdots,\widetilde G_T)^{^\intercal}$ a $T\times q$ matrix consisting of the eigenvectors scaled by $T$, corresponding to the $q$ largest eigenvalues of $\widetilde{\boldsymbol\Omega}$. The functional factor loadings are subsequently estimated as
\begin{equation}\label{eq3.3}
\widetilde\Lambda_i=\left(\widetilde\Lambda_i(u): u\in{\mathbb C}_{i}\right)=\frac{1}{T^2}\sum_{t=1}^TZ_{it}\widetilde G_t,\ \ i=1,\cdots,N,
\end{equation}
using least squares and the first restriction in (\ref{eq3.1}).

\subsection{Mean square convergence of $\widetilde{G}_t$}\label{sec3.2}

The following assumptions are needed to develop the convergence theory of $\widetilde G_t$ and $\widetilde{\Lambda}_i$.

\begin{assumption}\label{ass:1}

{\em (i) The factor loading operator ${\cal B}_{ij}$ satisfies that $\Vert {\cal B}_{ij}\Vert\leq C_B$ uniformly over $i$ and $j$ with $C_B$ being a positive constant.}  

{\em (ii) There exists a positive definite matrix ${\boldsymbol\Sigma}_\Lambda$ with eigenvalues bounded away from zero and infinity such that, for some $\kappa>0$, 
\[
\left\Vert \frac{1}{N}\sum_{i=1}^N \int_{u\in{\mathbb C}_{i}}\Lambda_i(u)\Lambda_i(u)^{^\intercal}du-{\boldsymbol\Sigma}_\Lambda\right\Vert =o(q^{-\kappa})\quad {\it as}\ N\rightarrow\infty.
\]
}

{\em (iii) Let $\{G_t\}$ be a sequence of integrated random vectors satisfying (\ref{eq2.4}) and 
\[
\left\Vert \frac{1}{T^2}\sum_{s=1}^TG_sG_s^{^\intercal}- \int_0^1 B_\xi(u) B_\xi(u)^{^\intercal}du\right\Vert=o_P\left(q^{-\kappa}\right)\quad {\it as}\ T\rightarrow\infty,\]
where $B_\xi(\cdot)$ is a $q$-vector Brownian motion with positive definite covariance matrix ${\boldsymbol\Sigma}_\xi$, being the long-run variance matrix of $\xi_t$, i.e., ${\boldsymbol\Sigma}_\xi=\lim_{T\rightarrow\infty}\frac{1}{T}\sum_{t=1}^T\sum_{s=1}^T{\sf E}\left(\xi_t\xi_s^{^\intercal}\right)$.}
%\red{[PCB: (i) It seems that the matrix ${\boldsymbol\Sigma}_\xi$ becomes singular when $q \to \infty$ because the sample paths of a BM can be perfectly fitted by an infinite set of independent BMs, \cite[see][]{P98}, in which case the LR variance is asymptotically singular. (ii) Since $\frac{1}{T^2}\sum_{s=1}^TG_sG_s^{^\intercal} \rightsquigarrow \int_0^1 B_\xi(u) B_\xi(u)^{^\intercal}du$ involves only weak convergence, we need to enlarge the probability space to achieve the stated $o_P\left(q^{-\kappa}\right)$ error and invoke a strong approximation argument - e.g. Lemma 3.1 in \cite{P07}, which will require suitable moment conditions on the innovations (at least, 4th moments and possibly more, depending on the rate of approximation). See further comments below.]} \blue{[Reply: (i) The positive definiteness on ${\boldsymbol\Sigma}_\xi$ is a key assumption here to ensure that $G_t$ is of full rank. Since $\{\xi_t\}$ is a sequence of stationary I(0) random vectors, it may be sensible to assume that its long-run variance matrix is positive definite as long as the dimension $q$ grows at a slow rate. Remark 1(i) gives sufficient conditions to verify this condition when $q$ diverges slowly. (ii) As you commented, we do need the multivariate strong approximation result. Many thanks for suggesting the reference \cite{P07}, which we need to verify the weak convergence condition in later remarks. Yes, suitable moment conditions on the innovations are (implicitly) required for this high-level condition.]}

{\em (iv) Let $\nu_{i,0}$ be the $i$-th largest eigenvalue of ${\boldsymbol\Sigma}_\Lambda^{1/2}(\int_0^1 B_\xi(u) B_\xi(u)^{^\intercal}du){\boldsymbol\Sigma}_\Lambda^{1/2}$ and $\iota_{q,0}=\min_{1\leq i\leq q-1} (\nu_{i,0} - \nu_{i+1,0})$. There exist deterministic $\overline{\nu}_{q}$, $\underline{\nu}_{q}$, and $\underline{\iota}_{q}$ such that 
\[
{\sf P}\left(\underline{\nu}_{q}\leq\nu_{q,0}<\nu_{1,0}\leq\overline{\nu}_{q},\; \iota_{q,0}\geq\underline{\iota}_{q}\right)\rightarrow1,\quad q^{-\kappa}\left({\overline\nu}_q/\underline\nu_q\right)^2=O(1),\]
and 
\[{\underline\iota}_q^{-1}q^{1-\kappa}\overline{\nu}_q^{3/2}{\underline\nu}_q^{-1/2}=O(1).\]
In addition, $q=O(T^{1/(2\kappa+1)})$.}

\end{assumption}

\begin{assumption}\label{ass:2}

{\em (i) Let the idiosyncratic components $\{\varepsilon_{it}\}$ be independent of $\{\xi_{t}\}$ and $\{\eta_{jt}\}$. In addition, $\varepsilon_{it}$ are mean-zero random functions and $\max_{1\leq i\leq N}\max_{1\leq t\leq T}{\sf E}\left[\Vert\varepsilon_{it}\Vert^4\right]<\infty$.} 

{\em (ii) There exist $\delta_q$ and $\{\delta_{t,q}\}$ such that
\[
\max\limits_{1\leq j\leq k}{\sf E}\left[\Vert \eta_{jt}\Vert^2\right]\leq\delta_{t,q}^2,\quad\max_{1\leq t\leq T}\delta_{t,q}\rightarrow0\quad \text{and}\quad
\frac{1}{T}\sum_{t=1}^T \delta_{t,q}^2=O\left(\delta_{q}^2\right).\]}

{\em (iii) Letting $\zeta_N(s,t)=\frac{1}{N}\sum_{i=1}^N{\sf E}\left[\langle \varepsilon_{it},\varepsilon_{is}\rangle\right]$, there exists a constant $C_\varepsilon>0$ such that 
\[
\max_{1\leq t\leq T}|\zeta_N(t,t)|\leq C_\varepsilon\quad\text{and}\quad\sum_{s=1}^T \left\vert\zeta_N(s,t)\right\vert\leq C_\varepsilon\quad \forall\ 1\leq t\leq T.
\]
In addition,
\[{\sf E}\left(\sum_{i=1}^N \left\{\langle \varepsilon_{it},\varepsilon_{is}\rangle-{\sf E}\left[\langle \varepsilon_{it},\varepsilon_{is}\rangle\right]\right\}\right)^2\leq C_\varepsilon N\quad \forall\ 1\leq s,t\leq T.
\]
}

{\em (iv) For $h_i\in{\mathscr H}_{i}$, any deterministic function defined on ${\mathbb C}_{i}$, we have
\[{\sf E}\left(\sum_{i=1}^N\langle h_i,\varepsilon_{it}\rangle\right)^2 \propto \sum_{i=1}^N \Vert h_i\Vert_2^2.
\]
}

\end{assumption}

\begin{remark}\label{re:1}

(i) Assumption 1 imposes some fundamental conditions on the functional loadings ${\cal B}_{ij}$ and $\Lambda_i$ and the integrated factors $G_t$. Similar assumptions commonly appear in the literature for (functional) factor model estimation \citep[e.g.,][]{BN02, B04, TNH23b}. Assumption \ref{ass:1}(i) imposes uniform boundedness on the factor loading operators whereas Assumption \ref{ass:1}(ii) indicates that the $q$-dimensional nonstationary factors are full rank and pervasive in the limit. The high-level convergence condition in Assumption \ref{ass:1}(iii) may be justified via a strong approximation form of the weak invariance principle for stationary processes in a suitably expanded probability space. In fact, Assumption \ref{ass:1}(iii) is satisfied if
\begin{equation}\label{eq3.4}
\max_{1\leq t\leq T}\left\Vert \frac{1}{T^{1/2}}\sum_{s=1}^t\xi_s-B_\xi(t/T)\right\Vert=o_P\left(q^{-(\kappa+1/2)}\right).
\end{equation} 
When $q$ is fixed, (\ref{eq3.4}) can be further replaced by 
\[
\frac{1}{T^{1/2}}\sum_{s=1}^{\lfloor Tu\rfloor}\xi_s\Rightarrow B_\xi(u),\quad 0\leq u\leq 1,
\]
where ``$\Rightarrow$" denotes weak convergence, $\lfloor\cdot\rfloor$ is the floor function, and Assumption \ref{ass:1}(iii) may be replaced by
\[
\frac{1}{T^2}\sum_{s=1}^TG_sG_s^{^\intercal}\rightsquigarrow\int_0^1 B_\xi(u) B_\xi(u)^{^\intercal}du,
\]
see, for example, \cite{B04}, where ``$\rightsquigarrow$" denotes convergence in distribution. Following the argument given in the proof of \cite[Lemma 3.1]{P07} we next verify (\ref{eq3.4}) by considering
\[
\xi_t={\boldsymbol A}(L) e_t,\ \ {\boldsymbol A}(L)=\sum_{j=0}^\infty {\boldsymbol A}_jL^j,
\]
where $\{{\boldsymbol A}_j\}$ is a sequence of $q\times q$ coefficient matrices, and $\{e_t\}$ is a sequence of independent and identically distributed (i.i.d.) random vectors with mean zero. Without loss of generality, we assume that the components of $e_t=(e_{1t},\cdots,e_{qt})^{^\intercal}$ are i.i.d., ${\sf E}[e_{it}^2]=1$ and ${\sf E}[|e_{it}|^4]<\infty$. If $q\propto T^{\tau}$ with $\tau<1/[2(2\kappa+3)]$, following the strong approximation theory in \cite[e.g.,][]{CR81,Z98,P07}, we may show that
\begin{equation}\label{eq3.5}
\max_{1\leq t\leq T}\left\Vert \frac{1}{T^{1/2}}\sum_{s=1}^t e_s-B_0(t/T)\right\Vert=o_P\left(q^{-(\kappa+1/2)}\right),
\end{equation} 
where $B_0(\cdot)$ is a $q$-dimensional vector of standard Brownian motions with identity covariance matrix. Then, using the BN decomposition as in \cite{PS92} gives the representation
\[
\frac{1}{T^{1/2}}\sum_{s=1}^t\xi_s=\frac{1}{T^{1/2}}\sum_{s=1}^t\overline\xi_s+\frac{1}{T^{1/2}}\left(\check\xi_0-\check\xi_t+G_0\right),\ \ \overline\xi_s={\boldsymbol A}(1)e_s,\ \ \check\xi_s=\sum_{j=0}^\infty \check{\boldsymbol A}_je_{s-j},
\]
where $\check{\boldsymbol A}_j=\sum_{k=j+1}^\infty {\boldsymbol A}_k$. Hence, assuming that ${\boldsymbol A}(1)$ is of full rank and $\sum_{j=0}^\infty j\Vert {\boldsymbol A}_j\Vert<C<\infty$, where $C$ is a constant that does not depend on $q$, we have
\begin{equation}\label{eq3.6}
\max_{1\leq t\leq T}\left\Vert\frac{1}{T^{1/2}}\sum_{s=1}^t\xi_s-{\boldsymbol A}(1)\frac{1}{T^{1/2}}\sum_{s=1}^t e_s\right\Vert=o_P\left(q^{-(\kappa+1/2)}\right).
\end{equation}
With (\ref{eq3.5}) and (\ref{eq3.6}), we prove (\ref{eq3.4}) by setting $B_\xi(\cdot)={\boldsymbol A}(1)B_0(\cdot)$ and ${\boldsymbol\Sigma}_\xi={\boldsymbol A}(1){\boldsymbol A}(1)^{^\intercal}$. Assumption \ref{ass:1}(iv) allows $\nu_{1,0}$ to diverge to infinity and $\nu_{q,0}$ to converge to zero (at appropriate rates) as $q$ tends to infinity, since $\int_0^1 B_\xi(u) B_\xi(u)^{^\intercal}du$ is a $q\times q$ random matrix. This contrasts with the commonly-used assumption that the eigenvalues should be bounded away from zero and infinity when the number of factors is fixed \citep[e.g.,][]{BN02, B04, BLL21}. Assumption \ref{ass:1}(iv) further restricts the gaps between consecutive eigenvalues. When $q$ grows to infinity, the minimum eigenvalue gap is allowed to be $o_P(1)$. When $q$ is fixed, this restriction can be relaxed to the simple requirement of distinct eigenvalues with probability one.

(ii) Assumption \ref{ass:2}(ii) restricts the convergence rate of the series approximation error $\eta_{jt}$. Furthermore, a combination of Assumptions \ref{ass:1}(i) and \ref{ass:2}(ii) leads to the same convergence rate for $\chi_{it}^{\eta}$ in (\ref{eq2.6}), facilitating the asymptotic derivation. Assumption \ref{ass:2}(iii)(iv) contains some high-level moment conditions on the functional idiosyncratic components, indicating that $\varepsilon_{it}$ can be weakly cross-sectional dependent and temporally correlated. In particular, we allow for temporal heterogeneity on $\varepsilon_{it}$ when deriving the mean squared convergence in Proposition \ref{prop:3.1} below. These high-level conditions are comparable to those used in \cite{BN02} and \cite{B04}. 

%\red{[PCB: it seems also from Assumption 2(i) that you want to allow for temporal heterogeneity. That may cause issues because the cross section and temporal (CS+temp) covariance kernel will need to accommodate this. At least, we would need a requirement that the CS covariance kernel is homogeneous over t, which allows for CS heterogeneity and temporal heterogeneity that stabilizes as $T \to \infty$ arising from time series heterogeneity and stability in the (BM) component of the limiting two parameter Gaussian process. See \ref{eq:covkern} below. ]}\blue{[Reply: Sorry for the confusion. We allow for temporal heterogeneity on $\varepsilon_{it}$ when deriving the mean square convergence in Proposition \ref{prop:3.1}. However, a stronger temporal homogeneity condition is required for the asymptotic distribution result, say \eqref{eq:covkern}. I have given the clarification in the revised Assumption 4(ii). ]}

\end{remark}

Let ${\boldsymbol V}_{NT}$ be a $q\times q$ diagonal matrix with its diagonal elements being the $q$ largest eigenvalues of $\frac{1}{T^2}\widetilde{\boldsymbol\Omega}$ (arranged in the decreasing order), and define the following $q\times q$ rotation matrix
\begin{equation}\label{eq3.7}
{\boldsymbol H}_{NT}={\boldsymbol V}_{NT}^{-1}\left(\frac{1}{T^2}\widetilde{\boldsymbol G}^{^\intercal}{\boldsymbol G}\right)\left[\frac{1}{N}\sum_{i=1}^N \int_{u\in{\mathbb C}_i}\Lambda_i(u)\Lambda_i(u)^{^\intercal}du\right],
\end{equation}
where ${\boldsymbol G}=\left(G_1,\cdots,G_T\right)^{^\intercal}$. By Proposition \ref{prop:4.1} and Lemma \ref{le:B.5}, the limits as $\{N,T\} \to \infty$ of ${\boldsymbol V}_{NT}$ and ${\boldsymbol H}_{NT}$ are ${\boldsymbol V}_{0}$ and ${\boldsymbol H}_{0}$, respectively, where ${\boldsymbol V}_0$ is a $q\times q$ diagonal matrix with the diagonal elements being the eigenvalues of ${\boldsymbol\Sigma}_\Lambda^{1/2}(\int_0^1 B_\xi(u) B_\xi(u)^{^\intercal}du){\boldsymbol\Sigma}_\Lambda^{1/2}$ (arranged in decreasing order) and ${\boldsymbol H}_0={\boldsymbol V}_0^{-1/2}{\boldsymbol W}_0^{^\intercal}{\boldsymbol \Sigma}_\Lambda^{1/2}$ where ${\boldsymbol W}_0$ is a matrix consisting of the eigenvectors of ${\boldsymbol\Sigma}_\Lambda^{1/2}(\int_0^1 B_\xi(u) B_\xi(u)^{^\intercal}du){\boldsymbol\Sigma}_\Lambda^{1/2}$.
% \red{[PCB: It seems in the above that $q$ is taken as finite/fixed, although this was not mentioned. I have added that $q$ is finite and that limits are taken as $\{N,T\} \to \infty$ above to clarify and distinguish this from the following result where all of ${N,T,q}$ pass to infinity. The issue is whether sequential convergence is used with $q \to \infty$ after $\{N,T\} \to \infty$ or whether joint convergence holds. The following proposition provides conditions for joint convergence but it is not entirely clear whether this has been proved, i.e., whether the conditions for joint convergence in \cite{PM99} are satisfied. To be sure, we might add them to the assumptions.]}\blue{[Reply: In this paper we let $(N,T,q)$ to jointly pass to infinity, see (3.8), Assumption 3(i) and Assumption 4(i). We have clarified this joint divergence in some of the following assumptions. As a result, our mean square convergence rate in Proposition 3.1 depends on $N,T$ and $q$ and we state that $\left\Vert {\boldsymbol H}_{NT}-{\boldsymbol H}_0\right\Vert=o_P({\overline \nu}_q^{-1/2})$ in Lemma \ref{le:B.5}. For the asymptotic distribution properties in Theorems \ref{thm:3.2} and \ref{thm:3.3}, we use a projection (or rotation) matrix ${\boldsymbol R}$ to reduce the dimension from $q$ to $q_0$ (a fixed positive integer) and derive a sensible limit distribution in the low-dimensional setup via the joint limit approach (i,e, $\{N,T\} \to \infty$) as in \cite{PM99}.]} 
The following proposition derives the mean square convergence property for $\widetilde{G}_t$.

\smallskip

\renewcommand{\theprop}{3.\arabic{prop}}
\setcounter{prop}{0}

\begin{prop}\label{prop:3.1}

\em Suppose that Assumptions \ref{ass:1} and \ref{ass:2} are satisfied. If
\begin{equation}\label{eq3.8}
\underline\nu_{q}^{-2}q\left(T^{-2}+q{\overline\nu}_qN^{-1}+q{\overline\nu}_q\delta_q^2\right)=o(1),
\end{equation}
the following mean square convergence holds for the functional PCA estimate
\begin{equation}\label{eq3.9}
\frac{1}{T}\sum_{t=1}^T\left\Vert \widetilde G_t-{\boldsymbol H}_{NT} G_t\right\Vert^2=O_P\left(\underline\nu_{q}^{-2}q\left(T^{-2}+q{\overline\nu}_qN^{-1}+q{\overline\nu}_q\delta_q^2\right)\right).
\end{equation}

\end{prop}

\smallskip

\begin{remark}\label{re:2}

The mean square convergence rate in (\ref{eq3.9}) depends on $N,T$ and $q$, as the convergence property is derived in a high-dimensional dual functional factor model framework that allows the number of common trends to diverge slowly to infinity. In particular, the lower and upper orders of the eigenvalues $\underline\nu_q$ and $\overline\nu_q$, and the approximation rate to the low-rank structure $\delta_q$, all affect the mean square convergence rate. If the rate due to the series approximation error satisfies $\delta_q^2=O(N^{-1}\vee (q\overline\nu_qT^2)^{-1})$, the mean square convergence rate can be simplified to $\underline\nu_q^{-2}q(T^{-2}+q{\overline\nu}_qN^{-1})$. Furthermore, if $q$ is fixed and 
\begin{equation}\label{eq3.10}
0<\underline\nu\leq \underline\nu_q\leq\overline\nu_q\leq\overline\nu<\infty,
\end{equation} for some constants $\underline\nu$ and $\overline\nu$, 
the rate becomes $T^{-2}+N^{-1}$, the same as that in Lemma 1 of \cite{B04} which considers high-dimensional real-valued time series with common $I(1)$ trends. As $G_t$ is integrated, our rate is faster than the one derived in Proposition 4.1 of \cite{LLSX24}. In particular, when $\delta_q\equiv0$, $q$ is fixed and (\ref{eq3.10}) is satisfied, our rate is faster than the rate $T^{-1}+N^{-1}$ derived by \cite{TNH23b}.

\end{remark}

\subsection{Limit distribution of $\widetilde G_t$ and $\widetilde\Lambda_i$}\label{sec3.3}

To conduct inference on the estimated common trends and functional loadings, limit theory is needed, for which the following additional conditions are employed. 

\begin{assumption}\label{ass:3}

{\em (i) Let $N, T$ and $q$ diverge to infinity jointly and satisfy $\underline\nu_{q}^{-2}qNT^{-3}=o(1)$ and ${\overline\nu}_q^{1/2}\underline\nu_{q}^{-1}q\delta_{t,q}^\dagger=o(N^{-1/2})$ for each $t$, where $\delta_{t,q}^\dagger=\delta_{t,q}\vee \delta_q$. In addition,
\begin{equation}\label{eq3.11}
{\underline\nu}_{q}^{-2} q^{3}\left(T^{-1}+{\overline\nu}_{q}^{1/2}q^{1/2}N^{-1/2}+{\overline\nu}_{q}^{1/2}q^{1/2}\delta_q\right)=o(1).
\end{equation}
}

{\em (ii) For each $t$ and a $q_0\times q$ deterministic rotation matrix ${\boldsymbol R}$,
\begin{equation}\label{eq3.12}
{\boldsymbol R}{\boldsymbol\Psi}_t^{-1/2}\left(\frac{1}{\sqrt{N}}\sum_{i=1}^N \langle \Lambda_i,\varepsilon_{it}\rangle\right)\rightsquigarrow {\cal N}\left({\bf 0}, {\boldsymbol\Upsilon}\right)\quad{\it as\ N\rightarrow\infty},
\end{equation}
where $q_0 \le q$ is a fixed integer, and}
\[{\boldsymbol\Psi}_t=\lim_{N\rightarrow\infty}\frac{1}{N}\sum_{i=1}^N\sum_{j=1}^N {\sf E}\left[\left(\langle \Lambda_i,\varepsilon_{it}\rangle\right)\left(\langle \Lambda_j,\varepsilon_{jt}\rangle\right)^{^\intercal}\right], \quad \Vert{\boldsymbol\Psi}_t\Vert=O(1),\quad\text{\em and }\quad
{\boldsymbol R}{\boldsymbol R}^{^\intercal}\rightarrow {\boldsymbol\Upsilon}.
\]

{\em (iii) The following high-level convergence results hold 
\begin{equation}\label{eq3.13}
\max_{1\leq t\leq T}\Vert G_t\Vert=O_P\left((qT)^{1/2}\right),\quad\left\Vert \sum_{s=1}^T\sum_{i=1}^NG_s\langle \varepsilon_{is},\Lambda_i^{^\intercal}\rangle\right\Vert=O_P\left( N^{1/2}Tq\right).
\end{equation}
}

\end{assumption}

\begin{remark}\label{re:3}

Assumption \ref{ass:3}(i) implies that $N$ and $T$ diverge jointly to infinity but satisfying $N=o(T^3)$ and $\delta_{t,q}^\dagger$ decays to zero at a sufficiently fast rate. The condition (\ref{eq3.11}) slightly strengthens (\ref{eq3.8}) in Proposition \ref{prop:3.1}. The rotation matrix ${\boldsymbol R}$ in Assumption \ref{ass:3}(ii) is required as $q$ may be divergent \citep[e.g.,][]{LLL15}. If $q$ is fixed, (\ref{eq3.12}) can be replaced by 
\[
\frac{1}{\sqrt{N}}\sum_{i=1}^N \langle \Lambda_i,\varepsilon_{it}\rangle\rightsquigarrow {\cal N}\left({\bf 0}, {\boldsymbol\Psi}_t\right)\quad{\it as\ N\rightarrow\infty},
\]
which is the same as Assumption G in \cite{B04}. The first high-level condition in (\ref{eq3.13}) is implied by (\ref{eq3.4}) and its sufficiency is discussed in Remark \ref{re:1}. When $q$ is fixed, a sufficient condition for the second high-level condition in (\ref{eq3.13}) is
\[
\frac{1}{T^{1/2}}G_{\lfloor Tu\rfloor}\Rightarrow B_\xi(u),\quad \frac{1}{(NT)^{1/2}}\sum_{t=1}^{\lfloor Tu\rfloor}\sum_{i=1}^N\langle \varepsilon_{it},\Lambda_i\rangle\Rightarrow B_{\varepsilon\Lambda}(u),\quad 0\leq u\leq 1,
\]
where $B_\xi(\cdot)$ is defined in Assumption \ref{ass:1}(iii) and $B_{\varepsilon\Lambda}(\cdot)$ is a $q$-vector Brownian motion independent of $B_\xi(\cdot)$. 

\end{remark}

\smallskip

\renewcommand{\thetheorem}{3.\arabic{theorem}}
\setcounter{theorem}{1}

\begin{theorem}\label{thm:3.2}

Suppose that Assumptions \ref{ass:1}--\ref{ass:3} are satisfied. The following asymptotic distribution theory holds for the functional PCA estimate $\widetilde G_t$
\begin{equation}\label{eq3.14}
\sqrt{N}{\boldsymbol R}{\boldsymbol\Psi}_t^{-1/2}{\mathbf Q}_{NT}^{-1} \left(\widetilde G_t-{\boldsymbol H}_{NT} G_t\right)\rightsquigarrow {\cal N}\left({\bf 0}, {\boldsymbol\Upsilon}\right),
\end{equation}
where ${\mathbf Q}_{NT}={\boldsymbol V}_{NT}^{-1}\left(\frac{1}{T^2}\sum_{s=1}^{T}\widetilde{G}_s G_s^{^\intercal}\right)$ satisfying
\begin{equation}\label{eq3.15}
\left\Vert {\mathbf Q}_{NT}-{\mathbf Q}_0\right\Vert=o_P(1),\quad {\mathbf Q}_0={\boldsymbol V}_0^{-1/2}{\boldsymbol W}_0^{^\intercal}{\boldsymbol\Sigma}_\Lambda^{-1/2}.
\end{equation}
\end{theorem}

\smallskip

\begin{remark}\label{re:4}

The asymptotic normal distribution in (\ref{eq3.14}) is derived via the joint limit approach \citep[e.g.,][]{PM99}, letting $N$ and $T$ tend to infinity jointly. Theorem \ref{thm:3.2} is more general than Theorem 2 in \cite{B04} as we allow $q$ to diverge slowly to infinity. When $q$ is fixed, we set ${\boldsymbol R}$ as an identity matrix and write the limit distribution theory as
\[
\sqrt{N}\left(\widetilde G_t-{\boldsymbol H}_{NT} G_t\right)\rightsquigarrow {\mathbf Q}_0\cdot{\cal N}\left({\bf 0}, {\boldsymbol\Psi}_t\right),
\]
where ${\mathbf Q}_0$ is independent of the normal vector ${\cal N}\left({\bf 0}, {\boldsymbol\Psi}_t\right)$, giving stable convergence to a mixed normal limit \citep[e.g.,][]{HH80}. 

\end{remark}

We next turn to the limit distribution theory of the functional factor loading estimate $\widetilde\Lambda_i$, which requires the following conditions.

\begin{assumption}\label{ass:4}

{\em (i) Let $N, T$ and $q$ tend to infinity jointly, satisfying 
\[({\overline\nu}_q/{\underline\nu}_{q})^2 {\overline\nu}_q^{1/2}qT^{1/2}\delta_{q}=o(1),\quad 
({\overline\nu}_q/{\underline\nu}_{q})^2q^{1/2}\left[T^{-1/2}+{\overline\nu}_q^{1/2}q^{1/2}(T/N)^{1/2}\right]=o(1).\]
}

{\em (ii) For each $i$, $\{\varepsilon_{it}\}$ is a sequence of stationary random functions (over $t$), and there exists a $q_0\times q$ deterministic rotation matrix ${\boldsymbol R}$ such that
\begin{equation}\label{eq3.16}
{\boldsymbol R}\left(\frac{1}{T}\sum_{t=1}^T G_t \varepsilon_{it}(u)\right)\rightsquigarrow \int_0^1 B_{\xi}^R(r) dB_{\varepsilon}^{(i)}(r,u)\quad{\it as}\ T\rightarrow\infty,
\end{equation}
where $B_{\xi}^R(\cdot)$ is a $q_0$-dimensional vector Brownian motion with variance matrix ${\boldsymbol R}{\boldsymbol\Sigma}_\xi{\boldsymbol R}^{^\intercal}$ and $B_{\varepsilon}^{(i)}(r,u)$ is a two parameter Gaussian process in $C[0,1]\times{\mathscr H}_i$ with arguments $r \in [0,1], u \in  {\mathbb C}_i$ and covariance kernel function 
\begin{equation}\label{eq:covkern}
{\sf E}\left[B_{\varepsilon}^{(i)}(r,u)B_{\varepsilon}^{(i)}(s,v)\right]=(r\wedge s) \;\sigma_\varepsilon^{(i)}(u,v),
\end{equation}
$\sigma_\varepsilon^{(i)}(u,v)=\sum_{h=-\infty}^{\infty}\lim_{T\rightarrow\infty}\frac1{T}\sum_{t=1}^T{\sf E}[\varepsilon_{it}(u)\varepsilon_{it+h}(v)]$.}

%\red{[PCB: note the changes above. We need to allow for temporal heterogeneity that stabilizes upon summation so that the limiting Gaussian process $B_{\varepsilon}^{(i)}(r,u)$ has the usual BM covariance kernel $r\wedge s$ over limiting temporal coordinates $(r,s)$ and that this is scaled by the cross section covariance kernel $\sigma_\varepsilon^{(i)}(u,v)$ over spatial coordinates $(u,v)$.]} \blue{[Reply: Peter, following your comment, we have added ``$\{\varepsilon_{it}\}$ is a sequence of stationary random functions (over $t$)" in the assumption. The above covariance kernel function is slightly amended as $\varepsilon_{it}(r,u)$ is not defined.]}  

\end{assumption}

\begin{remark}\label{re:5}

Assumption \ref{ass:4}(i) can be removed if $q$ is fixed and $\delta_q\equiv0$. Assumption \ref{ass:4}(ii) requires weak convergence of a sample covariance of a nonstationary time series ($G_t$) with a curve time series function ($\varepsilon_{it}(\cdot)$) to the stochastic integral on the right side of \eqref{eq3.16}. Functional limit theory of this type involves several components: (a) weak convergence of the partial sum curve process $\frac1{\sqrt T}\sum_{t=1}^{\lfloor Tr\rfloor} \varepsilon_{it}(u) \rightsquigarrow B_{i}(r,u)$, a two-parameter Gaussian process with covariance kernel \eqref{eq:covkern}, which is shown in \citet[Lemma A]{PJ25b}; (b) weak convergence of the standardized partial sum $\frac1{\sqrt T}\sum_{t=1}^{\lfloor Tr\rfloor} G_t \rightsquigarrow B_{\xi}(r)$, which follows by conventional limit theory; and (c) convergence to the stochastic integral limit form in \eqref{eq3.16}, which can be justified as in the proof of \citet[Theorem 1]{PJ25b} and by using  martingale convergence methods as in \citet{IP08}. 

%\red{[PCB: Note the changes above: (i) to accommodate a two parameter Gaussian process limit; (ii) to avoid the double usage of the index $u$ as an index over $[0,1]$ and as an index of a curve in ${\mathscr H}_i$; (iii) references to my UR paper where a version of the result is proved; and (iv) the complexities of proving weak convergence to a stochastic integral limit involving the process $B_{\varepsilon}^{(i)}(u)$. The stochastic integral is simpler than in \citet[Lemma A]{PJ25b} because $B_{\xi}^R(u)$ is simple single parameter vector BM process. But the proof does not follow by simple continuous mapping because the stochastic integral cannot be represented in terms of simple partial sums as there are two different components in the stochastic integral. There is no formal proof for this type of stochastic integral in \citet[Lemma A]{PJ25b} but a rigorous development can be obtained using martingale convergence methods as in that paper and in \citet{IP08}. So, I have referenced both.]}  \blue{[Reply: Peter, many thanks for adding the two references and making the remark more rigorous.]} 

\end{remark}

\begin{theorem}\label{thm:3.3}

Suppose that Assumptions \ref{ass:1}, \ref{ass:2}, and \ref{ass:4} are satisfied. The following limit theory holds for the functional factor loading estimate $\widetilde \Lambda_i$
\begin{equation}\label{eq3.17}
T\left({\boldsymbol R}{\boldsymbol H}_{NT}^{-1}\right) \left[\widetilde \Lambda_i(u)-({\boldsymbol H}_{NT}^{-1})^{^\intercal} \Lambda_i(u)\right]\rightsquigarrow \int_0^1 B_{\xi}^R(r) dB_{\varepsilon}^{(i)}(r,u).
\end{equation}

\end{theorem}

%\red{[PCB: Corrections made here on the indices are due to similar issues of notation and limit theory as those in the above. We might also standardize the weak convergence notation to $\rightsquigarrow$ (as weak convergence in the relevant probability space), which I think is more elegant notation.]} \blue{[Reply: All sounds good, thanks Peter.]}

\begin{remark}\label{re:6}

The normalization rate $T$ in (\ref{eq3.17}) shows that super-fast convergence is achieved for the functional factor loading estimates, since the common factors $G_t$ are integrated. As is shown in the proof, it further follows that  
\[
T\left({\boldsymbol R}{\boldsymbol H}_{NT}^{-1}\right) \left[\widetilde \Lambda_i-({\boldsymbol H}_{NT}^{-1})^{^\intercal} \Lambda_i\right]={\boldsymbol R}\left(\frac{1}{T}\sum_{t=1}^T\varepsilon_{it} G_t\right)+o_P(1),
\]
indicating that the limit distribution is derived as if the common stochastic trends were known (ignoring the rotation matrix ${\boldsymbol H}_{NT}^{-1}$).

\end{remark}

%%%%%%%%%%%%%%%%%%%

\section{Estimation of the factor number}\label{sec4}
\renewcommand{\theequation}{4.\arabic{equation}}
\setcounter{equation}{0}

In practice, the number of latent nonstationary factors is unknown. Implementation of the functional PCA proposed in Section \ref{sec3} requires a consistent estimation of $q$. There have been extensive studies on determining the number of factors in the conventional factor model for real-valued time series. \cite{BN02} and \cite{B04} propose some information criteria to consistently estimate the factor number for a large panel of stationary and nonstationary time series; \cite{LY12} and \cite{AH13} recommend an easy-to-implement ratio criterion where ratios of consecutive estimated eigenvalues are compared; \cite{T18} and \cite{BT22} estimate factor numbers by randomised sequential testing using estimated eigenvalues. For the present setting, we here employ an easy-to-implement information criterion to consistently estimate the number of common stochastic trends. 

\smallskip

Let $\widetilde\nu_i=\nu_i(\widetilde{\boldsymbol\Omega}/T^2)$ denote the $i$-th eigenvalue of $\widetilde{\boldsymbol\Omega}/T^2$. We start with the following proposition on the asymptotic orders of $\widetilde\nu_i$, which motivate the selection criterion.

\smallskip

\renewcommand{\theprop}{4.\arabic{prop}}
\setcounter{prop}{0}

\begin{prop}\label{prop:4.1}

{\em Suppose that Assumptions \ref{ass:1} and \ref{ass:2} are satisfied. As $N, T, q \to \infty$ jointly,
{\small
\begin{eqnarray}{}
\vert\widetilde\nu_i-\nu_{i,0}\vert&=&O_P\left({\overline \nu}_q^{1/2}q^{1/2}T^{-1/2}\left(N^{-1/2}+\delta_q\right)+T^{-3/2}\right)+o_P\left(q^{-\kappa}\overline{\nu}_q\right),\ \ \ 1\leq i\leq q,\label{eq4.1}\\
\widetilde\nu_i&=&O_P\left({\overline \nu}_q^{1/2}q^{1/2}T^{-1/2}\left(N^{-1/2}+\delta_q\right)+T^{-3/2}\right),\ \ \ q+1\leq i\leq N\wedge T.\label{eq4.2}
\end{eqnarray}
}}

\end{prop}

\smallskip

\begin{remark}\label{re:7}

This proposition shows that the gap between $\widetilde\nu_q$ and $\widetilde\nu_{q+1}$ is strictly larger than $\nu_{q,0}/2$ {\em w.p.a.1} if ${\overline \nu}_q^{1/2}q^{1/2}T^{-1/2}\left(N^{-1/2}+\delta_q\right)+T^{-3/2}=o(\underline\nu_{q})$ and $q^{-\kappa}\overline\nu_q=O(\underline\nu_{q})$, which are implied by (\ref{eq3.8}) and Assumption \ref{ass:1}(iv). These conditions allow for feasible consistent estimation of the latent factor number. 

\end{remark}

\smallskip

In particular, define the criterion
\begin{equation}\label{eq4.3}
\widetilde q=\argmin_{1\leq j\leq q_{\max}}\left\{\widetilde\nu_j+j\cdot \rho_{NT}\right\}-1,
\end{equation}
where the penalty parameter $\rho_{NT}$ satisfies some mild restrictions (see Theorem \ref{thm:4.2}) and $q_{\max}$ is a user-specified upper bound of the factor number. This criterion was used in \cite{AX17} in the context of factor models for high-dimensional and high-frequency financial data. It can be viewed as a modification of the information criterion proposed in \cite{BN02} and \cite{B04}, which replaces $\widetilde\nu_j$ by summation of $\widetilde\nu_i$ over $i>j$ and which does not require a ``-1" adjustment. The modified information criterion (\ref{eq4.2}) is easier to implement and proof of consistency is simpler. 

\smallskip

\renewcommand{\thetheorem}{4.\arabic{theorem}}
\setcounter{theorem}{1}

\begin{theorem}\label{thm:4.2}

Suppose that Assumptions \ref{ass:1}, \ref{ass:2} and (\ref{eq3.8}) are satisfied.  In addition, 
 the tuning parameter $\rho_{NT}$ satisfies that 
\begin{equation}\label{eq4.4}
\rho_{NT}=o(\underline\nu_{q}),\quad {\overline \nu}_q^{1/2}q^{1/2}T^{-1/2}\left(N^{-1/2}+\delta_q\right)+T^{-3/2}=o(\rho_{NT}).
\end{equation}
Then we have ${\sf P}\left(\widetilde{q}= q\right)\rightarrow1$. 

\end{theorem}

\begin{remark}\label{re:8}

Condition (\ref{eq4.4}) is crucial to ensure selection consistency. When $q$ is fixed, $\delta_q\equiv0$ and (\ref{eq3.10}) is satisfied, it can be simplified to $\rho_{NT}\rightarrow0$ and $N^{-1}+T^{-1}=o(\rho_{NT})$, which is slightly weaker than the restriction in \cite{B04} and \cite{TNH23b}. %It is worthwhile pointing out that the consistency property holds when $q$ is zero, positive but fixed, or divergent at a slow rate.

\end{remark}

%%%%%%%%%%%%%%%%%%%

\section{Estimation with cointegrated factors}\label{sec5}
\renewcommand{\theequation}{5.\arabic{equation}}
\setcounter{equation}{0}

This section considers the case where $G_t$ is cointegrated, i.e., ${\boldsymbol\Sigma}_\xi$ has reduced rank $q^\dagger$ with $1\leq q^\dagger\leq q-1$. Following \cite{PP88, PP89}, there exists a $q\times q$ orthogonal matrix ${\boldsymbol P}=({\boldsymbol P}_1, {\boldsymbol P}_2)$ with ${\boldsymbol P}_1$ and ${\boldsymbol P}_2$ dimensioned $q\times q^\dagger$ and $q\times q^{\ddag}$, such that
\begin{equation}\label{eq5.1}
{\boldsymbol P}^{^\intercal}G_t=\left(
\begin{array}{c}
{\boldsymbol P}_1^{^\intercal}G_t\\
{\boldsymbol P}_2^{^\intercal}G_t
\end{array}
\right)=\left(
\begin{array}{c}
G_{t1}\\
G_{t2}
\end{array}
\right)=G_t^\dagger,
\end{equation}
where $q^{\ddag}=q-q^\dagger$ is called the cointegrating rank, $G_{t1}$ is a $q^\dagger$-dimensional vector of full-rank integrated variables and $G_{t2}$ is a $q^{\ddag}$-dimensional vector of stationary time series. The rotation (\ref{eq5.1}) successfully separates out stationary and nonstationary components, the latter of which drive the common stochastic trends for large-scale curve time series. We may use the functional PCA  method as in Section \ref{sec3.1} to estimate both the integrated factors $G_{t1}$ and stationary factors $G_{t2}$ \citep[e.g.,][]{B04}. However, it would require consistent estimation of $q$ and $q^\dag$ (or $q^{\ddag}$), and different normalization rates for $G_{t1}$ and $G_{t2}$. In this section, we propose a different approach which is a functional version of the PANIC method. PANIC was introduced by \cite{BN04} for high-dimensional real-valued time series, allowing some idiosyncratic components to be nonstationary \citep[e.g.,][]{BLL21}. %In fact, the asymptotic theorems in Sections \ref{sec3.2} and \ref{sec3.3} would be invalid when the functional idiosyncratic components $\varepsilon_{it}$ are not stationary over $t$.

\smallskip

As in \cite{BN04}, taking differences on both sides of (\ref{eq2.6}) gives
\begin{equation}\label{eq5.2}
z_{it}=\Lambda_{i}^{^\intercal}\xi_{t}+\varepsilon_{it}^\dagger,\ \ i=1,\cdots,N,\ \ t=2,\cdots,T,
\end{equation} 
where $z_{it}=\Delta Z_{it}$, $\xi_t=\Delta G_t$ and $\varepsilon_{it}^\dagger=\Delta(\chi_{it}^{\eta}+\varepsilon_{it})$. Model (\ref{eq5.2}) can be seen as a functional factor model for high-dimensional stationary curve time series \citep[e.g.,][]{LLSX24}. However, weak cross-section dependence of $\varepsilon_{it}^\dagger$ over $i$ may not be satisfied due to the presence of series approximation errors in the functional common components. Throughout this section, we only require $\Delta\varepsilon_{it}$ to be stationary over $t$, which implies that $\varepsilon_{it}$ may be integrated.

\smallskip

We next estimate the functional factor loadings $\Lambda_i$ and stationary factor vector $\xi_t$. Since $\Lambda_i$ and $\xi_t$ are not identifiable in the functional factor model (\ref{eq5.2}), we employ the following identification restrictions in the functional PCA algorithm
\begin{equation}\label{eq5.3}
\frac{1}{T-1}\sum_{t=2}^T\xi_t\xi_t^{^\intercal}={\boldsymbol I}_{q}\ \ \ \text{and}\ \ \ \frac{1}{N}\sum_{i=1}^N \int_{u\in{\mathbb C}_{i}}\Lambda_i(u)\Lambda_i(u)^{^\intercal}du\ \ {\rm is\ diagonal}.
\end{equation}
Unlike (\ref{eq3.1}), the adjusted normalization rate $T-1$ is used for the stationary factors $\xi_t$. Define
\begin{equation}\label{eq5.4}
\widehat{\boldsymbol\Omega}=\left(\widehat\Omega_{ts}\right)_{(T-1)\times (T-1)}\ \ \text{with}\ \ \widehat\Omega_{ts}=\frac{1}{N}\sum_{i=1}^N \int_{u\in{\mathbb C}_{i}}z_{it}(u)z_{is}(u)du.
\end{equation}
Conducting the eigenanalysis of $\widehat{\boldsymbol\Omega}$, we obtain $\widehat {\boldsymbol \xi}=(\widehat \xi_2,\cdots,\widehat \xi_T)^{^\intercal}$ as a matrix consisting of the eigenvectors scaled by $\sqrt{T-1}$, corresponding to the $q$ largest eigenvalues of $\widehat{\boldsymbol\Omega}$. It follows from Proposition 4.1 in \cite{LLSX24} that, under some mild conditions,
\begin{equation}\label{eq5.5}
\frac{1}{T-1}\sum_{t=2}^T\left\Vert \widehat \xi_t-{\boldsymbol H}_{NT}^\dagger \xi_t\right\Vert^2=O_P\left(q\left(T^{-1}+q^2N^{-1}+q^2\delta_q^2\right)\right),
\end{equation}
where
\[
{\boldsymbol H}_{NT}^\dagger=\left({\boldsymbol V}_{NT}^{\dagger}\right)^{-1}\left(\frac{1}{T-1}\widehat{\boldsymbol \xi}^{^\intercal}{\boldsymbol \xi}\right)\left[\frac{1}{N}\sum_{i=1}^N \int_{u\in{\mathbb C}_i}\Lambda_i(u)\Lambda_i(u)^{^\intercal}du\right],
\]
 ${\boldsymbol V}_{NT}^\dagger={\sf diag}\{\widehat{\lambda}_1,\cdots,\widehat{\lambda}_q\}$ with $\widehat\lambda_j$ being the $j$-th largest eigenvalue of $\frac{1}{T-1}\widehat{\boldsymbol\Omega}$, and ${\boldsymbol\xi}=\left(\xi_2,\cdots,\xi_T\right)^{^\intercal}$. Evidently, the mean square convergence rate in (\ref{eq5.5}) is slower than that of (\ref{eq3.9}). Using the first restriction in (\ref{eq5.3}), the factor loading functions are estimated as
\begin{equation}\label{eq5.6}
\widehat\Lambda_i=\left(\widehat\Lambda_i(u): u\in{\mathbb C}_{i}\right)=\frac{1}{T-1}\sum_{t=2}^Tz_{it}\widehat\xi_t,\ \ i=1,\cdots,N,
\end{equation}
via least squares. Furthermore, we can estimate the (original) cointegrated factors by
\begin{equation}\label{eq5.7}
\widehat G_t=\sum_{s=2}^t \widehat \xi_s,\quad t=2,\cdots,T.
\end{equation} 

\smallskip 

We next discuss estimation of $q$ and $q^\ddag$. The information criterion (\ref{eq4.2}) needs modification to consistently estimate $q$. Specifically, we define
\begin{equation}\label{eq5.8}
\widehat q=\argmin_{1\leq j\leq q_{\max}}\left\{\widehat\lambda_j+j\cdot \rho_{NT}^\dagger\right\}-1,
\end{equation}
 where $\rho_{NT}^\dagger$ satisfies
\[
\rho_{NT}^\dagger\rightarrow0,\quad q\left(N^{-1/2}+\delta_q\right)+T^{-1/2}=o\left(\rho_{NT}^\dagger\right),
\]
which differ from (\ref{eq4.3}) in Theorem \ref{thm:4.2}. It follows from Proposition 5.1 in \cite{LLSX24} that ${\sf P}(\widehat q=q)\rightarrow1$. To estimate the cointegrating rank $q^\ddag$, we adopt the information criterion introduced by \cite{CP09, CP12} and modified for a curve time series context in \cite{P25}, which is robust to weak dependence and time-varying variances in the errors. Assume the following VECM structure:
\begin{equation}\label{eq5.9}
\xi_t=\Delta G_t=\alpha_0\beta_0^{^\intercal} G_{t-1}+v_t, 
\end{equation}
where $\alpha_0$ and $\beta_0$ are two $q\times q^{\ddag}$ matrices, and $v_t$ is stationary satisfying the conditions in \cite{CP09} or heterogeneously distributed as assumed in \cite{CP12}. For each $j = 1, \cdots, \widehat q-1$, we estimate the $\widehat{q}\times j$ matrices $\alpha_0$ and $\beta_0$ via reduced-rank regression (RRR), giving $\widehat\alpha(j)$ and $\widehat\beta(j)$, and subsequently define 
\[
\widehat{\boldsymbol\Sigma}(j)=\frac{1}{T-1}\sum_{t=2}^T\left[\widehat\xi_t-\widehat\alpha(j)\widehat\beta(j)^{^\intercal} \widehat G_t\right]\left[\widehat\xi_t-\widehat\alpha(j)\widehat\beta(j)^{^\intercal} \widehat G_t\right]^{^\intercal},
\]
as the residual covariance matrix, with $\widehat{\boldsymbol\Sigma}(0)=\frac{1}{T-1}\sum_{t=2}^T\widehat\xi_t\widehat\xi_t^{^\intercal}$. Cointegrating rank is selected as
\begin{equation}\label{eq5.10}
\widehat q^{\ddag} = \argmin_{0\leq j\leq \widehat q-1}\ \left\{\log\left({\sf det}(\widehat{\boldsymbol\Sigma}(j))\right) +\frac{\rho_{T}^{\ddagger}}{T}\left(2\widehat q j-j^2\right)\right\},
\end{equation}
with $\rho_{T}^{\ddagger}=\log T$ corresponding to the Bayesian information criterion (BIC) and $\rho_{T}^{\ddagger}=2\log \log T$ corresponding to the HQ criterion \citep{HQ79}. Limit theory and consistency for these criteria, as well as the inconsistency of the related AIC criterion, are provided in \cite{P25}.

%%%%%%%%%%%%%%%%%%%

\section{Simulations}\label{sec6}
\renewcommand{\theequation}{6.\arabic{equation}}
\setcounter{equation}{0}

This section reports the findings of two simulation studies designed to examine the finite-sample performance of our proposed methods. In each example, we first assess the estimation performance given that the number of common stochastic trends (or the cointegrating rank) is known and then examine the performance of various information criteria defined in Sections \ref{sec4} and \ref{sec5}. To quantify the assessment of functional PCA, we compute the approximation errors for factors and factor loadings as follows
\begin{eqnarray}
&&{\sf AE}(\widetilde G)=\min_{{\boldsymbol H}\in {\mathbb R}^{q\times q}}\frac{1}{qT}\sum_{t=1}^T\left\Vert \widetilde G_t-{\boldsymbol H} G_t\right\Vert^2,\notag\\
&&{\sf AE}(\widetilde \xi)=\min_{{\boldsymbol H}\in {\mathbb R}^{q\times q}}\frac{1}{q(T-1)}\sum_{t=2}^T\left\Vert \widetilde\xi_t-{\boldsymbol H} \xi_t\right\Vert^2,\notag\\
&&{\sf AE}(\widetilde\Lambda)=\min_{{\boldsymbol H}\in {\mathbb R}^{q\times q}}\frac{1}{qN}\sum_{i=1}^N\left\Vert \widetilde \Lambda_i-{\boldsymbol H} \Lambda_i\right\Vert^2,\notag
\end{eqnarray} 
where $\widetilde\xi_t=\widetilde G_t-\widetilde G_{t-1}$ for $t=2,\cdots,T$. Similarly, for functional PANIC estimation the measurements are defined as
\begin{eqnarray}
&&{\sf AE}(\widehat G)=\min_{{\boldsymbol H}\in {\mathbb R}^{q\times q}}\frac{1}{q(T-1)}\sum_{t=2}^T\left\Vert \widehat G_t-{\boldsymbol H} (G_t-G_1)\right\Vert^2,\notag\\
&&{\sf AE}(\widehat\xi)=\min_{{\boldsymbol H}\in {\mathbb R}^{q\times q}}\frac{1}{q(T-1)}\sum_{t=2}^T\left\Vert \widehat\xi_t-{\boldsymbol H} \xi_t\right\Vert^2,\notag\\
&&{\sf AE}(\widehat\Lambda)=\min_{{\boldsymbol H}\in {\mathbb R}^{q\times q}}\frac{1}{qN}\sum_{i=1}^N\left\Vert \widehat \Lambda_i-{\boldsymbol H} \Lambda_i\right\Vert^2.\notag
\end{eqnarray} 

\medskip

\noindent {\bf Example 6.1}.\ \ Consider $G_t=\sum_{s=1}^t\xi_s$ with $\xi_t$ following a VAR(1) model given by
\[
\xi_t = \mathbf{A}\xi_{t-1} +\epsilon^{\xi}_t,
\]
where $\mathbf{A}$ is a $q\times q$ diagonal companion matrix and $\epsilon^{\xi}_t$'s denote the innovations. As in \cite{TNH23b}, the diagonal entries of $\mathbf{A}$ were randomly drawn from a uniform distribution ${\cal U}[-1,1]$ and the matrix rescaled to have operator norm $0.8$. The innovations were independently drawn from a $q$-variate standard normal distribution. The initial 100 observations of the VAR(1) process $\{\xi_t\}$ were discarded to ensure data stability and independence of initial conditions. Letting $\phi_1,\cdots,\phi_{51}$ be $51$ orthonormal basis functions on $[0,1]$, we generated 
\[
\eta_{t1}(u)=\sum_{j=1}^{51}b^{\eta}_{tj}\phi_j(u)/j^2,\quad t=1,\cdots,T,
\]
where 
\[
b^{\eta}_{tj}=\frac{1}{\sqrt{T}}\left[\sum_{s=1}^t\epsilon_{tj}^{\eta}-\left(\frac{t}{T}\right)\sum_{s=1}^T\epsilon_{tj}^{\eta}\right]
\] 
and the $\epsilon_{tj}^{\eta}$'s were independently drawn from the standard normal distribution. Latent factor curves were generated via \eqref{eq2.3} with $k=1$ and ${\Phi}_1(u)=[\phi_1(u),\cdots,\phi_q(u)]^{^\intercal}$, i.e.,
\begin{equation}\label{eq6.1}
{F}_{t1}(u) = {\Phi}_1(u)^{^\intercal} {G}_t + \frac{{\eta}_{t1}(u)}{q},
\end{equation}
where the factor $1/q$ in the series approximation errors serves the purpose of enhancing the signal-to-noise ratio. Factor loading functions were simulated as
\begin{equation}\label{eq6.2}
 B_{i1}(u, v) = \sum_{j_1=1}^{51}\sum_{j_2=1}^{51}b_{i,j_1j_2}\phi_{j_1}(u)\phi_{j_2}(v)/(j_1-j_2+1)^2,
\end{equation}
where the $b_{i,j_1j_2}$'s were independently generated from the uniform distribution ${\cal U}[0,3]$ over $i$, $j_1$ and $j_2$.

\smallskip

The functional idiosyncratic components $\varepsilon_{it}(u)$ were generated by 
\begin{equation}\label{eq6.3}
\varepsilon_{it}(u)=\sum_{j=1}^{51}b^{\varepsilon}_{it,j}\phi_j(u),
\end{equation}
where, for each $t$, $\mathbf{b}_t^{\varepsilon}=(b^{\varepsilon}_{1t,1},b^{\varepsilon}_{1t,2},\cdots,b^{\varepsilon}_{Nt,51})^{^\intercal}\in {\mathbb R}^{51N}$ were independently drawn from ${\cal N}({\bf 0},\ {\boldsymbol\Sigma}_{b})$ with $\ {\boldsymbol\Sigma}_{b}$ a block covariance matrix with  ($i,j$)-block
\[
\Sigma_{b,ij}=\max\{0,(1-|i-j|/10)\}\cdot {\sf diag}(1^{-2},\cdots,51^{-2}),\quad 1\leq i,j\leq N.
\]

\smallskip

Combining \eqref{eq6.1}--\eqref{eq6.3}, the nonstationary curve observations $Z_{it}$ were generated as 
$$
\begin{aligned}
Z_{it}(u) & =\int_0^1 B_{i 1}(u, v) F_{t 1}(v) \mathrm{d} v+\varepsilon_{i t}(u) \\
& =\int_0^1 B_{i 1}(u, v)\left[\Phi_1(v)^{^\intercal} G_{t}+\frac{\eta_{t 1}(v)}{q}\right] \mathrm{d} v+\varepsilon_{i t}(u) \\
& =\Lambda_i(u)^{^\intercal}G_{t}+\frac{1}{q} \int_0^1 B_{i 1}(u, v) \eta_{t 1}(v) \mathrm{d} v+\varepsilon_{i t}(u), 
\end{aligned}
$$
where 
\begin{equation}\label{eq6.4}
\Lambda_i(u)=\int_0^1 B_{i 1}(u, v) \Phi_1(v) \mathrm{d} v=\sum_{j_1=1}^{51}\phi_{j_1}(u)\left[
\frac{b_{i,j_11}}{j_1^2},\ \frac{b_{i,j_12}}{(j_1-1)^2},\cdots, \frac{b_{i,j_1q}}{(j_1-q+1)^2}\right]^{^\intercal}.
\end{equation}

\smallskip

Table \ref{tab1} reports the logarithms of the approximation errors for common stochastic trends obtained by functional PCA and PANIC, i.e., $\log({\sf AE}(\widetilde G))$ and $\log({\sf AE}(\widehat G))$, respectively. In Table \ref{tab1a} for functional PCA estimation performance, a consistent reduction in the approximation errors is observed with an increase in $N$ across all values of $T$ and $q$. A slight increase in the approximation errors is noted with an increase in $T$ but the magnitude of increase is insignificant compared with standard deviations. Focusing on the three diagonal cases in each block of the table, i.e., ($N=100$, $T=200$), ($N=200$, $T=300$), and ($N=300$, $T=400$), we observe that the approximation errors diminish as both $T$ and $N$ approach infinity, confirming the joint convergence of functional PCA (see Proposition \ref{prop:3.1}). In addition, as $q$ increases, the logarithms of the approximation errors decrease generally, which may be partly attributed to dominant loadings on the first few basis functions in the definitions of $\varepsilon_{i t}(u)$ and $\Lambda_i(u)$, see \eqref{eq6.3} and \eqref{eq6.4}, and reduction of the sieve approximation errors (when $q$ increases). The results of functional PANIC follow a similar pattern, as evident in Table \ref{tab1b}. The approximation errors decrease as $N$ increases, but are insensitive to increasing $T$. Although the functional PANIC estimates converge when $N$ and  $T$ jointly diverge to infinity, they generally exhibit higher approximation errors than functional PCA, indicating that the latter is more efficient when $G_t$ is of full rank.

\begin{table}
\scriptsize
\centering 
\begin{subtable}{.4\textwidth}
\caption{Functional PCA}
\label{tab1a}
\begin{tabular}{cc|ccc}
\hline
$N$ &  $q$ & $T=200$ &$T=300$& $T=400$ \\
\hline
100 & 5 &-4.140 &-4.129 &-4.125\\
&&(0.076) &(0.063) &(0.055)\\
200 & 5 &-4.795 &-4.787 &-4.784\\
&&(0.080) &(0.065) &(0.057)\\
300 & 5 &-5.127 &-5.118 &-5.114\\
&&(0.085) &(0.071) &(0.064)\\\hline
100 & 10 &-4.810 &-4.794 &-4.788\\
&&(0.073) &(0.058) &(0.050)\\
200 & 10 &-5.485 &-5.471 &-5.463\\
&&(0.073) &(0.059) &(0.051)\\
300 & 10 &-5.833 &-5.820 &-5.813\\
&&(0.072) &(0.058) &(0.051)\\\hline
100 & 15 &-5.212 &-5.196 &-5.186\\
&&(0.068) &(0.055) &(0.048)\\
200 & 15 &-5.890 &-5.873 &-5.864\\
&&(0.072) &(0.057) &(0.048)\\
300 & 15 &-6.243 &-6.226 &-6.217\\
&&(0.072) &(0.058) &(0.050)\\\hline
\hline
\end{tabular}
\end{subtable}\quad\quad 
\begin{subtable}{.4\textwidth}
\caption{Functional PANIC}
\label{tab1b}
\begin{tabular}{cc|ccc}
\hline
$N$ &  $q$ & $T=200$ &$T=300$& $T=400$ \\
\hline
100 & 5 &-3.948 &-3.957 &-3.961\\
&&(0.139) &(0.130) &(0.127)\\
200 & 5 &-4.648 &-4.662 &-4.668\\
&&(0.127) &(0.119) &(0.116)\\
300 & 5 &-4.994 &-5.013 &-5.020\\
&&(0.125) &(0.116) &(0.113)\\\hline
100 & 10 &-4.622 &-4.638 &-4.645\\
&&(0.094) &(0.083) &(0.076)\\
200 & 10 &-5.322 &-5.341 &-5.349\\
&&(0.086) &(0.076) &(0.071)\\
300 & 10 &-5.670 &-5.694 &-5.705\\
&&(0.088) &(0.077) &(0.074)\\\hline
100 & 15 &-4.990 &-5.019 &-5.030\\
&&(0.090) &(0.082) &(0.076)\\
200 & 15 &-5.688 &-5.720 &-5.733\\
&&(0.090) &(0.082) &(0.077)\\
300 & 15 &-6.042 &-6.078 &-6.092\\
&&(0.090) &(0.082) &(0.079)\\\hline
\hline
\end{tabular}
\end{subtable}
\caption{Logarithms of the approximation errors for common stochastic trends by the functional PCA and PANIC, i.e., $\log({\sf AE}(\widetilde G))$ and $\log({\sf AE}(\widehat G))$, respectively, averaged over 1000 replications in Example 6.1. The standard deviations are reported in parentheses.}
\label{tab1}
\end{table}

Table \ref{tab2} reports logarithms of approximation errors for factor loading functions obtained through functional PCA and PANIC, i.e., $\log({\sf AE}(\widetilde \Lambda))$ and $\log({\sf AE}(\widehat \Lambda))$. In Table \ref{tab2a} for functional PCA, a consistent decrease is evident in the approximation errors as $T$ increases across all combinations of $N$ and $q$. In contrast when $N$ varies the approximation errors remain relatively stable. This reversal of roles between $N$ and $T$ in comparison to Table \ref{tab1} reveals an interesting pattern, which was also observed in \cite{TNH23b} for stationary curve time series. When $N$ and $T$ are fixed, the approximation errors increase as $q$ increases, which differs from the evolving pattern observed in Table \ref{tab1}. In Table \ref{tab2b} for functional PANIC, the approximation errors decrease when $T$ and $N$ increase. As in Table \ref{tab1}, functional PANIC also exhibits higher approximation errors than functional PCA, which again shows that functional PCA converges faster than functional PANIC in this example.

\begin{table}
\scriptsize
\centering 
\begin{subtable}{.4\textwidth}
\caption{Functional PCA}
\label{tab2a}
\begin{tabular}{cc|ccc}
\hline
$N$ &  $q$ & $T=200$ &$T=300$& $T=400$ \\
\hline
100 & 5 &-6.578 &-7.388 &-7.946\\
&&(0.334) &(0.332) &(0.331)\\
200 & 5 &-6.560 &-7.368 &-7.926\\
&&(0.310 &(0.307) &(0.307)\\
300 & 5 &-6.554 &-7.360 &-7.916\\
&&(0.299) &(0.301) &(0.300)\\\hline
100 & 10 &-5.973 &-6.747 &-7.300\\
&&(0.213) &(0.223) &(0.234)\\
200 & 10 &-5.971 &-6.742 &-7.293\\
&&(0.179) &(0.194) &(0.205)\\
300 & 10 &-5.967 &-6.737 &-7.289\\
&&(0.166) &(0.183) &(0.193)\\\hline
100 & 15 &-5.582 &-6.344 &-6.899\\
&&(0.144) &(0.150) &(0.150)\\
200 & 15 &-5.592 &-6.349 &-6.897\\
&&(0.117 &(0.124) &(0.129)\\
300 & 15 &-5.591 &-6.346 &-6.894\\
&&(0.108) &(0.117) &(0.122)\\\hline
\hline
\end{tabular}
\end{subtable}\quad\quad 
\begin{subtable}{.4\textwidth}
\caption{Functional PANIC}
\label{tab2b}
\begin{tabular}{cc|ccc}
\hline
$N$ &  $q$ & $T=200$ &$T=300$& $T=400$ \\
\hline
100 & 5 &-3.928 &-4.317 &-4.587\\
&&(0.140) &(0.132) &(0.129)\\
200 & 5 &-3.988 &-4.396 &-4.680\\
&&(0.105) &(0.094) &(0.091)\\
300 & 5 &-4.000 &-4.411 &-4.698\\
&&(0.091) &(0.082) &(0.078)\\\hline
100 & 10 &-4.031 &-4.454 &-4.741\\
&&(0.116) &(0.107) &(0.103)\\
200 & 10 &-4.092 &-4.522 &-4.819\\
&&(0.084) &(0.077) &(0.075)\\
300 & 10 &-4.107 &-4.539 &-4.838\\
&&(0.075) &(0.066) &(0.065)\\\hline
100 & 15 &-3.937 &-4.374 &-4.671\\
&&(0.095) &(0.090) &(0.084)\\
200 & 15 &-4.001 &-4.442 &-4.743\\
&&(0.070) &(0.064) &(0.061)\\
300 & 15 &-4.017 &-4.459 &-4.762\\
&&(0.062) &(0.056) &(0.053)\\\hline
\hline
\end{tabular}
\end{subtable}
\caption{Logarithm of the approximation errors for factor loading functions by the functional PCA and PANIC, i.e., $\log({\sf AE}(\widetilde \Lambda))$ and $\log({\sf AE}(\widehat \Lambda))$, respectively, averaged over 1000 replications in Example 6.1. The standard deviations are reported in parentheses.}
\label{tab2}
\end{table}

Table \ref{tab3} reports the numbers for underestimation (in square brackets), correct-estimation, and over-estimation (in round brackets) for $q$ (the number of full-rank stochastic trends) over $1000$ replications. For functional PCA the number of stochastic trends is correctly estimated in most trials. The underestimation numbers are zero across all combinations of $(N,T,q)$, whereas overestimation numbers are generally small (i.e., $<3\%$).
For functional PANIC, the correct estimation numbers are again close to $1000$. The underestimation numbers are zero except when $q$ is large but $N$ and $T$ are small ($q=15$, $N=100$, $T=200$). The overestimation numbers are small, decreasing rapidly when $N$ and $T$ increase. These outcomes suggest that the two information criteria (\ref{eq4.2}) and (\ref{eq5.8}) perform accurately in determining the number of common stochastic trends when either functional PCA or PANIC is adopted. 

\begin{table}
\footnotesize	
\centering 
\begin{subtable}{\textwidth}
\centering 
\caption{Functional PCA: the number of common stochastic trends is determined by \eqref{eq4.2} with $\rho_{NT}=4\log(N\wedge T)(1/T+1/N)$.}
\label{tab3a}
\begin{tabular}{cc|ccc}
\hline
$T$ &  $q$ & $N=100$ &$N=200$& $N=300$ \\
\hline
200 & 5 &   [0] 973 (27) & [0] 988 (12)&[0] 985 (15)\\
300 & 5 &   [0] 990 (10) & [0] 979 (21)&[0] 984 (16)\\
400 & 5 &   [0] 990 (10) & [0]  978 (22)&[0] 979 (21)\\\hline
200 & 10 &   [0] 973 (27) & [0] 988 (12)&[0] 987 (13)\\
300 & 10 &   [0] 976 (24) & [0] 990 (10)&[0] 986 (14)\\
400 & 10 &   [0] 988 (12) & [0] 977 (23)&[0] 981 (19)\\
\hline
200 & 15 &   [0] 985 (15) & [0] 988 (12)&[0] 986 (14)\\
300 & 15 &   [0] 986 (14) & [0] 980 (20)&[0] 980 (20)\\
400 & 15 &   [0] 986 (14) & [0] 985 (15)&[0] 982 (18)\\\hline
\hline
\end{tabular}
\end{subtable}\\
\begin{subtable}{\textwidth}
\centering 
\caption{Functional PANIC: the number of common stochastic trends is determined by \eqref{eq5.8} with $\rho^{\dagger}_{NT}=0.6\log(\sqrt{N}\wedge \sqrt{T})(1/\sqrt{T}+1/\sqrt{N})$.}
\label{tab3b}
\begin{tabular}{cc|ccc}
\hline
$T$ &  $q$ & $N=100$ &$N=200$& $N=300$ \\
\hline
200 & 5 &   [0] 940 (60) & [0] 996 (4)&[0] 995 (5)\\
300 & 5 &   [0] 971 (29) & [0] 993 (7)&[0] 1000 (0)\\
400 & 5 &   [0] 976 (24) & [0] 996 (4)&[0] 999 (1)\\\hline
200 & 10 &   [0] 926 (74) & [0] 993 (7)&[0] 996 (4)\\
300 & 10 &   [0] 955 (45) & [0] 998 (2)&[0] 999 (1)\\
400 & 10 &   [0] 966 (34) & [0] 996 (4)&[0] 1000 (0)\\
\hline
200 & 15 &   [10] 953 (47) & [0] 995 (5)&[0] 997 (3)\\
300 & 15 &   [0] 960 (40) & [0] 998 (2)&[0] 998 (2)\\
400 & 15 &   [0] 965 (35) & [0] 997 (3)&[0] 997 (3)\\\hline
\hline
\end{tabular}
\end{subtable}
\caption{Numbers of under-estimation (in square brackets), correct-estimation,  and over-estimation (in round brackets) of $q$ over 1000 replications in Example 6.1.}\label{tab3}
\end{table}

\medskip

\noindent {\bf Example 6.2}.\ \ The next example has $G_t$ generated from the VECM \eqref{eq5.9}, where $\alpha_0$ and $\beta_0$ are  $4\times q^{\ddag}$ matrices to be defined later, and $v_t$ follows a VARMA(1,1) process\footnote{The initial 100 observations of the VARMA(1,1) process are discarded to ensure data stability over time.}:
\[
v_t=0.4v_{t-1}+\epsilon^{v}_t+0.4\epsilon^{v}_{t-1}
\]
with $\epsilon^{v}_{t}$ independently drawn from ${\cal N}({\bf0},{\sf diag}(1.25,0.75,1.4,0.6))$. Similar to \cite{CP09}, we consider the following four scenarios for $(\alpha_0,\beta_0,q^\ddag)$:

\begin{itemize}

\item $q^{\ddag}=0$ and $\alpha_0\beta_0^{^\intercal}={\boldsymbol O}$,
 
\item $q^{\ddag}=1$ and $\alpha_0\beta_0^{^\intercal}={\sf diag}({\boldsymbol R}_2,0,0)$,

\item $q^{\ddag}=2$ and $\alpha_0\beta_0^{^\intercal}={\sf diag}({\boldsymbol R}_3,0,0)$,

\item $q^{\ddag}=3$ and $\alpha_0\beta_0^{^\intercal}={\sf diag}({\boldsymbol R}_1, {\boldsymbol R}_2)$,

\end{itemize}
where ${\boldsymbol O}$ is a $4\times 4$ null matrix,
\[
{\boldsymbol R}_1=
\begin{pmatrix}
-0.5& 0.1\\
0.2& -0.4
\end{pmatrix},\quad
{\boldsymbol R}_2=
\begin{pmatrix}
2\\
0.5
\end{pmatrix}\begin{pmatrix}
-1 & 1
\end{pmatrix},\quad {\rm and}\quad
{\boldsymbol R}_3=
\begin{pmatrix}
-0.7& 0.1\\
0.2& -0.6
\end{pmatrix}.
\]
The functional idiosyncratic components are generated by $\varepsilon_{it}=\sum_{s=1}^t\varepsilon_{is}^\dagger$ with $\varepsilon_{it}^\dagger$ simulated according to (\ref{eq6.3}) where $\varepsilon_{it}(u)$ is replaced by $\varepsilon_{it}^\dagger(u)$. Finally, we generate the nonstationary curve observations: 
\[
Z_{it}(u)=\Lambda_i(u)^{^\intercal}G_{t}+\varepsilon_{i t}(u),
\]
where $\Lambda_i(u)$ is generated in the same way as in \eqref{eq6.4} with $q=4$.

\begin{table}
\scriptsize	
\centering 
\begin{subtable}{.4\textwidth}
\caption{Functional PCA}
\label{tab4a}
\begin{tabular}{cc|ccc}
\hline
$N$ &  $q^{\ddagger}$ & $T=200$ &$T=300$& $T=400$ \\
\hline
100 & 0 &-0.915 &-0.482 &-0.217\\
&&(0.707) &(0.714) &(0.684)\\
200 & 0 &-1.233 &-0.792 &-0.534\\
&&(0.793) &(0.808) &(0.773)\\
300 & 0 &-1.300 &-0.853 &-0.606\\
&&(0.854) &(0.870) &(0.832)\\\hline
100 & 1 &-0.983 &-0.834 &-0.709\\
&&(0.162) &(0.194) &(0.219)\\
200 & 1 &-1.082 &-0.967 &-0.872\\
&&(0.150) &(0.171) &(0.194)\\
300 & 1 &-1.115 &-1.013 &-0.931\\
&&(0.149) &(0.168) &(0.186)\\\hline
100 & 2 &-0.134 &0.186 &0.378\\
&&(0.303) &(0.300) &(0.222)\\
200 & 2 &-0.188 &0.162 &0.382\\
&&(0.317) &(0.326) &(0.235)\\
300 & 2 &-0.185 &0.183 &0.402\\
&&(0.334) &(0.327) &(0.221)\\\hline
100 & 3 &0.034 &0.170 &0.283\\
&&(0.146) &(0.136) &(0.181)\\
200 & 3 &0.016 &0.131 &0.226\\
&&(0.139) &(0.124) &(0.166)\\
300 & 3 &0.023 &0.121 &0.213\\
&&(0.131) &(0.121) &(0.168)\\\hline
\hline
\end{tabular}
\end{subtable}\quad\quad 
\begin{subtable}{.4\textwidth}
\caption{Functional PANIC}
\label{tab4b}
\begin{tabular}{cc|ccc}
\hline
$N$ &  $q^{\ddagger}$ & $T=200$ &$T=300$& $T=400$ \\
\hline
100 & 0 &-1.503 &-1.080 &-0.786\\
&&(0.516) &(0.521) &(0.510)\\
200 & 0 &-2.187 &-1.764 &-1.470\\
&&(0.501) &(0.524) &(0.526)\\
300 & 0 &-2.523 &-2.099 &-1.812\\
&&(0.517) &(0.528) &(0.525)\\
\hline
100 & 1 &-1.575 &-1.257 &-1.031\\
&&(0.407) &(0.383) &(0.367)\\
200 & 1 &-2.144 &-1.792 &-1.554\\
&&(0.435) &(0.420) &(0.400)\\
300 & 1 &-2.442 &-2.080 &-1.835\\
&&(0.469) &(0.444) &(0.418)\\\hline
100 & 2 &-1.310 &-0.983 &-0.764\\
&&(0.453) &(0.400) &(0.376)\\
200 & 2 &-1.872 &-1.512 &-1.274\\
&&(0.502) &(0.470) &(0.437)\\
300 & 2 &-2.171 &-1.787 &-1.543\\
&&(0.531) &(0.504) &(0.468)\\\hline
100 & 3 &-1.011 &-0.698 &-0.495\\
&&(0.490) &(0.429) &(0.395)\\
200 & 3 &-1.553 &-1.201 &-0.971\\
&&(0.545) &(0.508) &(0.471)\\
300 & 3 &-1.844 &-1.469 &-1.239\\
&&(0.576) &(0.550) &(0.508)\\\hline
\hline
\end{tabular}
\end{subtable}
\caption{Logarithms of approximation errors for $G_t$ by functional PCA and PANIC, i.e., $\log({\sf AE}(\widetilde G))$ and $\log({\sf AE}(\widehat G))$, averaged over 1000 replications in Example 6.2. Standard deviations are reported in parentheses.}
\label{tab4}
\end{table}

\begin{table}
\scriptsize		
\centering 
\begin{subtable}{.4\textwidth}
\caption{Functional PCA}
\label{tab5a}
\begin{tabular}{cc|ccc}
\hline
$N$ &  $q^{\ddagger}$ & $T=200$ &$T=300$& $T=400$ \\
\hline
100 & 0 &-3.820 &-3.812 &-3.823\\
&&(0.215) &(0.269) &(0.222)\\
200 & 0 &-4.467 &-4.467 &-4.486\\
&&(0.258) &(0.282) &(0.195)\\
300 & 0 &-4.778 &-4.779 &-4.810\\
&&(0.288) &(0.340) &(0.224)\\\hline
100 & 1 &-1.686 &-1.655 &-1.646\\
&&(0.417) &(0.400) &(0.394)\\
200 & 1 &-1.662 &-1.636 &-1.628\\
&&(0.399) &(0.379) &(0.372)\\
300 & 1 &-1.634 &-1.611 &-1.594\\
&&(0.368) &(0.356) &(0.348)\\\hline
100 & 2 &-1.575 &-1.407 &-1.290\\
&&(0.501) &(0.437) &(0.409)\\
200 & 2 &-1.545 &-1.401 &-1.278\\
&&(0.474) &(0.444) &(0.394)\\
300 & 2 &-1.528 &-1.387 &-1.262\\
&&(0.457) &(0.424) &(0.381)\\\hline
100 & 3 &-1.029 &-0.957 &-0.923\\
&&(0.226) &(0.216) &(0.207)\\
200 & 3 &-1.031 &-0.944 &-0.906\\
&&(0.228) &(0.217) &(0.208)\\
300 & 3 &-1.001 &-0.919 &-0.881\\
&&(0.226) &(0.217) &(0.200)\\\hline
\hline
\end{tabular}
\end{subtable}\quad\quad 
\begin{subtable}{.4\textwidth}
\caption{Functional PANIC}
\label{tab5b}
\begin{tabular}{cc|ccc}
\hline
$N$ &  $q^{\ddagger}$ & $T=200$ &$T=300$& $T=400$ \\
\hline
100 & 0 &-3.954 &-3.950 &-3.948\\
&&(0.076) &(0.061) &(0.053)\\
200 & 0 &-4.639 &-4.632 &-4.630\\
&&(0.077) &(0.062) &(0.052)\\
300 & 0 &-4.984 &-4.978 &-4.976\\
&&(0.079) &(0.064) &(0.055)\\\hline
100 & 1 &-3.949 &-3.945 &-3.943\\
&&(0.075) &(0.061) &(0.053)\\
200 & 1 &-4.636 &-4.629 &-4.627\\
&&(0.076) &(0.061) &(0.052)\\
300 & 1 &-4.982 &-4.975 &-4.974\\
&&(0.079) &(0.064) &(0.055)\\\hline
100 & 2 &-3.957 &-3.952 &-3.951\\
&&(0.075) &(0.062) &(0.054)\\
200 & 2 &-4.640 &-4.633 &-4.631\\
&&(0.077) &(0.062) &(0.052)\\
300 & 2 &-4.985 &-4.978 &-4.977\\
&&(0.080) &(0.064) &(0.056)\\\hline
100 & 3 &-3.955 &-3.951 &-3.950\\
&&(0.075) &(0.062) &(0.053)\\
200 & 3 &-4.639 &-4.632 &-4.630\\
&&(0.077) &(0.062) &(0.052)\\
300 & 3 &-4.984 &-4.977 &-4.976\\
&&(0.079) &(0.064) &(0.055)\\\hline
\hline
\end{tabular}
\end{subtable}
\caption{Logarithms of the approximation errors for $\xi_t$ by functional PCA and PANIC, i.e., $\log({\sf AE}(\widetilde \xi))$ and $\log({\sf AE}(\widehat \xi))$, averaged over 1000 replications in Example 6.2. The standard deviations are reported in parentheses.}
\label{tab5}
\end{table}

\begin{table}
\scriptsize		
\centering 
\begin{subtable}{.4\textwidth}
\caption{Functional PCA}
\label{tab6a}
\begin{tabular}{cc|ccc}
\hline
$N$ &  $q^{\ddagger}$ & $T=200$ &$T=300$& $T=400$ \\
\hline
100 & 0 &-2.742 &-2.736 &-2.749\\
&&(0.468) &(0.463) &(0.444)\\
200 & 0 &-2.701 &-2.688 &-2.704\\
&&(0.455) &(0.453) &(0.436)\\
300 & 0 &-2.688 &-2.676 &-2.695\\
&&(0.448) &(0.452) &(0.432)\\\hline
100 & 1 &-1.369 &-1.365 &-1.364\\
&&(0.065) &(0.060) &(0.054)\\
200 & 1 &-1.348 &-1.345 &-1.345\\
&&(0.050) &(0.045) &(0.042)\\
300 & 1 &-1.409 &-1.407 &-1.407\\
&&(0.048) &(0.045) &(0.041)\\\hline
100 & 2 &-1.002 &-0.760 &-0.609\\
&&(0.285) &(0.297) &(0.236)\\
200 & 2 &-0.978 &-0.715 &-0.541\\
&&(0.286) &(0.303) &(0.225)\\
300 & 2 &-1.006 &-0.727 &-0.553\\
&&(0.302) &(0.301) &(0.205)\\\hline
100 & 3 &-0.745 &-0.700 &-0.657\\
&&(0.082) &(0.072) &(0.107)\\
200 & 3 &-0.694 &-0.657 &-0.621\\
&&(0.062) &(0.059) &(0.097)\\
300 & 3 &-0.718 &-0.688 &-0.650\\
&&(0.051) &(0.057) &(0.101)\\\hline
\hline
\end{tabular}
\end{subtable}\quad\quad 
\begin{subtable}{.4\textwidth}
\caption{Functional PANIC}
\label{tab6b}
\begin{tabular}{cc|ccc}
\hline
$N$ &  $q^{\ddagger}$ & $T=200$ &$T=300$& $T=400$ \\
\hline
100 & 0 &-5.341 &-5.754 &-6.038\\
&&(0.143) &(0.135) &(0.130)\\
200 & 0 &-5.341 &-5.760 &-6.053\\
&&(0.112) &(0.103) &(0.095)\\
300 & 0 &-5.338 &-5.757 &-6.052\\
&&(0.101) &(0.091) &(0.083)\\\hline
100 & 1 &-5.367 &-5.772 &-6.048\\
&&(0.159) &(0.147) &(0.145)\\
200 & 1 &-5.381 &-5.794 &-6.083\\
&&(0.125) &(0.112) &(0.109)\\
300 & 1 &-5.381 &-5.797 &-6.086\\
&&(0.112) &(0.099) &(0.093)\\\hline
100 & 2 &-5.212 &-5.617 &-5.894\\
&&(0.138) &(0.131) &(0.126)\\
200 & 2 &-5.222 &-5.635 &-5.926\\
&&(0.109) &(0.101) &(0.095)\\
300 & 2 &-5.221 &-5.636 &-5.929\\
&&(0.097) &(0.087) &(0.080)\\\hline
100 & 3 &-5.272 &-5.677 &-5.956\\
&&(0.152) &(0.141) &(0.136)\\
200 & 3 &-5.287 &-5.699 &-5.990\\
&&(0.118) &(0.109) &(0.103)\\
300 & 3 &-5.288 &-5.700 &-5.995\\
&&(0.105) &(0.094) &(0.088)\\\hline
\hline
\end{tabular}
\end{subtable}
\caption{Logarithms of the approximation errors of factor loading functions by the functional PCA and PANIC, i.e., $\log({\sf AE}(\widetilde \Lambda))$ and $\log({\sf AE}(\widehat \Lambda))$, averaged over 1000 replications in Example 6.2. Standard deviations are reported in parentheses.}
\label{tab6}
\end{table}

Table \ref{tab4} reports logarithms of approximation errors for the stochastic trends in levels obtained by functional PCA and PANIC. In Table \ref{tab4a} for functional PCA, we observe a significant increase in approximation errors with the expansion of the time series length ($T$) across all values of $N$ and $q^{\ddagger}$. When $N$ increases, the pattern for approximation errors is not the same in different settings. Focusing on the three diagonal cases in each block of the table, i.e., ($N=100$, $T=200$), ($N=200$, $T=300$) and ($N=300$, $T=400$), we observe that the approximation errors still increase as both $T$ and $N$ diverge, suggesting that functional PCA estimates are inconsistent due to violation of the full-rank condition in Assumption \ref{ass:1}. In Table \ref{tab4b} for functional PANIC, we observe an increase in approximation errors when $T$ increases, but decreases in approximation errors when $N$ increases. Furthermore, the decreasing approximation errors in the three diagonal cases as $N$ and $T$ increase indicates that functional PANIC consistently estimates the stochastic trends in levels.

\smallskip

Table \ref{tab5} reports logarithms of approximation errors for the stochastic trends in differences obtained by functional PCA and PANIC. Functional PCA, as observed in Table \ref{tab5a}, is inconsistent when $q^{\ddagger}>0$. However, when $q^{\ddagger}=0$, the approximation errors decrease as $N$ grows but are stable with respect to $T$, and consequently decrease as both $N$ and $T$ tend to infinity. Functional PANIC, reported in Table \ref{tab5b}, has decreasing approximation errors with expansion of $N$ across all values of $T$ and $q^\ddag$. The approximation errors slightly increase as $T$ increases. Focusing on the three diagonal cases in each block of the table, we observe that the approximation errors generally diminish as both $T$ and $N$ increase, suggesting that functional PANIC can consistently estimate increments of the stochastic trends.

\smallskip

Table \ref{tab6} reports logarithms of approximation errors for the functional factor loadings, i.e., $\log({\sf AE}(\widetilde \Lambda))$ and $\log({\sf AE}(\widehat \Lambda))$. For the functional PCA results in Table \ref{tab6a}, the patterns of approximation errors evolving with $N$ and $T$ observed within each block of the table, indicate that the factor loading estimates via functional PCA are inconsistent. This may be due to inconsistency of the functional PCA in estimating the cointegrated factors. For the functional PANIC results in Table \ref{tab6b}, the approximation errors decrease as $T$ increases and remain stable when $N$ varies. Consequently, the approximation errors via the functional PANIC decrease as $N$ and $T$ jointly diverge.

\smallskip

Table \ref{tab7} reports numbers of underestimation, correct-estimation, and overestimation of cointegrating ranks under the BIC and HQ criteria. When the sample size is small ($N=100$ or $200$ and $T=200$ or $300$) and $q^{\ddag}=2$ or 3,  BIC tends to underestimate cointegrating rank and thereby choose more parsimonious models. This observation aligns with the finding of \cite{CP09}. It is worth pointing out that $N$ and $T$ play different roles in cointegrating rank estimation. An increase in $N$ results in reduced approximation errors for the stochastic trends and consequently increases the numbers of correct estimation (of $q^\ddag$) in most scenarios. In contrast, an increase in $T$ leads to larger approximation errors, as seen in Table \ref{tab4b}, but simultaneously contributes to more accurate cointegrating rank estimation (when the cointegrated factors were known), which is assured by theorems in \cite{CP09,CP12}. When $T$ increases to $400$, the performance of BIC improves significantly. It follows from Table \ref{tab7b} that the HQ criterion exhibits a notable tendency to over-estimate cointegrating rank, especially when $q^{\ddag}$ is small. The HQ criterion outperforms BIC only when $q^{\ddag}=3$.  

\begin{table}
\scriptsize	
\centering 
\begin{subtable}{.4\textwidth}
%\centering 
\caption{BIC}
\label{tab7a}
\begin{tabular}{cc|ccc}
\hline
$T$ &  $q^{\ddag}$ & $N=100$ &$N=200$& $N=300$ \\
\hline
200 & 0 &   [0] 738 (262) & [0] 738 (262)&[0] 741 (259)\\
300 & 0 &   [0] 804 (196) & [0] 803 (197)&[0] 798 (202)\\
400 & 0 &   [0] 833 (167) & [0] 827 (173)&[0] 823 (177)\\\hline
200 & 1 &   [0] 847 (153) & [0] 849 (151)&[0] 848 (152)\\
300 & 1 &   [2] 882 (116) & [0] 881 (119)&[0] 881 (119)\\
400 & 1 &   [3] 898 (99) & [0] 899 (101)&[0] 896 (104)\\\hline
200 & 2 &   [773] 191 (36) & [715] 240 (45)&[689] 264 (47)\\
300 & 2 &   [440] 505 (55) & [292] 643 (65)&[244] 783 (73)\\
400 & 2 &   [216] 723 (61) & [70] 865 (65)&[50] 877 (73)\\\hline
200 & 3 &   [864] 36 (0) & [828]  72 (0)&[808] 192 (0)\\
300 & 3 &   [563] 437 (0) & [443] 557 (0)&[402] 598 (0)\\
400 & 3 &   [280] 720 (0) & [123] 877 (0)&[94] 906 (0)\\
\hline
\end{tabular}
\end{subtable}\quad\quad \quad \quad
\begin{subtable}{.4\textwidth}
%\centering .
\caption{HQ}
\label{tab7b}
\begin{tabular}{cc|ccc}
\hline
$T$ &  $q^{\ddag}$ & $N=100$ &$N=200$& $N=300$ \\
\hline
200 & 0 &   [0] 321 (679) & [0] 326 (674)&[0] 392 (608)\\
300 & 0 &   [0] 344 (656) & [0] 339 (661)&[0] 381 (619)\\
400 & 0 &   [0] 392 (608) & [0] 345 (654)&[0] 378 (622)\\\hline
200 & 1 &   [0] 516 (484) & [0] 519 (481)&[0] 518 (482)\\
300 & 1 &   [0] 544 (456) & [0] 549 (451)&[0] 550 (450)\\
400 & 1 &   [1] 577 (423) & [0] 573 (427)&[0] 566 (434)\\\hline
200 & 2 &   [94] 650 (256) & [48] 670 (272)&[38] 684 (278)\\
300 & 2 &   [24] 729 (247) & [2] 742 (256)&[0] 741 (259)\\
400 & 2 &   [12] 748 (240) & [0] 758 (242)&[0] 756 (244)\\\hline
200 & 3 &   [191] 809 (0) & [135] 865 (0)&[121] 879 (0)\\
300 & 3 &   [32] 968 (0) & [10] 990 (0)&[7] 993 (0)\\
400 & 3 &   [5] 995 (0) & [1] 999 (0)&[0] 1000 (0)\\
\hline
\end{tabular}
\end{subtable}
\caption{Numbers of underestimation (in square brackets), correct-estimation,  and overestimation (in round brackets) of the cointegrating rank using BIC and HQ criterion over 1000 replications in Example 6.2.}
\label{tab7}
\end{table}

%In conclusion, Example 6.1 shows when the idiosyncratic component is stationary, the functional PCA method outperforms the functional PANIC method. However,  in cases where the idiosyncratic component is nonstationary and integrated, the functional PANIC method performs better. These findings emphasise the importance of understanding the characteristics of the dataset and choosing the appropriate dimensionality reduction method based on the specific problem and dataset at hand.

%%%%%%%%%%%%%%%%%%%%%%%%%%

\section{Real data analysis}\label{sec7}
\renewcommand{\theequation}{7.\arabic{equation}}
\setcounter{equation}{0}

This section presents two empirical applications of our methods. The first example studies a dataset of temperature curves and functional PCA is used. The second example studies a dataset of stock price curves and functional PANIC is employed. A notable distinction between the two applications lies in the properties of the nonstationary idiosyncratic components. For the temperature data collected from different weather stations, it is less likely for the idiosyncratic components to be nonstationary, given that temperature records at individual locations generally do not deviate significantly from the global or regional temperature patterns -- see \cite{OU03} and the references therein. In contrast, for the stock price data, the presence of nonstationary idiosyncratic components is highly probable due to the dynamic nature of financial markets, influenced by the occurrence of firm-specific information and announcements. It would be unrealistic to expect stock prices to exhibit a nonstationary factor structure with stationary idiosyncratic components. Such scenarios could create numerous hedging opportunities among randomly selected stocks, a phenomenon not observed in real-world financial markets.

\subsection{Minimum temperatures in Australia}

We applied functional PCA to yearly minimum temperature curves in Australia. The initial dataset collected from the Australian Bureau of Meteorology at \url{http://www.bom.gov.au} comprised daily minimum temperature observations. This dataset was considered by \cite{ARS18} and \cite{NSS23} in studying nonstationarity in temperature dynamics. For each calendar year, we employed a smoothing algorithm of \cite{Rpackagefda} using 51 Fourier basis functions to create a curve representing minimum temperatures throughout the year\footnote{This smoothing process was applied only if the number of observations exceeded 200 days in a year. Otherwise, the data for the weather station was removed for that year.}. To deal with missing observations in an entire year, we adjust the calculation of $\widetilde{\Omega}_{ts}$ in \eqref{eq3.2}, only using weather stations that have observations in both years $t$ and $s$. Our analysis focuses on weather stations that initiated observations before 1943 and continued beyond 2022. Consequently, we have 36 stations with 80 years of observations and $Z_{it}(u)$ denotes the minimum temperature of weather station $i$ on year $t$ at day $u$.

\smallskip

\begin{figure}
  \centering
  \includegraphics[width=0.8\textwidth,height=0.25\textheight]{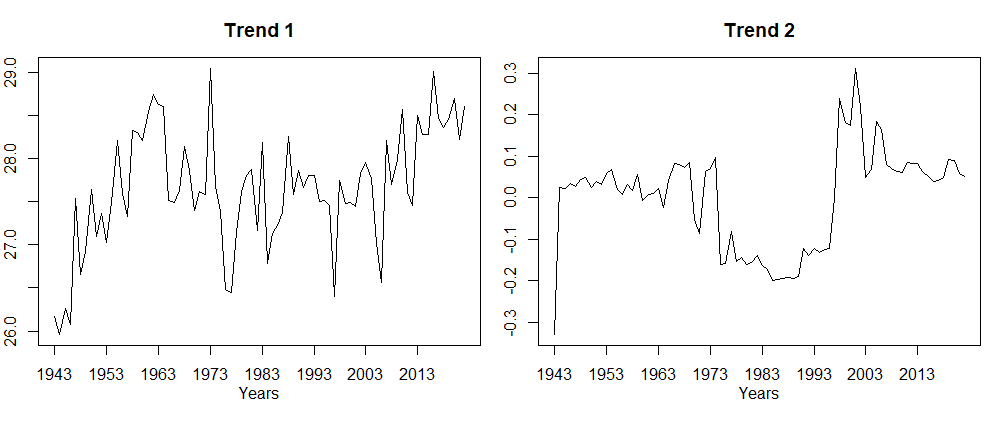}
  \includegraphics[width=0.8\textwidth,height=0.25\textheight]{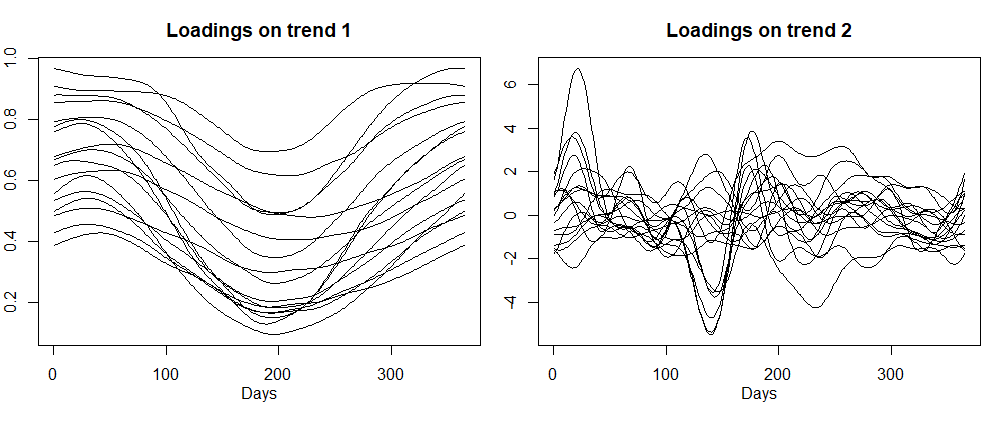}  
  \caption{Stochastic trends and factor loading functions for minimum temperature curves in Australia from 1943 to 2022.}
  \label{fig:1}
\end{figure}

Figure \ref{fig:1} shows estimates of the stochastic trends and their loadings. For illustration, the estimates of stochastic trends were scaled by the corresponding eigenvalues, while the loading functions were normalized by dividing them by the corresponding eigenvalues. Determination of the number of stochastic trends was based on the information criterion (\ref{eq4.2}), which resulted in two common stochastic trends. The first stochastic trend reveals an upward trajectory, and its loading functions depict a temperature profile characteristic of Australia, with higher temperatures observed at the beginning and end of the year. This stochastic trend, consistent with findings in several other studies such as \cite{NSS23}, signifies a stochastic trend in the mean temperature, suggesting the presence of global warming. The second stochastic trend exhibits a substantial negative value in 1943 and deviates from zero during the period from 1973 to 1993, suggesting significant temperature fluctuations within those years. Its loading functions display diverse patterns across different weather stations, highlighting their ability to capture station-specific intra-year temperature dynamics.

\subsection{High-frequency stock prices of S\&P 500 index constituents}

We next applied functional PANIC to intraday log-prices of S\&P 500  stocks. We selected the time period from 3 January 2023 to 1 November 2023, containing 209 trading days after removing a half trading day on 3 July 2023. The sample included $N = 209$ stocks. We adopted the 5-min frequency rather than 1-min frequency in data collection to minimize the impact of microstructure noise effects. Since all stocks trade from 9:30 a.m. to 4:00 p.m., 79 measurements were available per day. Asynchronous missing observations were interpolated by the linear algorithm of \cite{Rpackageforecast}. The discrete data were converted to a continuous function using \cite{Rpackagefda}'s algorithm, and the resulting curves denoted by ${Z_{1t}(u), Z_{2t}(u),\cdots, Z_{Nt}}(u)$, with $N=209$ and where the index $u$ lies in the time interval between 9:30 a.m. and 4:00 p.m.

\begin{figure}
  \centering
  \includegraphics[width=0.8\textwidth]{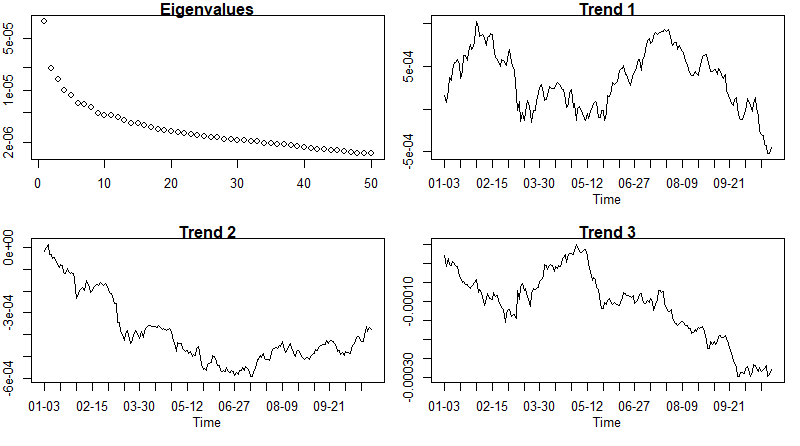}
  \caption{Sample eigenvalues and stochastic trends in S\&P stock log-prices from 3 January 2023 to 1 November 2023.}
  \label{fig:2}
\end{figure}

The scree plot in Figure \ref{fig:2} shows the first 50  sample eigenvalues in log-scale. The first three eigenvalues are relatively large, leading to the selection of three stochastic trends based on the proposed information criterion. The cointegrating rank determined by BIC is zero, signifying the absence of cointegration among the estimated stochastic trends. In fact, the lack of a cointegrating relation aligns with the expectation of an efficient market, where hedging opportunities arising from such a relation should not exist.

\smallskip

Further, increments of the three estimated stochastic trends were regressed on the Fama-French five factors\footnote{Data are collected from \url{https://mba.tuck.dartmouth.edu/pages/faculty/ken.french/Data_Library/f-f_factors.html}.}, i.e., market ($r_m-r_f$), size (SMB), value (HML), profitability (RMW), and investment (CMA). The results are reported in Table \ref{tab8}. For the first stochastic trend, the market factor is significant, whereas for the second stochastic trend, both the market and size factors are significant at the 10\%  level. No significant factors are identified for the third stochastic trend. The relatively low $R^2$ values indicate that the Fama-French factors may not fully explain the movements of the three common stochastic trends. 

\vspace{3mm}

\begin{table}[H]
\footnotesize	
\centering
\begin{tabular}{c|cccccc}
\hline
 & Trend 1 &Trend 2& Trend 3\\\hline
Intercept &-0.037&-0.063&-0.091\\
 &(0.070)&(0.071)&(0.071)\\
 $r_m-r_f$&0.342***&-0.183*&0.064\\
 &(0.096)&(0.098)&(0.098)\\
SMB &-0.177&-0.049&-0.210\\
&(0.132)&(0.135)&(0.135) \\
HML &-0.007&0.215&0.071
\\
 &(0.132)&(0.135)&(0.134) \\
RMW &-0.083&-0.276*&0.036\\
&(0.162)&(0.166)&(0.165)\\
CMA &0.137&-0.080&0.253\\ 
&(0.210)&(0.215)&(0.214)\\
\hline
$R^2$&0.076&0.040&0.037\\
$F$-statistic&3.271&1.668&1.548\\
p-value&0.007&0.144&0.177\\\hline
\hline
\end{tabular}
\caption{Increments of the estimated stochastic trends were regressed on the Fama-French factors with standard errors reported in parentheses. ***, ** and * indicate the regression parameters are significant at the 99\%, 95\% and 90\% confidence levels, respectively. } \label{tab8}
\end{table}

%%%%%%%%%%%%%%%%%%%

\section{Conclusion}\label{sec8}
\renewcommand{\theequation}{8.\arabic{equation}}
\setcounter{equation}{0}

The emergence and growth of vast cross section and time series datasets has substantially increased interest in the development of high-dimensional methods in econometrics. This paper contributes to this growing body of literature by introducing a general dual functional factor model for large-scale nonstationary curve time series. The approach involves the construction of a high-dimensional factor model for the observed curve time series that allows both factors and factor loadings to lie in function spaces with a low-dimensional factor model structure obtained by way of sieve approximation. An important feature of this framework is that both the dimension and time series length diverge to infinity. For the case of full-rank integrated factor curves and stationary functional idiosyncratic components, we employ functional PCA methodology to estimate the common stochastic trends and functional factor loadings and establish mean square convergence and asymptotic distribution theory. An easy-to-implement information criterion is proposed to consistently select the number of common stochastic trends. A functional PANIC methodology is introduced to handle the more general setting with cointegrated factors and possibly nonstationary functional idiosyncratic components. The simulation results reveal that functional PCA outperforms functional PANIC when factors are full-rank integrated and functional idiosyncratic components are stationary, whereas functional PANIC is more reliable when the integrated factors are rank-reduced and functional idiosyncratic components are nonstationary. Two empirical case studies are provided from climatological and financial data, each demonstrating the existence of common stochastic trends for these high dimensional curve time series.

%%%%%%%%%%%%%%%%%%%

\section*{Appendix A:\ Proofs of the main results}
\renewcommand{\theequation}{A.\arabic{equation}}
\setcounter{equation}{0}

Recall that
\[
{\boldsymbol G}=(G_1,\cdots,G_T)^{^\intercal},\quad \widetilde {\boldsymbol G}=(\widetilde G_1,\cdots,\widetilde G_T)^{^\intercal},
\]
and ${\boldsymbol V}_{NT}$ is a $q\times q$ diagonal matrix with the diagonal elements being the $q$ largest eigenvalues of $\frac{1}{T^2}\widetilde{\boldsymbol\Omega}$. We next provide the detailed proofs of the main theorems stated in Sections \ref{sec3} and \ref{sec4}. Throughout the proofs, we let $C$ denote a generic positive constant whose value may change from line to line.

\medskip

\noindent{\bf Proof of Proposition \ref{prop:3.1}}.\ \ Writing $\varepsilon_{it}^\ast(u)=\chi_{it}^{\eta}(u)+\varepsilon_{it}(u)$, it follows from (\ref{eq2.6}) and (\ref{eq3.2}) that
\begin{eqnarray}
\widetilde\Omega_{ts}&=&\frac{1}{N}\sum_{i=1}^N \int_{u\in{\mathbb C}_{i}}\left[\Lambda_i(u)^{^\intercal}G_t+\varepsilon_{it}^\ast(u)\right] \left[\Lambda_i(u)^{^\intercal}G_s+\varepsilon_{is}^\ast(u)\right]du\nonumber\\
&=&G_t^{^\intercal}\left[\frac{1}{N}\sum_{i=1}^N \int_{u\in{\mathbb C}_{i}}\Lambda_i(u) \Lambda_i(u)^{^\intercal}du\right] G_s+\frac{1}{N}\sum_{i=1}^N \int_{u\in{\mathbb C}_{i}}G_s^{^\intercal}\Lambda_i(u)\varepsilon_{it}^\ast(u)du+\nonumber\\
&&\frac{1}{N}\sum_{i=1}^N \int_{u\in{\mathbb C}_{i}} G_t^{^\intercal}\Lambda_i(u)\varepsilon_{is}^\ast(u)du+\frac{1}{N}\sum_{i=1}^N \int_{u\in{\mathbb C}_{i}} \varepsilon_{it}^\ast(u)\varepsilon_{is}^\ast(u)du.\label{eqA.1}
\end{eqnarray}
Let ${\boldsymbol H}_{NT}$ be the $q\times q$ rotation matrix defined in (\ref{eq3.7}). By virtue of its definition, PCA estimation yields 
\begin{equation}\label{eqA.2}
\left(\frac{1}{T^2}\widetilde{\boldsymbol\Omega}\right)\widetilde{\boldsymbol G}=\widetilde{\boldsymbol G}{\boldsymbol V}_{NT}.
\end{equation}
Combining (\ref{eqA.1}) and (\ref{eqA.2}), we have
\begin{equation}\label{eqA.3}
\widetilde G_t-{\boldsymbol H}_{NT} G_t={\boldsymbol V}_{NT}^{-1}\frac{1}{NT^2}\sum_{s=1}^{T}\sum_{i=1}^N\int_{u\in{\mathbb C}_{i}}\left[\widetilde{G}_s G_s^{^\intercal}\Lambda_i(u)\varepsilon_{it}^\ast(u)+\widetilde{G}_s \varepsilon_{is}^\ast(u)\Lambda_i(u)^{^\intercal}G_t+ \widetilde{G}_s\varepsilon_{is}^\ast(u)\varepsilon_{it}^\ast(u)\right]du,
\end{equation}
which has the following matrix form:
\begin{equation}\label{eqA.4}
\widetilde{\boldsymbol G}^{^\intercal}-{\boldsymbol H}_{NT}{\boldsymbol G}^{^\intercal}
={\boldsymbol V}_{NT}^{-1}\frac{1}{NT^2}\left[\widetilde{\boldsymbol G}^{^\intercal}{\boldsymbol G}\langle{\boldsymbol\Lambda}^{^\intercal},{\boldsymbol\varepsilon}^\ast\rangle+\widetilde{\boldsymbol G}^{^\intercal}\langle({\boldsymbol\varepsilon}^\ast)^{^\intercal},{\boldsymbol\Lambda}\rangle {\boldsymbol G}^{^\intercal}+ \widetilde{\boldsymbol G}^{^\intercal}\langle({\boldsymbol\varepsilon}^\ast)^{^\intercal},{\boldsymbol\varepsilon}^\ast\rangle\right],
\end{equation}
where ${\boldsymbol\Lambda}=(\Lambda_{ij})_{N\times q}$ and ${\boldsymbol\varepsilon}^\ast=(\varepsilon_{it}^\ast)_{N\times T}$.

By Proposition \ref{prop:4.1}, ${\boldsymbol V}_{NT}$ is positive definite with the minimum eigenvalue larger than $\underline\nu_{q}/2$ {\em w.p.a.1}. Hence, the inverse of ${\boldsymbol V}_{NT}$ exists and $\Vert{\boldsymbol V}_{NT}^{-1}\Vert=O_P(\underline\nu_{q}^{-1})$. By the triangle inequality and Lemma \ref{le:B.1} in Appendix B, we have
\begin{eqnarray}\label{eqA.5}
&&\frac{1}{T} \left\Vert\widetilde{\boldsymbol G}-{\boldsymbol G}{\boldsymbol H}_{NT}^{^\intercal}\right\Vert^2 \nonumber \\
&\leq&\Vert{\boldsymbol V}_{NT}^{-1}\Vert^2\frac{1}{N^2T^5}\left(\left\Vert\widetilde{\boldsymbol G}^{^\intercal}{\boldsymbol G}\langle{\boldsymbol\Lambda}^{^\intercal},{\boldsymbol\varepsilon}^\ast\rangle\right\Vert +\left\Vert\widetilde{\boldsymbol G}^{^\intercal}\langle({\boldsymbol\varepsilon}^\ast)^{^\intercal},{\boldsymbol\Lambda}\rangle {\boldsymbol G}^{^\intercal}\right\Vert+ \left\Vert\widetilde{\boldsymbol G}^{^\intercal}\langle({\boldsymbol\varepsilon}^\ast)^{^\intercal},{\boldsymbol\varepsilon}^\ast\rangle\right\Vert\right)^2\nonumber\\
&=&O_P(\underline\nu_{q}^{-2})\cdot \left[O_P\left({\overline\nu}_qq\left(N^{-1}+\delta_{q}^2\right)\right)+O_P\left(T^{-2}+N^{-1}T^{-1}+T^{-1}\delta_q^4\right)\right]\nonumber\\
&=&O_P\left(\underline\nu_{q}^{-2}\left(T^{-2}+q{\overline\nu}_qN^{-1}+q{\overline\nu}_q\delta_q^2\right)\right),\label{eqA.5}
\end{eqnarray}
which, together with the inequality
\[
\frac{1}{T}\sum_{t=1}^T\Vert\widetilde G_t-{\boldsymbol H}_{NT} G_t\Vert^2 \leq\frac{q}{T} \left\Vert\widetilde{\boldsymbol G}-{\boldsymbol G}{\boldsymbol H}_{NT}^{^\intercal}\right\Vert^2,
\]
completes the proof of (\ref{eq3.9}).\hfill$\Box$

\medskip

\noindent{\bf Proof of Theorem \ref{thm:3.2}}.\ By (\ref{eqA.3}) and Lemmas \ref{le:B.2} and \ref{le:B.6} in Appendix B, to prove (\ref{eq3.14}), we need to show 
\begin{equation}\label{eqA.6}
\sqrt{N}{\boldsymbol R}{\boldsymbol\Psi}_t^{-1/2}{\mathbf Q}_{NT}^{-1}\left({\boldsymbol V}_{NT}^{-1}\frac{1}{NT^2}\sum_{s=1}^{T}\sum_{i=1}^N\widetilde{G}_s G_s^{^\intercal}\langle\Lambda_i,\varepsilon_{it}^\ast\rangle\right) \rightsquigarrow {\cal N}\left({\bf 0}, {\boldsymbol\Upsilon}\right),
\end{equation}
where ${\boldsymbol R}$, ${\boldsymbol\Psi}_t$ and ${\boldsymbol\Upsilon}$ are defined in Assumption \ref{ass:3}(ii), and ${\mathbf Q}_{NT}$ is defined in (\ref{eq3.14}). Similar to the proof of (\ref{eqB.1}) in Appendix B, noting that $({\overline\nu}_q/\underline\nu_{q})^{1/2}q^{1/2}\delta_{t,q}^\dagger=o(N^{-1/2})$ by Assumption \ref{ass:3}(i), we may show that
\begin{equation}\label{eqA.7}
\left\Vert\frac{1}{NT^2}\sum_{s=1}^{T}\sum_{i=1}^N\widetilde{G}_s G_s^{^\intercal}\langle\Lambda_i,\varepsilon_{it}^\ast\rangle
-\frac{1}{NT^2}\sum_{s=1}^{T}\sum_{i=1}^N\widetilde{G}_s G_s^{^\intercal}\langle\Lambda_i,\varepsilon_{it}\rangle\right\Vert
=o_P\left({\underline\nu}_{q}^{1/2}N^{-1/2}\right).
\end{equation}
With (\ref{eqA.7}) and $\Vert{\mathbf Q}_{NT}^{-1}{\boldsymbol V}_{NT}^{-1}\Vert=O_P({\underline\nu}_{q}^{-1/2})$ by Lemma \ref{le:B.6}, to prove (\ref{eqA.6}), it is sufficient to show that
\begin{equation}\label{eqA.8}
\sqrt{N}{\boldsymbol R}{\boldsymbol\Psi}_t^{-1/2}{\mathbf Q}_{NT}^{-1}\left({\boldsymbol V}_{NT}^{-1}\frac{1}{NT^2}\sum_{s=1}^{T}\widetilde{G}_s G_s^{^\intercal}\sum_{i=1}^N\langle\Lambda_i,\varepsilon_{it}\rangle\right)\rightsquigarrow {\cal N}\left({\bf 0}, {\boldsymbol\Upsilon}\right),
\end{equation}
which follows from Assumption \ref{ass:3}(ii). Finally, by (\ref{eqB.45}) in Lemma \ref{le:B.6}, we have
\begin{equation}\label{eqA.9}
\left\Vert{\boldsymbol V}_{NT}^{-1}\left(\frac{1}{T^2}\sum_{s=1}^{T}\widetilde{G}_s G_s^{^\intercal}\right)-{\mathbf Q}_0\right\Vert=o_P(1),
\end{equation}
completing the proof of (\ref{eq3.15}).\hfill$\Box$

\medskip

\noindent{\bf Proof of Theorem \ref{thm:3.3}}.\ \ It follows from (\ref{eq2.6}), (\ref{eq3.1}) and (\ref{eq3.3}) that
\begin{eqnarray}
\widetilde \Lambda_i&=&\frac{1}{T^2}\sum_{t=1}^T\left(\Lambda_i^{^\intercal}G_t+\chi_{it}^{\eta}+\varepsilon_{it}\right)\widetilde G_t\notag\\
&=&\frac{1}{T^2}\sum_{t=1}^T\widetilde G_t G_t^{^{\intercal}}\Lambda_i+\frac{1}{T^2}\sum_{t=1}^T\chi_{it}^{\eta}\widetilde G_t+\frac{1}{T^2}\sum_{t=1}^T\varepsilon_{it}\widetilde G_t\notag\\
&=&\left({\boldsymbol H}_{NT}^{-1}\right)^{^\intercal}\Lambda_i+\frac{1}{T^2}\sum_{t=1}^T\widetilde G_t\left({\boldsymbol H}_{NT}G_t-\widetilde G_t\right)^{^\intercal}\left({\boldsymbol H}_{NT}^{-1}\right)^{^\intercal}\Lambda_i+\frac{1}{T^2}\sum_{t=1}^T\chi_{it}^{\eta}\widetilde G_t+\notag\\
&&{\boldsymbol H}_{NT}\frac{1}{T^2}\sum_{t=1}^T\varepsilon_{it} G_t+\frac{1}{T^2}\sum_{t=1}^T\varepsilon_{it}\left(\widetilde G_t-{\boldsymbol H}_{NT}G_t\right).\label{eqA.10}
\end{eqnarray}
By (\ref{eqB.6}), (\ref{eqB.10}), Lemma \ref{le:B.7} and Assumption \ref{ass:4}(i), we have
\begin{equation}\label{eqA.11}
\left\Vert \frac{1}{T^2}\sum_{t=1}^T\chi_{it}^{\eta}\widetilde G_t\right\Vert\leq T^{-1/2} \left(\frac{1}{T}\left\Vert \widetilde {\boldsymbol G}\right\Vert\right)\left(\frac{1}{T}\sum_{t=1}^T \left\Vert \chi_{it}^{\eta}\right\Vert^2\right)^{1/2}=O_P\left(T^{-1/2}\delta_q\right)=o_P\left({\overline\nu}_q^{-1/2}T^{-1}\right).
\end{equation}
By Assumption \ref{ass:2}(i), \eqref{eqA.5}, Lemma \ref{le:B.7} and Assumption \ref{ass:4}(i), we have
\begin{eqnarray}
\left\Vert \frac{1}{T^2}\sum_{t=1}^T\varepsilon_{it}\left(\widetilde G_t-{\boldsymbol H}_{NT}G_t\right)\right\Vert &\leq&T^{-1}\left(\frac{1}{T^{1/2}} \left\Vert\widetilde{\boldsymbol G}-{\boldsymbol G}{\boldsymbol H}_{NT}^{^\intercal}\right\Vert\right)\left(\frac{1}{T}\sum_{t=1}^T \left\Vert \varepsilon_{it}\right\Vert^2\right)^{1/2}\nonumber\\
&=&O_P\left(T^{-1}\underline\nu_{q}^{-1}\left(T^{-1}+{\overline\nu}_q^{1/2}q^{1/2}N^{-1/2}+{\overline\nu}_q^{1/2}q^{1/2}\delta_q\right)\right)\notag\\
&=&o_P\left({\overline\nu}_q^{-1/2}T^{-1}\right).\label{eqA.12}
\end{eqnarray}
By (\ref{eqA.5}), (\ref{eqB.10}), Lemmas \ref{le:B.4} and \ref{le:B.5}, we find that
\begin{eqnarray}
&&\frac{1}{T^2}\left\Vert\sum_{t=1}^T\widetilde G_t\left(\widetilde G_t-{\boldsymbol H}_{NT}G_t\right)^{^\intercal}\left({\boldsymbol H}_{NT}^{-1}\right)^{^\intercal}\Lambda_i\right\Vert\notag\\
&\leq&\frac{1}{T^2} \left\Vert\widetilde{\boldsymbol G}^{^\intercal}\left(\widetilde{\boldsymbol G}-{\boldsymbol G}{\boldsymbol H}_{NT}^{^\intercal}\right) \right\Vert\cdot\left\Vert{\boldsymbol H}_{NT}^{-1}\Vert\cdot\Vert\Lambda_i\right\Vert\notag\\
&=&o_P\left({\overline\nu}_{q}^{-1}q^{-1/2}T^{-1}\right)\cdot
O_P\left({\overline\nu}_q^{1/2}\right)\cdot O_P\left(q^{1/2}\right)
=o_P\left({\overline\nu}_q^{-1/2}T^{-1}\right),\label{eqA.13}
\end{eqnarray}
and with (\ref{eqA.10})--(\ref{eqA.13}), we have 
\[
\widetilde\Lambda_i-\left({\boldsymbol H}_{NT}^{-1}\right)^{^\intercal}\Lambda_i={\boldsymbol H}_{NT}\frac{1}{T^2}\sum_{t=1}^T\varepsilon_{it} G_t+o_P\left({\overline\nu}_q^{-1/2}T^{-1}\right),
\]
so that, using $\Vert{\boldsymbol H}_{NT}^{-1}\Vert=O_P({\overline\nu}_q^{1/2})$ by Lemma \ref{le:B.5}, 
\begin{align}
T({\boldsymbol R}{\boldsymbol H}_{NT}^{-1})\left(\widetilde\Lambda_i-\left({\boldsymbol H}_{NT}^{-1}\right)^{^\intercal}\Lambda_i\right)
={\boldsymbol R}\frac1{T}\sum_{t=1}^T\varepsilon_{it} G_t	+o_P\left(1\right),
\end{align}	
which, together with Assumption \ref{ass:4}(ii), completes the proof of Theorem \ref{thm:3.3}.
\hfill$\Box$
\medskip

\noindent{\bf Proof of Proposition \ref{prop:4.1}}.\ \ Let $\varepsilon_{i\bullet}^\ast=(\varepsilon_{i1}^\ast,\cdots,\varepsilon_{iT}^\ast)^{^\intercal}$ with $\varepsilon_{it}^\ast=\chi_{it}^{\eta}+\varepsilon_{it}$. By (\ref{eq2.6}) and the definition of $\widetilde{\boldsymbol\Omega}$ in (\ref{eq3.2}), we readily have that
\begin{eqnarray}
\widetilde{\boldsymbol\Omega}&=&{\boldsymbol G}\left[\frac{1}{N}\sum_{i=1}^N \int_{u\in{\mathbb C}_{i}}\Lambda_i(u)\Lambda_i(u)^{^\intercal}du\right]{\boldsymbol G}^{^\intercal}+\frac{1}{N}{\boldsymbol G}\left[\sum_{i=1}^N \int_{u\in{\mathbb C}_{i}}\Lambda_i(u)\varepsilon_{i\bullet}^\ast(u)^{^\intercal}du\right]\nonumber\\
&&\frac{1}{N}\left[\sum_{i=1}^N \int_{u\in{\mathbb C}_{i}}\varepsilon_{i\bullet}^\ast(u)\Lambda_i(u)^{^\intercal}du\right]{\boldsymbol G}^{^\intercal}+\frac{1}{N}\sum_{i=1}^N \int_{u\in{\mathbb C}_{i}}\varepsilon_{i\bullet}^\ast(u)\varepsilon_{i\bullet}^\ast(u)^{^\intercal}du\nonumber\\
&=:&{\boldsymbol\Pi}_1+{\boldsymbol\Pi}_2+{\boldsymbol\Pi}_3+{\boldsymbol\Pi}_4.\label{eqA.14}
\end{eqnarray}
As in the proof of Lemma \ref{le:B.1}, we have
\begin{eqnarray}
\frac{1}{T^2}\Vert {\boldsymbol\Pi}_2\Vert=\frac{1}{T^2}\Vert {\boldsymbol\Pi}_3\Vert&=&\left\Vert\frac{1}{NT^2}{\boldsymbol G}\langle{\boldsymbol\Lambda}^{^\intercal},{\boldsymbol\varepsilon}^\ast\rangle\right\Vert
=O_P\left({\overline \nu}_q^{1/2}q^{1/2}T^{-1/2}\left(N^{-1/2}+\delta_q\right)\right),\label{eqA.15}\\
\frac{1}{T^2}\Vert {\boldsymbol\Pi}_4\Vert&=&\left\Vert \frac{1}{NT^2}\langle({\boldsymbol\varepsilon}^\ast)^{^\intercal},{\boldsymbol\varepsilon}^\ast\rangle\right\Vert
=O_P\left(T^{-1/2}(T^{-1}+(NT)^{-1/2}+T^{-1/2}\delta_q^2)\right).\label{eqA.16}
\end{eqnarray}
With (\ref{eqA.15}) and (\ref{eqA.16}), we prove that 
\begin{equation}\label{eqA.17}
\max_{1\leq i\leq q}\left\vert\widetilde\nu_i-\nu_i({\boldsymbol\Pi}_1/T^2)\right\vert=O_P\left({\overline \nu}_q^{1/2}q^{1/2}T^{-1/2}\left(N^{-1/2}+\delta_q\right)+T^{-3/2}\right).
\end{equation}

Recall that $\nu_{i,0}$ is the $i$-th largest eigenvalue of ${\boldsymbol\Sigma}_\Lambda^{1/2}\left(\int_0^1 B_\xi(r) B_\xi(r)^{^\intercal}dr\right){\boldsymbol\Sigma}_\Lambda^{1/2}$. As the eigenvalues of ${\boldsymbol\Sigma}_\Lambda$ are bounded away from zero and infinity, by Assumption \ref{ass:1}(iv), the eigenvalues of $\int_0^1 B_\xi(r) B_\xi(r)^{^\intercal}dr$ must be have order between $\underline\nu_q$ and $\overline\nu_q$ {\em w.p.a.1}. Then, by Assumption \ref{ass:1}(ii)(iii), we have
\begin{equation}\label{eqA.18}
\max_{1\leq i\leq q}\left\vert \nu_i({\boldsymbol\Pi}_1/T^2)-\nu_{i,0}\right\vert=o_P\left(q^{-\kappa}+q^{-\kappa}\overline{\nu}_q\right)=o_P\left(q^{-\kappa}\overline{\nu}_q\right).
\end{equation}
Hence, by (\ref{eqA.17}) and (\ref{eqA.18}), we prove 
\[
\max_{1\leq i\leq q}|\widetilde\nu_i-\nu_{i,0}|=O_P\left({\overline \nu}_q^{1/2}q^{1/2}T^{-1/2}\left(N^{-1/2}+\delta_q\right)+T^{-3/2}\right)+o_P\left(q^{-\kappa}\overline{\nu}_q\right),
\] 
completing the proof of (\ref{eq4.1}). On the other hand, ${\boldsymbol\Pi}_1$ is a low-rank matrix with $\nu_i({\boldsymbol\Pi}_1)\equiv0$ when $i>q$. Then, with (\ref{eqA.15})--(\ref{eqA.17}), we prove 
\[
\widetilde\nu_i=O_P\left({\overline \nu}_q^{1/2}q^{1/2}T^{-1/2}\left(N^{-1/2}+\delta_q\right)+T^{-3/2}\right),\quad q+1\leq i\leq N\wedge T,
\]
completing the proof of (\ref{eq4.2}). The proof of Proposition \ref{prop:4.1} is completed. \hfill$\Box$

\medskip

\noindent{\bf Proof of Theorem \ref{thm:4.2}}.\ \ By (\ref{eq3.8}), Assumption \ref{ass:1}(iv) and  (\ref{eq4.1}) in Proposition \ref{prop:4.1}, we may show that 
\[
\max_{1\leq i\leq q}|\widetilde\nu_i-\nu_{i,0}|=O_P\left({\overline \nu}_q^{1/2}q^{1/2}T^{-1/2}\left(N^{-1/2}+\delta_q\right)+T^{-3/2}\right)+o_P\left(q^{-\kappa}\overline{\nu}_q\right)=o_P(\underline\nu_{q}),
\] 
which, together with $\rho_{NT}=o(\underline\nu_{q})$ in (\ref{eq4.4}), indicates that $\widetilde\nu_j$ is the leading term and decreasing over $1\leq j\leq q$. On the other hand, by the second condition in (\ref{eq4.4}), the penalty term dominates $\widetilde\nu_j$ and is increasing over $q+1\leq j\leq q_{\max}$. Hence the objective function $\widetilde\nu_j+j\rho_{NT}$ is minimized at $j=q+1$ and $\widetilde{q}=q$ {\em w.p.a.1}. 
%By Proposition \ref{prop:4.1} and the definition of $\widetilde\nu_0$, we may show there exists a positive constant $\epsilon_0$ such that
%\begin{equation}\label{eqA.20}
%{\sf P}\left(\left\vert\frac{\widetilde\nu_{j+1}}{\widetilde\nu_{j}}\right\vert>\epsilon_0\right)\rightarrow1,\ \ j=0,\cdots,q-1.
%\end{equation}
%On the other hand, for $j=q+1,\cdots,q_{\max}$, by Proposition \ref{prop:4.1} and (\ref{eq4.5}), $\widetilde\nu_j=o_P(\rho_{NT})$ and $\widetilde\nu_j$ is thus set as $0$ in the ratio criterion. Hence,
%\begin{equation}\label{eqA.21}
%{\sf P}\left(\left\vert\frac{\widetilde\nu_{q+1}}{\widetilde\nu_{q}}\right\vert \geq \epsilon\right)\rightarrow0,
%\end{equation}
%and
%\begin{equation}\label{eqA.22}
%{\sf P}\left(\left\vert\frac{\widetilde\nu_{j+1}}{\widetilde\nu_{j}}\right\vert=\frac{0}{0}=1
%\right)\rightarrow1
%\end{equation}
%for $j=q+1,\cdots,q_{\max}$. Combining (\ref{eqA.20})--(\ref{eqA.22}), we prove $\widehat{q}=q$ {\em w.p.a.1}. 
\hfill$\Box$

%%%%%%%%%%%%%%%%%%%%

\section*{Appendix B:\ Technical lemmas}
\renewcommand{\theequation}{B.\arabic{equation}}
\setcounter{equation}{0}

In this appendix, we present some technical lemmas and their proofs.

\renewcommand{\thelemma}{B.\arabic{lemma}}
\setcounter{lemma}{0}

\begin{lemma}\label{le:B.1}

Suppose that the assumptions of Proposition \ref{prop:3.1} are satisfied. Then we have
\begin{eqnarray}
&&\frac{1}{N^2T^5}\left\Vert\widetilde{\boldsymbol G}^{^\intercal}{\boldsymbol G}\langle{\boldsymbol\Lambda}^{^\intercal},{\boldsymbol\varepsilon}^\ast\rangle\right\Vert^2=O_P\left(q{\overline \nu}_q\left(N^{-1}+\delta_{q}^2\right)\right),\label{eqB.1}\\
&&\frac{1}{N^2T^5}\left\Vert\widetilde{\boldsymbol G}^{^\intercal}\langle({\boldsymbol\varepsilon}^\ast)^{^\intercal},{\boldsymbol\Lambda}\rangle {\boldsymbol G}^{^\intercal}\right\Vert^2=O_P\left(q{\overline \nu}_q\left(N^{-1}+\delta_q^2\right)\right),\label{eqB.2}\\
&&\frac{1}{N^2T^5} \left\Vert\widetilde{\boldsymbol G}^{^\intercal}\langle({\boldsymbol\varepsilon}^\ast)^{^\intercal},{\boldsymbol\varepsilon}^\ast\rangle\right\Vert^2=O_P\left(T^{-2}+(NT)^{-1}+T^{-1}\delta_q^4\right),
\label{eqB.3}
\end{eqnarray}
where $\delta_q$ is defined in Assumption \ref{ass:2}(ii).
\end{lemma}

\noindent{\bf Proof of Lemma \ref{le:B.1}}.\ \ We start with the proof of (\ref{eqB.1}). Using the definition of $\varepsilon_{it}^\ast(u)$, we have 
\begin{eqnarray}
\frac{1}{N^2T}\left\Vert\langle{\boldsymbol\Lambda}^{^\intercal},{\boldsymbol\varepsilon}^\ast\rangle\right\Vert^2
&\leq& 
\frac{1}{N^2T}\left\Vert\langle{\boldsymbol\Lambda}^{^\intercal},{\boldsymbol\varepsilon}^\ast\rangle\right\Vert_F^2\nonumber\\
&\leq&\frac{2}{N^2T}\sum_{t=1}^T\left\Vert\sum_{i=1}^N\langle\Lambda_{i}, \chi_{it}^{\eta}\rangle\right\Vert^2+\frac{2}{N^2T}\sum_{t=1}^T\left\Vert\sum_{i=1}^N\langle\Lambda_{i}, \varepsilon_{it}\rangle\right\Vert^2.\label{eqB.4}
\end{eqnarray}
Letting $\Lambda_{ij}(\cdot)$ be the $j$-th element of $\Lambda_i(\cdot)$, by Assumptions \ref{ass:1}(i) and \ref{ass:2}(ii), we have
\begin{eqnarray}
&&\max_{1\leq i\leq N}\max_{1\leq j\leq q}\Vert \Lambda_{ij}\Vert\leq \max\limits_{1\leq i\leq N}\sum_{j=1}^k\Vert {\cal B}_{ij}\Vert\leq kC_B,\label{eqB.5}\\
&&\max_{1\leq i\leq N}{\sf E}\left[\Vert \chi_{it}^{\eta}\Vert^2\right]\leq \left(\max\limits_{1\leq i\leq N}\max\limits_{1\leq j\leq k}\Vert {\cal B}_{ij}\Vert^2\right)\sum_{j=1}^k {\sf E}\left[\Vert \eta_{jt}\Vert^2\right]=O\left(\delta_{t,q}^2\right).\label{eqB.6}
\end{eqnarray}
Then, with (\ref{eqB.5}), (\ref{eqB.6}) and the Cauchy-Schwarz inequality, we may show that 
\begin{eqnarray}
\frac{1}{N^2T}\sum_{t=1}^T{\sf E}\left[ \left\Vert\sum_{i=1}^N\langle\Lambda_{i}, \chi_{it}^{\eta}\rangle\right\Vert^2\right]
&\leq& \left(\frac{1}{N}\sum_{i=1}^N\sum_{j=1}^q  \Vert \Lambda_{ij}\Vert^2\right)\left(\frac{1}{NT}\sum_{t=1}^T\sum_{i=1}^N {\sf E}\left[\Vert \chi_{it}^{\eta}\Vert^2\right]\right)\notag\\
&=&O(q) \cdot O\left(\frac{1}{T}\sum_{t=1}^T\delta_{t,q}^2\right)=O\left(q\delta_q^2\right).\label{eqB.7}
\end{eqnarray}
By (\ref{eqB.5}) and Assumption \ref{ass:2}(iv), 
\begin{equation}\label{eqB.8}
\frac{1}{N^2T}\sum_{t=1}^T{\sf E}\left[\left\Vert\sum_{i=1}^N\langle\Lambda_{i}, \varepsilon_{it}\rangle\right\Vert^2\right]\leq \frac{C}{N^2T}\sum_{t=1}^T\sum_{i=1}^N \sum_{j=1}^q \Vert \Lambda_{ij}\Vert^2=O\left(qN^{-1}\right).
\end{equation}
Combining (\ref{eqB.4}), (\ref{eqB.7}) and (\ref{eqB.8}), we can prove 
\begin{equation}\label{eqB.9}
\frac{1}{N^2T}\left\Vert\langle{\boldsymbol\Lambda}^{^\intercal},{\boldsymbol\varepsilon}^\ast\rangle\right\Vert^2=O_P\left(q(N^{-1}+\delta_q^2)\right).
\end{equation}
By the identification restriction (\ref{eq3.1}) and Assumption \ref{ass:1}(iii)(iv), we have
\begin{equation}\label{eqB.10}
\frac{1}{T^2}\left\Vert\widetilde{\boldsymbol G}\right\Vert^2=1,\quad\frac{1}{T^2}\left\Vert{\boldsymbol G}\right\Vert^2=O_P\left({\overline \nu}_q\right).
\end{equation}
Combining  (\ref{eqB.9}) and (\ref{eqB.10}) and using the submultiplicativity property of the operator norm completes the proof of (\ref{eqB.1}). Noting that $\left\Vert\langle({\boldsymbol\varepsilon}^\ast)^{^\intercal},{\boldsymbol\Lambda}\rangle\right\Vert=\left\Vert\langle{\boldsymbol\Lambda}^{^\intercal},{\boldsymbol\varepsilon}^\ast\rangle\right\Vert$ leads to (\ref{eqB.2}). 

We finally turn to the proof of (\ref{eqB.3}). Note that
\begin{eqnarray}
&&\frac{1}{N^2T^5} \left\Vert\widetilde{\boldsymbol G}^{^\intercal}\langle({\boldsymbol\varepsilon}^\ast)^{^\intercal},{\boldsymbol\varepsilon}^\ast\rangle\right\Vert^2\nonumber\\
&\leq&\frac{1}{T^2}\left\Vert\widetilde{\boldsymbol G}\right\Vert^2\cdot \frac{1}{N^2T^3}\sum_{t=1}^T\sum_{s=1}^T\left\vert\sum_{i=1}^N\langle\varepsilon_{is}^\ast, \varepsilon_{it}^\ast\rangle\right\vert^2\nonumber\\
&\leq&\frac{1}{N^2T^3} \sum_{t=1}^T\sum_{s=1}^T\left\vert\sum_{i=1}^N\langle\varepsilon_{is}, \varepsilon_{it}\rangle\right\vert^2+\frac{1}{N^2T^3}\sum_{t=1}^T \sum_{s=1}^T\left\vert\sum_{i=1}^N\langle\chi_{is}^\eta, \varepsilon_{it}\rangle\right\vert^2+\nonumber\\
&&\frac{1}{N^2T^3} \sum_{t=1}^T\sum_{s=1}^T\left\vert\sum_{i=1}^N\langle\varepsilon_{is}, \chi_{it}^{\eta}\rangle\right\vert^2+\frac{1}{N^2T^3} \sum_{t=1}^T \sum_{s=1}^T\left\vert\sum_{i=1}^N\langle\chi_{is}^\eta, \chi_{it}^{\eta}\rangle\right\vert^2.\label{eqB.11}
\end{eqnarray}
By Assumption \ref{ass:2}(iii) and the Markov inequality, we have
\begin{eqnarray}
&&\sum_{t=1}^T \sum_{s=1}^T\left\vert\sum_{i=1}^N\langle\varepsilon_{is}, \varepsilon_{it}\rangle\right\vert^2\nonumber\\
&\leq&2\sum_{t=1}^T \sum_{s=1}^T\left\vert\sum_{i=1}^N\langle\varepsilon_{is}, \varepsilon_{it}\rangle-{\sf E}\left[\langle\varepsilon_{is}, \varepsilon_{it}\rangle\right] \right\vert^2+2\sum_{t=1}^T \sum_{s=1}^T\left\vert\sum_{i=1}^N{\sf E}\left[\langle\varepsilon_{is}, \varepsilon_{it}\rangle\right] \right\vert^2\nonumber\\
&=&O_P\left(T^2N\right)+O\left(N^2\right)\sum_{t=1}^T  \sum_{s=1}^T\zeta_N^2(s,t)\nonumber\\
&=&O_P\left(T^2N+TN^2\right).\label{eqB.12}
\end{eqnarray}
By Assumption \ref{ass:2}(i)(ii)(iv), we have
\begin{equation}\label{eqB.13}
\sum_{t=1}^T \sum_{s=1}^T\left\vert\sum_{i=1}^N\langle\chi_{is}^\eta, \varepsilon_{it}\rangle\right\vert^2+\sum_{t=1}^T \sum_{s=1}^T\left\vert\sum_{i=1}^N\langle\varepsilon_{is}, \chi_{it}^{\eta}\rangle\right\vert^2=O_P\left(T^2N\delta_q^2\right)
\end{equation}
and
\begin{equation}\label{eqB.14}
\sum_{t=1}^T \sum_{s=1}^T\left\vert\sum_{i=1}^N\langle\chi_{is}^\eta, \chi_{it}^{\eta}\rangle\right\vert^2\leq\left(\sum_{s=1}^T\sum_{i=1}^N\Vert \chi_{is}^\eta\Vert^2\right) \left(\sum_{t=1}^T \sum_{i=1}^N\Vert\chi_{it}^{\eta}\Vert^2\right)=O_P\left(T^2N^2
\delta_q^4\right).
\end{equation}
With (\ref{eqB.11})--(\ref{eqB.14}), we readily have (\ref{eqB.3}) and the proof of Lemma \ref{le:B.1} is complete. \hfill$\Box$

\medskip

\begin{lemma}\label{le:B.2}

Suppose that the assumptions of Theorem \ref{thm:3.2} are satisfied. Then we have
\begin{eqnarray}
&&\frac{1}{NT^2}\left\Vert\sum_{s=1}^{T}\sum_{i=1}^N\widetilde{G}_s \langle\varepsilon_{is}^\ast,\Lambda_i^{^\intercal}\rangle G_t\right\Vert=o_P\left(\underline\nu_{q}^{1/2}N^{-1/2}\right),\label{eqB.15}\\
&&\frac{1}{NT^2}\left\Vert\sum_{s=1}^{T}\sum_{i=1}^N\widetilde{G}_s\langle\varepsilon_{is}^\ast,\varepsilon_{it}^\ast\rangle\right\Vert=o_P\left(\underline\nu_{q}^{1/2}N^{-1/2}\right).\label{eqB.16}
\end{eqnarray}

\end{lemma}

\medskip

\noindent{\bf Proof of Lemma \ref{le:B.2}}.\ \ We first prove (\ref{eqB.15}). By the triangle inequality, we have
\begin{eqnarray}
&&\frac{1}{NT^2}\left\Vert\sum_{s=1}^{T}\sum_{i=1}^N\widetilde{G}_s \langle\varepsilon_{is}^\ast,\Lambda_i^{^\intercal}\rangle G_t\right\Vert\notag\\
&\leq&\frac{1}{NT^2}\left\Vert\sum_{s=1}^{T}\sum_{i=1}^N\left(\widetilde{G}_s-{\boldsymbol H}_{NT}G_s\right)\langle\varepsilon_{is}^\ast, \Lambda_i^{^\intercal}\rangle G_t\right\Vert+\frac{1}{NT^2}\left\Vert {\boldsymbol H}_{NT}\sum_{s=1}^{T}\sum_{i=1}^NG_s\langle\varepsilon_{is}^\ast, \Lambda_i^{^\intercal}\rangle G_t\right\Vert.\label{eqB.17}
\end{eqnarray}
By \eqref{eqA.5}, (\ref{eqB.7}), (\ref{eqB.8}) and (\ref{eq3.13}) in Assumption \ref{ass:3}(iii) and noting that 
\[{\underline\nu}_{q}^{-3/2}qT^{-1/2}\left(T^{-1}+{\overline\nu}_{q}^{1/2}q^{1/2}N^{-1/2}+{\overline\nu}_{q}^{1/2}q^{1/2}\delta_q\right)\rightarrow0
\]
implied by (\ref{eq3.11}) in Assumption \ref{ass:3}(i), we may show that 
\begin{eqnarray}
&&\frac{1}{NT^2}\left\Vert\sum_{s=1}^{T}\sum_{i=1}^N\left(\widetilde{G}_s-{\boldsymbol H}_{NT}G_s\right)\langle\varepsilon_{is}^\ast, \Lambda_i^{^\intercal}\rangle G_t\right\Vert\notag\\
&\leq&T^{-1/2}\left(\frac{1}{T^{1/2}} \left\Vert\widetilde{\boldsymbol G}-{\boldsymbol G}{\boldsymbol H}_{NT}^{^\intercal}\right\Vert\right)\left(\frac{1}{T^{1/2}}\left\Vert G_t\right\Vert\right) \left(\frac{1}{N^2T}\sum_{s=1}^T\left\Vert\sum_{i=1}^N\langle\Lambda_i, \varepsilon_{is}^\ast\rangle\right\Vert^2\right)^{1/2}\notag\\
&=&O_P\left({\underline\nu}_{q}^{-1}T^{-1/2}\left(T^{-1}+{\overline\nu}_q^{1/2}q^{1/2}N^{-1/2}+{\overline\nu}_q^{1/2}q^{1/2}\delta_q\right)\right)O_P\left(q^{1/2}\right)O_P\left(q^{1/2}\left(N^{-1/2}+\delta_q\right)\right)\notag\\
&=&o_P\left(\underline\nu_{q}^{1/2}(N^{-1/2}+\delta_q)\right).\label{eqB.18}
\end{eqnarray}
On the other hand, by Assumption \ref{ass:3}(i)(iii), (\ref{eqB.29}) and Lemma \ref{le:B.5}, we can prove that 
\begin{eqnarray}
&&\frac{1}{NT^2}\left\Vert {\boldsymbol H}_{NT}\sum_{s=1}^{T}\sum_{i=1}^NG_s\langle\varepsilon_{is}^\ast, \Lambda_i^{^\intercal}\rangle G_t\right\Vert\notag\\
&\leq&\Vert {\boldsymbol H}_{NT}\Vert\cdot \frac{1}{NT^{3/2}}\left\Vert{\boldsymbol G}\langle{\boldsymbol\Lambda}^{^\intercal},{\boldsymbol\varepsilon}^\ast\rangle\right\Vert\cdot \frac{1}{T^{1/2}}\Vert G_t\Vert\nonumber\\
&=&O_P({\underline\nu}_{q}^{-1/2})
O_P(qN^{-1/2}T^{-1/2}+{\overline \nu}_q^{1/2}q^{1/2}\delta_{q})O_P(q^{1/2})\nonumber\\
&=&O_P\left(\underline\nu_{q}^{-1/2}q^{3/2}(NT)^{-1/2}+\underline\nu_{q}^{-1/2}{\overline\nu}_{q}^{1/2}q\delta_q\right)\notag\\
&=&o_P\left(\underline\nu_{q}^{1/2}N^{-1/2}\right).\label{eqB.19}
\end{eqnarray}
(\ref{eqB.18})--(\ref{eqB.19}) in  conjunction with \eqref{eqB.17} completes the proof of (\ref{eqB.15}).
 
We next turn to the proof of (\ref{eqB.16}). Using the definition of $\varepsilon_{it}^\ast$ and the triangle inequality, 
\begin{eqnarray}
&&\frac{1}{NT^2}\left\Vert\sum_{s=1}^{T}\sum_{i=1}^N\widetilde{G}_s\langle\varepsilon_{is}^\ast,\varepsilon_{it}^\ast\rangle\right\Vert\notag\\
&\leq&\frac{1}{NT^2}\left\Vert\sum_{s=1}^{T}\sum_{i=1}^N\widetilde{G}_s\langle\varepsilon_{is},\varepsilon_{it}\rangle\right\Vert+\frac{1}{NT^2}\left\Vert\sum_{s=1}^{T}\sum_{i=1}^N\widetilde{G}_s\langle\chi_{is}^\eta,\varepsilon_{it}\rangle\right\Vert+\notag\\
&&\frac{1}{NT^2}\left\Vert\sum_{s=1}^{T}\sum_{i=1}^N\widetilde{G}_s\langle\varepsilon_{is},\chi_{it}^{\eta}\rangle\right\Vert+\frac{1}{NT^2}\left\Vert\sum_{s=1}^{T}\sum_{i=1}^N\widetilde{G}_s\langle\chi_{is}^{\eta},\chi_{it}^{\eta}\rangle\right\Vert. \label{eqB.20}
\end{eqnarray}
Let $\zeta_N^\ast(s,t)=\frac{1}{N}\sum_{i=1}^N\langle\varepsilon_{is}, \varepsilon_{it}\rangle-\zeta_{N}(s,t),$
where $\zeta_{N}(s,t)$ is defined in Assumption \ref{ass:2}(iii). Then, by  the triangle inequality, Assumptions \ref{ass:2}(iii) and \ref{ass:3}(iii), $\Vert{\boldsymbol H}_{NT}\Vert=O_P(\underline\nu_{q}^{-1/2})$ in Lemma \ref{le:B.5}, (\ref{eqA.5}), and
\[
\sum_{s=1}^T\left\vert \zeta_N^\ast(s,t)\right\vert^2=\sum_{s=1}^T\left\vert\frac{1}{N}\sum_{i=1}^N\langle\varepsilon_{is}, \varepsilon_{it}\rangle-\zeta_{N}(s,t)\right\vert^2=O_P\left(TN^{-1}\right), \] 
we have 
\begin{eqnarray}
&&\frac{1}{N}\left\Vert\sum_{s=1}^{T}\sum_{i=1}^N\widetilde{G}_s\langle\varepsilon_{is},\varepsilon_{it}\rangle\right\Vert\nonumber\\
&\leq&\left\Vert \sum_{s=1}^{T}{\boldsymbol H}_{NT}G_s\zeta_N(s,t)\right\Vert+\left\Vert \sum_{s=1}^{T}(\widetilde G_s-{\boldsymbol H}_{NT}G_s)\zeta_N(s,t)\right\Vert+\left\Vert \sum_{s=1}^{T}\widetilde G_s\zeta_N^\ast(s,t)\right\Vert\nonumber\\
&\leq&\Vert{\boldsymbol H}_{NT}^{^\intercal}\Vert\max_{1\leq s\leq T}\Vert G_s\Vert \sum_{s=1}^{T}\vert\zeta_N(s,t)\vert+\left\Vert\widetilde{\boldsymbol G}-{\boldsymbol G}{\boldsymbol H}_{NT}^{^\intercal}\right\Vert\left(\sum_{s=1}^{T}\zeta_N(s,t)^2\right)^{1/2}+\left\Vert\widetilde{\boldsymbol G}\right\Vert \left(\sum_{s=1}^{T}|\zeta_N^\ast(s,t)|^2\right)^{1/2}\nonumber \\
&=&O_P\left(\underline\nu_{q}^{-1/2}q^{1/2}T^{1/2}\right)+o_P(T^{1/2})+O_P(T^{3/2}N^{-1/2}).\label{eqB.21}
\end{eqnarray}
By Assumption \ref{ass:2}, the Markov inequality and following the proofs of (\ref{eqB.13}) and (\ref{eqB.14}), we have
\begin{eqnarray}
&&\sum_{s=1}^T \left\vert\sum_{i=1}^N\langle\chi_{is}^\eta, \varepsilon_{it}\rangle\right\vert^2+\sum_{s=1}^T\left\vert\sum_{i=1}^N\langle\varepsilon_{is}, \chi_{it}^{\eta}\rangle\right\vert^2=O_P\left(TN
(\delta_q^2+\delta_{t,q}^2)\right),\nonumber\\
&&\sum_{s=1}^T\left\vert\sum_{i=1}^N\langle\chi_{is}^\eta, \chi_{it}^{\eta}\rangle\right\vert^2=O_P\left(TN^2\delta_{t,q}^2\delta_q^2\right),\nonumber
\end{eqnarray}
for each $t$, which implies
\begin{eqnarray}
&&\frac{1}{T^2}\left\Vert\sum_{s=1}^{T}\sum_{i=1}^N\widetilde{G}_s\langle\chi_{is}^\eta,\varepsilon_{it}\rangle\right\Vert
^2\leq\frac{1}{T^2}\Vert\widetilde{\boldsymbol G}\Vert^2 \sum_{s=1}^T \left\vert\sum_{i=1}^N\langle\chi_{is}^\eta, \varepsilon_{it}\rangle\right\vert^2
=O_P\left(TN
(\delta_q^2+\delta_{t,q}^2)\right),\label{eqB.22}
\\
&&\frac{1}{T^2}\left\Vert\sum_{s=1}^{T}\sum_{i=1}^N\widetilde{G}_s\langle\varepsilon_{is},\chi_{it}^{\eta}\rangle\right\Vert^2\leq\frac{1}{T^2}\Vert\widetilde{\boldsymbol G}\Vert^2\sum_{s=1}^T\left\vert\sum_{i=1}^N\langle\varepsilon_{is}, \chi_{it}^{\eta}\rangle\right\vert^2=O_P\left(TN
(\delta_q^2+\delta_{t,q}^2)\right),\label{eqB.23}
\\
&&\frac{1}{T^2}\left\Vert\sum_{s=1}^{T}\sum_{i=1}^N\widetilde{G}_s\langle\chi_{is}^{\eta},\chi_{it}^{\eta}\rangle\right\Vert^2
\leq\frac{1}{T^2}\Vert\widetilde{\boldsymbol G}\Vert^2\sum_{s=1}^T\left\vert\sum_{i=1}^N\langle\chi_{is}^\eta, \chi_{it}^{\eta}\rangle\right\vert^2=O_P\left(TN^2\delta_{t,q}^2\delta_q^2\right). \label{eqB.24}
\end{eqnarray}
Combining \eqref{eqB.21}--\eqref{eqB.24} gives
\begin{eqnarray}
\frac{1}{NT^2}\left\Vert\sum_{s=1}^{T}\sum_{i=1}^N\widetilde{G}_s\langle\varepsilon_{is}^\ast,\varepsilon_{it}^\ast\rangle\right\Vert &=&O_P\left({\underline\nu}_{q}^{-1/2}q^{1/2}T^{-3/2}+(NT)^{-1/2}+T^{-1/2}
\delta_q\delta_{t,q}\right)\nonumber\\
&=&o_P\left(\underline\nu_{q}^{1/2}N^{-1/2}\right),\label{eqB.25}
\end{eqnarray}
where the last equality follows from $\underline\nu_{q}^{-2}qNT^{-3}=o(1)$ and $\delta_{t,q}^\dag=o(N^{-1/2})$ from Assumption \ref{ass:3}(i). This completes the proof of (\ref{eqB.16}).\hfill$\Box$ 

\medskip

The following lemma further improves the rates derived in (\ref{eqB.1}) and (\ref{eqB.2}).

\medskip

\begin{lemma}\label{le:B.3}

Suppose that the assumptions of Proposition \ref{prop:3.1} and Assumption \ref{ass:3}(iii) are satisfied. Then, 
\begin{eqnarray}
&&\frac{1}{N^2T^5}\left\Vert\widetilde{\boldsymbol G}^{^\intercal}{\boldsymbol G}\langle{\boldsymbol\Lambda}^{^\intercal},{\boldsymbol\varepsilon}^\ast\rangle\right\Vert^2=O_P\left(q^2(NT)^{-1}+q{\overline \nu}_q\delta_{q}^2\right),\label{eqB.26}\\
&&\frac{1}{N^2T^5}\left\Vert\widetilde{\boldsymbol G}^{^\intercal}\langle({\boldsymbol\varepsilon}^\ast)^{^\intercal},{\boldsymbol\Lambda}\rangle {\boldsymbol G}^{^\intercal}\right\Vert^2=O_P\left(q^2(NT)^{-1}+q{\overline \nu}_q\delta_{q}^2\right).\label{eqB.27}
\end{eqnarray}

\end{lemma}

\medskip

\noindent{\bf Proof of Lemma \ref{le:B.3}}.\ \ It follows from \eqref{eq3.13} that
\begin{equation}\label{eqB.28}
\left\Vert{\boldsymbol G}\langle{\boldsymbol\Lambda}^{^\intercal},{\boldsymbol\varepsilon}\rangle\right\Vert^2=\left\Vert\langle{\boldsymbol\Lambda}^{^\intercal},{\boldsymbol\varepsilon}\rangle{\boldsymbol G}\right\Vert^2=\left\Vert{\boldsymbol G}^{^\intercal}\langle{\boldsymbol\varepsilon}^{^\intercal},{\boldsymbol\Lambda}\rangle\right\Vert^2=O_P\left(q^2NT^2\right).
\end{equation}
Using the definition of $\varepsilon_{it}^\ast$, \eqref{eqB.7}, \eqref{eqB.10} and \eqref{eqB.28}, we have
\begin{eqnarray}
\frac{1}{N^2T^3}\left\Vert{\boldsymbol G}\langle{\boldsymbol\Lambda}^{^\intercal},{\boldsymbol\varepsilon}^\ast\rangle\right\Vert^2
&\leq&\frac{2}{N^2T^3}\left( \left\Vert{\boldsymbol G}\langle{\boldsymbol\Lambda}^{^\intercal},{\boldsymbol\varepsilon}\rangle\right\Vert^2+\left\Vert{\boldsymbol G}\right\Vert^2\left\Vert\langle{\boldsymbol\Lambda}^{^\intercal},{\boldsymbol\chi}^\eta\rangle\right\Vert^2\right)\nonumber\\
&=&O_P\left(q^2N^{-1}T^{-1}+{\overline \nu}_qq\delta_{q}^2\right),\label{eqB.29}
\end{eqnarray}
which proves \eqref{eqB.26}. In a similar way we can prove \eqref{eqB.27}  and the proof of Lemma \ref{le:B.3} is complete. \hfill$\Box$

\begin{lemma}\label{le:B.4}

Suppose that the assumptions of Theorem \ref{thm:3.3} are satisfied. Then we have
\begin{equation}\label{eqB.30}
\frac{1}{T^2}\left\Vert\sum_{t=1}^T\widetilde G_t\left(\widetilde G_t-{\boldsymbol H}_{NT}G_t\right)^{^\intercal}\right\Vert=o_P\left({\overline\nu}_{q}^{-1}q^{-1/2}T^{-1}\right).
\end{equation}

\end{lemma}

\medskip

\noindent{\bf Proof of Lemma \ref{le:B.4}}.\ \ By \eqref{eqA.5} and \eqref{eqB.10}, we have
\begin{eqnarray}
\frac{1}{T^2}\left\Vert\sum_{t=1}^T\widetilde G_t\left(\widetilde G_t-{\boldsymbol H}_{NT}G_t\right)^{^\intercal}\right\Vert
&=&\frac{1}{T^2}\left\Vert\widetilde{\boldsymbol G}^{^\intercal}\left(\widetilde{\boldsymbol G}-{\boldsymbol G}{\boldsymbol H}_{NT}^{^\intercal}\right)\right\Vert\nonumber\\
&=&O_P\left(T^{-1/2}\underline\nu_{q}^{-1}\left(T^{-1}+q^{1/2}{\overline\nu}_q^{1/2}N^{-1/2}+q^{1/2}{\overline\nu}_q^{1/2}\delta_q\right)\right)\nonumber\\
&=&o_P\left({\overline\nu}_{q}^{-1}q^{-1/2}T^{-1}\right),\nonumber
\end{eqnarray}
due to the fact that 
\[
T^{-1/2}q^{1/2}(\overline\nu_{q}/\underline\nu_{q})=o(1),\quad (T/N)^{1/2}\overline\nu_{q}^{1/2}q(\overline\nu_{q}/\underline\nu_{q})=o(1)\quad \text{and}\quad
(\overline\nu_{q}/\underline\nu_{q})\overline\nu_{q}^{1/2}qT^{1/2}\delta_q=o(1),
\]
implied by Assumption \ref{ass:4}(i).\hfill$\Box$

\medskip

\begin{lemma}\label{le:B.5}

Suppose that Assumptions \ref{ass:1} and \ref{ass:2} are satisfied and
\begin{equation}\label{eqB.31}
{\underline\nu}_{q}^{-2}\overline\nu_qT^{-1/2} \left(T^{-1}+{\overline\nu}_{q}^{1/2}q^{1/2}N^{-1/2}+{\overline\nu}_{q}^{1/2}q^{1/2}\delta_q\right)\rightarrow0.
\end{equation}
Then we have the following convergence results for the rotation matrix ${\boldsymbol H}_{NT}$ and its inverse ${\boldsymbol H}_{NT}^{-1}$:
\begin{equation}\label{eqB.32}
\left\Vert {\boldsymbol H}_{NT}-{\boldsymbol H}_0\right\Vert=o_P( {\overline \nu}_q^{-1/2}),\quad \left\Vert {\boldsymbol H}_{NT}\right\Vert=O_P({\underline \nu}_q^{-1/2}),
\end{equation}
\begin{equation}\label{eqB.33}
\left\Vert {\boldsymbol H}_{NT}^{-1}-{\boldsymbol H}_0^{-1}\right\Vert=o_P({\overline \nu}_q^{1/2}),\quad \left\Vert {\boldsymbol H}_{NT}^{-1}\right\Vert=O_P({\overline \nu}_q^{1/2}),
\end{equation}
where ${\boldsymbol H}_0={\boldsymbol V}_0^{-1/2}{\boldsymbol W}_0^{^\intercal}{\boldsymbol \Sigma}_\Lambda^{1/2}$, ${\boldsymbol V}_0={\sf diag}\{\nu_{1,0},\cdots,\nu_{q,0}\}$ and ${\boldsymbol W}_0$ is a matrix consisting of the eigenvectors of ${\boldsymbol\Sigma}_\Lambda^{1/2}(\int_0^1 B_\xi(r) B_\xi(r)^{^\intercal}dr){\boldsymbol\Sigma}_\Lambda^{1/2}$. 

\end{lemma}

\noindent{\bf Proof of Lemma \ref{le:B.5}}.\ \ Let
\[
{\boldsymbol\Sigma}_{\Lambda,N}=\frac{1}{N}\sum_{i=1}^N \int_{u\in{\mathbb C}_{i}}\Lambda_i(u)\Lambda_i(u)^{^\intercal}du,\ \ {\boldsymbol\Sigma}_{G,T}=\frac{1}{T^2} {\boldsymbol G}^{^\intercal}{\boldsymbol G},\ \ \widetilde{\boldsymbol\Sigma}_{G,T}=\frac{1}{T^2} {\boldsymbol G}^{^\intercal}\widetilde{\boldsymbol G}
\]
and 
\[\widetilde{\boldsymbol W}_{NT}={\boldsymbol W}_*{\boldsymbol D}_{W_*}^{-1},\ \ {\boldsymbol W}_*= {\boldsymbol\Sigma}_{\Lambda,N}^{1/2}\widetilde{\boldsymbol\Sigma}_{G,T},\ \ \ {\boldsymbol D}_{W_*}=\left({\sf diag}\left\{{\boldsymbol W}_*^{^\intercal}{\boldsymbol W}_*\right\}\right)^{1/2},\]
where ${\sf diag}\{\cdot\}$ denotes the diagonalization of a square matrix. Write
\[
{\boldsymbol\Delta}_{NT}= {\boldsymbol\Sigma}_{\Lambda,N}^{1/2}{\boldsymbol\Sigma}_{G,T}{\boldsymbol\Sigma}_{\Lambda,N}^{1/2},\ \ \ {\boldsymbol\Delta}_0={\boldsymbol\Sigma}_\Lambda^{1/2}\left(\int_0^1 B_\xi(u) B_\xi(u)^{^\intercal}du\right){\boldsymbol\Sigma}_\Lambda^{1/2},
\]
and
\[
{\boldsymbol\Delta}_{\ast}={\boldsymbol\Sigma}_{\Lambda,N}^{1/2}\frac{{\boldsymbol G}^{^\intercal}}{T}\left(\frac{1}{T^2}\widetilde{\boldsymbol\Omega}-\frac{1}{T^2}{\boldsymbol\Pi}_1\right)\frac{\widetilde 
{\boldsymbol G}}{T},
\]
where $\widetilde{\boldsymbol\Omega}$ is defined in (\ref{eq3.2}) and ${\boldsymbol\Pi}_1$ is defined in (\ref{eqA.15}). 

It follows from the definition of the functional PCA estimation that
\begin{equation}\label{eqB.34}
\left({\boldsymbol\Delta}_{NT}+{\boldsymbol\Delta}_{\ast} {\boldsymbol W}_*^{-1}\right)\widetilde{\boldsymbol W}_{NT}=\widetilde{\boldsymbol W}_{NT}{\boldsymbol V}_{NT}.
\end{equation}
Hence $\widetilde{\boldsymbol W}_{NT}$ consists of the eigenvectors of ${\boldsymbol\Delta}_{NT}+{\boldsymbol\Delta}_{\ast} {\boldsymbol W}_*^{-1}$. Write 
\[{\boldsymbol H}_{NT}={\boldsymbol V}_{NT}^{-1}{\boldsymbol D}_{W_*}\widetilde{\boldsymbol W}_{NT}^{^\intercal}{\boldsymbol\Sigma}_{\Lambda,N}^{1/2}.\] 
Note that the second assertion in (\ref{eqB.32}) follows from the first assertion and $\underline\nu_q\leq \overline\nu_q$. With the triangle inequality, 
\begin{eqnarray}
\left\Vert{\boldsymbol H}_{NT}-{\boldsymbol H}_0 \right\Vert
&\leq&\left\Vert\left({\boldsymbol V}_{NT}^{-1}-{\boldsymbol V}_0^{-1}\right){\boldsymbol D}_{W_*}\widetilde{\boldsymbol W}_{NT}^{^\intercal}{\boldsymbol\Sigma}_{\Lambda,N}^{1/2}\right\Vert+\left\Vert{\boldsymbol V}_0^{-1}\left({\boldsymbol D}_{W_*}-{\boldsymbol V}_0^{1/2}\right)\widetilde{\boldsymbol W}_{NT}^{^\intercal}{\boldsymbol\Sigma}_{\Lambda,N}^{1/2}\right\Vert+\nonumber\\
&&\left\Vert{\boldsymbol V}_0^{-1/2}\left(\widetilde{\boldsymbol W}_{NT}-{\boldsymbol W}_0\right)^{^\intercal}{\boldsymbol\Sigma}_{\Lambda,N}^{1/2}\right\Vert+\left\Vert{\boldsymbol V}_0^{-1/2}{\boldsymbol W}_0^{^\intercal}\left({\boldsymbol\Sigma}_{\Lambda,N}^{1/2}-{\boldsymbol \Sigma}_\Lambda^{1/2}\right)\right\Vert,\label{eqB.35}
\end{eqnarray}
we next only need to show that
\begin{alignat}{2}
\left\Vert {\boldsymbol V}_{NT}^{-1} - {\boldsymbol V}_0^{-1} \right\Vert &= o_P\left({\overline{\nu}}_q^{-1}\right), \quad & \left\Vert {\boldsymbol V}_{NT}^{-1} \right\Vert &= O_P({\underline{\nu}}_q^{-1}), \label{eqB.36} \\
\left\Vert {\boldsymbol D}_{W_*} - {\boldsymbol V}_0^{1/2} \right\Vert &= o_P({\underline\nu}_q{\overline\nu}_q^{-1/2}), \quad & \left\Vert {\boldsymbol D}_{W_*} \right\Vert &= O_P({\overline\nu}_q^{1/2}), \label{eqB.37} \\
\left\Vert \widetilde{\boldsymbol W}_{NT} - {\boldsymbol W}_0 \right\Vert &= o_P({\underline\nu}_q^{1/2}{\overline\nu}_q^{-1/2}), \quad & \left\Vert \widetilde{\boldsymbol W}_{NT} \right\Vert &= O_P(1), \label{eqB.38} \\
\left\Vert {\boldsymbol\Sigma}_{\Lambda,N}^{1/2} - {\boldsymbol\Sigma}_{\Lambda}^{1/2} \right\Vert &= o({\underline\nu}_q^{1/2}{\overline\nu}_q^{-1/2}), \quad & \left\Vert {\boldsymbol\Sigma}_{\Lambda,N}^{1/2} \right\Vert &= O(1). \label{eqB.39}
\end{alignat}
As $\underline\nu_q\leq \overline\nu_q$, it is easy to verify the second assertion in each of (\ref{eqB.36})--(\ref{eqB.39}) from the respective first one. Hence, we next only prove the first assertion in (\ref{eqB.36})--(\ref{eqB.39}).   

Note first that 
$$\left\Vert {\boldsymbol A}^{-1}\right\Vert-\left\Vert{\boldsymbol B}^{-1}\right\Vert\leq\left\Vert {\boldsymbol A}^{-1}-{\boldsymbol B}^{-1}\right\Vert\leq \left\Vert {\boldsymbol A}^{-1}\right\Vert\left\Vert {\boldsymbol A}-{\boldsymbol B}\right\Vert\left\Vert {\boldsymbol B}^{-1}\right\Vert$$
and thus
\begin{equation*}
\left\Vert {\boldsymbol A}^{-1}\right\Vert\leq \frac{\left\Vert {\boldsymbol B}^{-1}\right\Vert}{1-\left\Vert {\boldsymbol B}^{-1}\right\Vert\left\Vert {\boldsymbol A}-{\boldsymbol B}\right\Vert} \quad\text{when} \left\Vert {\boldsymbol B}^{-1}\right\Vert\left\Vert {\boldsymbol A}-{\boldsymbol B}\right\Vert=o(1).
\end{equation*}
Combining the two inequalities, we obtain
\begin{equation}\label{eqB.40}
\left\Vert {\boldsymbol A}^{-1}-{\boldsymbol B}^{-1}\right\Vert\leq \frac{\left\Vert {\boldsymbol B}^{-1}\right\Vert^2\left\Vert {\boldsymbol A}-{\boldsymbol B}\right\Vert}{1-\left\Vert {\boldsymbol B}^{-1}\right\Vert\left\Vert {\boldsymbol A}-{\boldsymbol B}\right\Vert} \quad\text{when} \left\Vert {\boldsymbol B}^{-1}\right\Vert\left\Vert {\boldsymbol A}-{\boldsymbol B}\right\Vert=o(1).
\end{equation}
Using \eqref{eqB.40},
Proposition \ref{prop:4.1}, \eqref{eqB.31} and Assumption \ref{ass:1}(iv), we readily have that 
\[\left\Vert {\boldsymbol V}_{NT}^{-1}-{\boldsymbol V}_0^{-1}\right\Vert=O_P\left({\underline\nu}_q^{-2}\right)\cdot \left[O_P\left({\overline{\nu}}_q^{1/2}q^{1/2}T^{-1/2}\left(N^{-1/2}+\delta_q\right)+T^{-3/2}\right)+o_P\left(q^{-\kappa}\overline{\nu}_q\right)\right]=o_P({\overline{\nu}}_q^{-1}),\] 
proving the first assertion in (\ref{eqB.36}).

With the triangle inequality and Proposition \ref{prop:4.1}, we have
\begin{eqnarray}
\left\Vert {\boldsymbol D}_{W_*}^2-{\boldsymbol V}_0\right\Vert&\leq& \left\Vert {\boldsymbol D}_{W_*}^2-{\boldsymbol V}_{NT}\right\Vert+\left\Vert {\boldsymbol V}_{NT}-{\boldsymbol V}_0\right\Vert\notag\\
&=&\left\Vert {\boldsymbol D}_{W_*}^2-{\boldsymbol V}_{NT}\right\Vert+O_P\left({\overline{\nu}}_q^{1/2}q^{1/2}T^{-1/2}\left(N^{-1/2}+\delta_q\right)+T^{-3/2}\right)+o_P\left(q^{-\kappa}\overline{\nu}_q\right).\notag
\end{eqnarray}
Note that
\begin{eqnarray}
\left\Vert {\boldsymbol D}_{W_*}^2-{\boldsymbol V}_{NT}\right\Vert&\leq&\left\Vert \frac{1}{T^2}\widetilde{\boldsymbol G}^{^\intercal}\left(\frac{1}{T^2}\widetilde{\boldsymbol\Omega}-\frac{1}{T^2}{\boldsymbol G} {\boldsymbol\Sigma}_{\Lambda,N}{\boldsymbol G}^{^\intercal}\right)\widetilde{\boldsymbol G}\right\Vert\notag\\
&=&\left\Vert\frac{1}{T^2}\widetilde{\boldsymbol\Omega}-\frac{1}{T^2}{\boldsymbol G} {\boldsymbol\Sigma}_{\Lambda,N}{\boldsymbol G}^{^\intercal}\right\Vert\notag\\
&=&\frac{1}{T^2}\left\Vert \widetilde{\boldsymbol\Omega}-{\boldsymbol\Pi}_1\right\Vert=O_P\left({\overline{\nu}}_q^{1/2}q^{1/2}T^{-1/2}\left(N^{-1/2}+\delta_q\right)+T^{-3/2}\right).\notag
\end{eqnarray}
Hence, we have
\begin{equation}\label{eqB.41}
\left\Vert {\boldsymbol D}_{W_*}^2-{\boldsymbol V}_0\right\Vert=O_P\left({\overline{\nu}}_q^{1/2}q^{1/2}T^{-1/2}\left(N^{-1/2}+\delta_q\right)+T^{-3/2}\right)+o_P\left(q^{-\kappa}\overline{\nu}_q\right)
\end{equation}
Using (\ref{eqB.31}), \eqref{eqB.41}, $\Vert{\boldsymbol V}_0^{-1/2}\Vert=O_P({\underline\nu}_q^{-1/2})$ and $q^{-\kappa}({\overline\nu}_q/\underline\nu_q)^{3/2}=O(1)$ indicated by Assumption \ref{ass:1}(iv), we have 
\begin{eqnarray}
\left\Vert {\boldsymbol D}_{W_*}-{\boldsymbol V}_0^{1/2}\right\Vert&\leq&\frac{\left\Vert {\boldsymbol D}_{W_*}^2-{\boldsymbol V}_0\right\Vert}{\left\Vert {\boldsymbol D}_{W_*}+{\boldsymbol V}_0^{1/2}\right\Vert}\notag\\
&=&O_P({\underline\nu}_q^{-1/2})\left[O_P\left({\overline{\nu}}_q^{1/2}q^{1/2}T^{-1/2}\left(N^{-1/2}+\delta_q\right)+T^{-3/2}\right)+o_P\left(q^{-\kappa}\overline{\nu}_q\right)\right]\notag\\
&=&o_P({\underline\nu}_q{\overline\nu}_q^{-1/2}),\notag
\end{eqnarray}
proving the first assertion in (\ref{eqB.37}).

Applying the $\sin \theta$ theorem in \cite{DK70} to each eigenvector of ${\boldsymbol\Delta}_{NT}+{\boldsymbol\Delta}_{\ast} {\boldsymbol W}_*^{-1}$ and the corresponding eigenvector of ${\boldsymbol\Delta}_0$, we have
\begin{eqnarray}
\left\Vert  \widetilde{\boldsymbol W}_{NT}-{\boldsymbol W}_0\right\Vert&\leq&\left\Vert  \widetilde{\boldsymbol W}_{NT}-{\boldsymbol W}_0\right\Vert_F\nonumber\\
&\leq& 2\sqrt{2}\cdot \left[(q-1){\underline\iota}_q^{-1}+{\underline\nu}_q^{-1}\right]\Vert {\boldsymbol\Delta}_{NT}+{\boldsymbol\Delta}_{\ast} {\boldsymbol W}_*^{-1}-{\boldsymbol\Delta}_0\Vert.\label{eqB.42}
\end{eqnarray}
It follows from (\ref{eqA.18}) that
we have 
\begin{equation}
\Vert {\boldsymbol\Delta}_{NT}-{\boldsymbol\Delta}_0\Vert=o_P(q^{-\kappa}\overline{\nu}_q).
\end{equation}
As in the proof of Proposition \ref{prop:4.1}, we readily have that
\begin{eqnarray}
\Vert {\boldsymbol\Delta}_{\ast}\Vert
&\leq& \left\Vert{\boldsymbol\Sigma}_{\Lambda,N}^{1/2}\right\Vert\cdot\frac{1}{T}\left\Vert{\boldsymbol G}^{^\intercal}\right\Vert\cdot\left\Vert\frac{1}{T^2}\widetilde{\boldsymbol\Omega}-\frac{1}{T^2}{\boldsymbol\Pi}_1\right\Vert\cdot\frac{1}{T}\Vert\widetilde 
{\boldsymbol G}\Vert\nonumber\\
&=&O_P\left({\overline\nu}_q^{1/2}\left[{\overline\nu}_q^{1/2}q^{1/2}T^{-1/2}\left(N^{-1/2}+\delta_q\right)+T^{-3/2}\right]\right).\notag
\end{eqnarray}
Since $\Vert {\boldsymbol W}_*^{-1}\Vert=\Vert {\boldsymbol W}_*^{-1}({\boldsymbol W}_*^{^\intercal})^{-1}\Vert^{1/2}$, we have 
\begin{equation}\label{eqB.44}
\Vert {\boldsymbol W}_*^{-1}\Vert
=\left\Vert\left({\boldsymbol\Sigma}_{\Lambda,N}^{1/2}\frac{1}{T^2} {\boldsymbol G}^{^\intercal}\widetilde{\boldsymbol G}\right)^{-1}\right\Vert
=\left\Vert\left({\boldsymbol\Sigma}_{\Lambda,N}\frac{1}{T^2} {\boldsymbol G}^{^\intercal}{\boldsymbol G}\right)^{-1}\right\Vert^{1/2}
=O_P({\underline\nu}_{q}^{-1/2}).
\end{equation}
Combining \eqref{eqB.42}--\eqref{eqB.44}, we have
\begin{eqnarray}
\left\Vert  \widetilde{\boldsymbol W}_{NT}-{\boldsymbol W}_0\right\Vert&=&O\left(q{\underline\iota}_q^{-1}+{\underline\nu}_q^{-1}\right)\left[o_P(q^{-\kappa}\overline{\nu}_q)+O_P\left(({\overline\nu}_q/\underline\nu_q)^{1/2}\left[{\overline\nu}_q^{1/2}q^{1/2}T^{-1/2}\left(N^{-1/2}+\delta_q\right)+T^{-3/2}\right]\right)\right]\notag\\
&=&o_P\left(q^{1-\kappa}{\underline\iota}_q^{-1}\overline{\nu}_q\right)+o_P\left(q^{-\kappa}\overline{\nu}_q/{\underline\nu}_q\right)+o_P\left({\underline\iota}_q^{-1}{\overline\nu}_q\underline\nu_q^{-1/2}q^{3/2}T^{-1/2}\right)+\notag\\
&&O_P\left({\underline\nu}_q^{-1}({\overline\nu}_q/\underline\nu_q)^{1/2}\left[{\overline\nu}_q^{1/2}q^{1/2}T^{-1/2}\left(N^{-1/2}+\delta_q\right)+T^{-3/2}\right]\right),\notag
\end{eqnarray}
which, together with Assumption \ref{ass:1}(iv) and (\ref{eqB.31}), leads to the first assertion in (\ref{eqB.38}). Using Assumption \ref{ass:1}(ii) and $\underline\nu_q\leq \overline\nu_q$, we readily have the first assertion in (\ref{eqB.39}).

Since $\left\Vert {\boldsymbol H}_0^{-1}\right\Vert\left\Vert {\boldsymbol H}_{NT}-{\boldsymbol H}_0\right\Vert=o_P(1)$ by (\ref{eqB.32}), using \eqref{eqB.40}, we can prove that
$$\left\Vert {\boldsymbol H}_{NT}^{-1}-{\boldsymbol H}_0^{-1}\right\Vert\leq \frac{\left\Vert {\boldsymbol H}_0^{-1}\right\Vert^2\left\Vert {\boldsymbol H}_{NT}-{\boldsymbol H}_0\right\Vert}{1-\left\Vert {\boldsymbol H}_0^{-1}\right\Vert\left\Vert {\boldsymbol H}_{NT}-{\boldsymbol H}_0\right\Vert}=O_P( {\overline \nu}_q)o_P( {\overline \nu}_q^{-1/2})=o_P( {\overline \nu}_q^{1/2}).$$
The proof of Lemma \ref{le:B.5} is completed.\hfill$\Box$

\begin{lemma}\label{le:B.6}

Suppose that the assumptions of Lemma \ref{le:B.5} are satisfied. We have the following convergence results for ${\boldsymbol Q}_{NT}$ and ${\mathbf Q}_{NT}^{-1}{\boldsymbol V}_{NT}^{-1}$:
\begin{equation}\label{eqB.45}
\Vert{\mathbf Q}_{NT}-{\mathbf Q}_0\Vert=o_P({\overline \nu}_q^{-1/2}),\quad \Vert{\mathbf Q}_{NT}\Vert=O_P({\underline\nu}_{q}^{-1/2}),
\end{equation}
\begin{equation}\label{eqB.46}
\Vert{\mathbf Q}_{NT}^{-1}{\boldsymbol V}_{NT}^{-1}-{\boldsymbol\Sigma}_\Lambda^{-1/2}{\boldsymbol W}_0{\boldsymbol V}_0^{-1/2}\Vert=o_P({\underline\nu}_{q}^{-1/2}),\quad \Vert{\mathbf Q}_{NT}^{-1}{\boldsymbol V}_{NT}^{-1}\Vert=O_P({\underline\nu}_{q}^{-1/2}),
\end{equation}
where ${\mathbf Q}_0={\boldsymbol V}_0^{-1/2}{\boldsymbol W}_0^{^\intercal}{\boldsymbol\Sigma}_\Lambda^{-1/2}$.

\end{lemma}

\medskip

\noindent{\bf Proof of Lemma \ref{le:B.6}}.\ \ With the triangle inequality, we obtain
\begin{eqnarray*}
\left\Vert{\mathbf Q}_{NT}-{\mathbf Q}_0 \right\Vert
&\leq&\left\Vert\left({\boldsymbol V}_{NT}^{-1}-{\boldsymbol V}_0^{-1}\right){\boldsymbol D}_{W_*}{\boldsymbol W}_{NT}^{^\intercal}{\boldsymbol\Sigma}_{\Lambda,N}^{-1/2}\right\Vert+\left\Vert{\boldsymbol V}_0^{-1}\left({\boldsymbol D}_{W_*}-{\boldsymbol V}_0^{1/2}\right){\boldsymbol W}_{NT}^{^\intercal}{\boldsymbol\Sigma}_{\Lambda,N}^{-1/2}\right\Vert+\\
&&\left\Vert{\boldsymbol V}_0^{-1/2}\left({\boldsymbol W}_{NT}-{\boldsymbol W}_0\right)^{^\intercal}{\boldsymbol\Sigma}_{\Lambda,N}^{-1/2}\right\Vert+\left\Vert{\boldsymbol V}_0^{-1/2}{\boldsymbol W}_0^{^\intercal}\left({\boldsymbol\Sigma}_{\Lambda,N}^{-1/2}-{\boldsymbol \Sigma}_\Lambda^{-1/2}\right)\right\Vert,
\end{eqnarray*}
as in \eqref{eqB.35}. Following the proofs of \eqref{eqB.36}--\eqref{eqB.39}, we can show the first assertion in \eqref{eqB.45}, and subsequently,
\[
\Vert{\mathbf Q}_{NT}\Vert\leq \left\Vert{\mathbf Q}_{NT}-{\mathbf Q}_0 \right\Vert+ \left\Vert {\mathbf Q}_0 \right\Vert=o_P({\overline \nu}_q^{-1/2})+O_P({\underline\nu}_{q}^{-1/2})=O_P({\underline\nu}_{q}^{-1/2}).
\]

Similarly, following the proofs of \eqref{eqB.37}--\eqref{eqB.39} again, we can prove that
\begin{eqnarray*}
\left\Vert{\boldsymbol V}_{NT}{\mathbf Q}_{NT}-{\boldsymbol V}_0{\mathbf Q}_0 \right\Vert
&\leq&\left\Vert\left({\boldsymbol D}_{W_*}-{\boldsymbol V}_0^{1/2}\right){\boldsymbol W}_{NT}^{^\intercal}{\boldsymbol\Sigma}_{\Lambda,N}^{-1/2}\right\Vert+\\
&&\left\Vert{\boldsymbol V}_0^{1/2}\left({\boldsymbol W}_{NT}-{\boldsymbol W}_0\right)^{^\intercal}{\boldsymbol\Sigma}_{\Lambda,N}^{-1/2}\right\Vert+\left\Vert{\boldsymbol V}_0^{1/2}{\boldsymbol W}_0^{^\intercal}\left({\boldsymbol\Sigma}_{\Lambda,N}^{-1/2}-{\boldsymbol \Sigma}_\Lambda^{-1/2}\right)\right\Vert\\
&=&o_P({\underline\nu}_q{\overline\nu}_q^{-1/2})+O_P({\overline\nu}_q^{1/2})o_P({\underline\nu}_q{\overline\nu}_q^{-1/2})+o_P({\overline\nu}_q^{1/2}q^{-\kappa})\\
&=&o_P({\underline\nu}_q^{1/2}),
\end{eqnarray*}
which, together with $\left\Vert ({\boldsymbol V}_0{\mathbf Q}_0 )^{-1}\right\Vert\left\Vert {\boldsymbol V}_{NT}{\mathbf Q}_{NT}-{\boldsymbol V}_0{\mathbf Q}_0 \right\Vert=o_P(1)$ and \eqref{eqB.40}, leads to the first assertion in \eqref{eqB.46}. Furthermore, we may show that
\[
\Vert{\mathbf Q}_{NT}^{-1}{\boldsymbol V}_{NT}^{-1}\Vert\leq \Vert{\mathbf Q}_{NT}^{-1}{\boldsymbol V}_{NT}^{-1}-{\boldsymbol\Sigma}_\Lambda^{-1/2}{\boldsymbol W}_0{\boldsymbol V}_0^{-1/2}\Vert+\Vert {\boldsymbol\Sigma}_\Lambda^{-1/2}{\boldsymbol W}_0{\boldsymbol V}_0^{-1/2}\Vert=o_P({\underline\nu}_{q}^{-1/2})+O_P({\underline\nu}_{q}^{-1/2})=O_P({\underline\nu}_{q}^{-1/2}).
\]
The proof of Lemma \ref{le:B.6} is completed.\hfill$\Box$

\medskip

\begin{lemma}\label{le:B.7}

Let $\boldsymbol A$ be an $M\times N$ matrix and ${\boldsymbol F} = (f_1, f_2, \cdots, f_N)^{^\intercal}$ be a vector of square integrable functions defined on ${\mathbb C}$. Then we have $\Vert {\boldsymbol A}{\boldsymbol F}\Vert\leq\Vert {\boldsymbol A}\Vert\Vert {\boldsymbol F}\Vert$.

\end{lemma}

\medskip

\noindent{\bf Proof of Lemma \ref{le:B.7}}.\ \ Note that 
$$\Vert {\boldsymbol A}{\boldsymbol F}\Vert^2=\int_\mathbb{C}  {\boldsymbol F}^{^\intercal}(u){\boldsymbol A}^{^\intercal}{\boldsymbol A}{\boldsymbol F}(u)du\leq \int_\mathbb{C} \Vert {\boldsymbol A}\Vert^2 {\boldsymbol F}^{^\intercal}(u){\boldsymbol F}(u)du=\Vert {\boldsymbol A}\Vert^2\Vert {\boldsymbol F}\Vert^2,$$
which completes the proof.\hfill$\Box$
 
\bigskip

%%%%%%%%%%%%%%%%%%%%%%%%%%%%%%%%%%%%%%%%%%%%%%%%%%%%%%%%%%%%%%%%%%%%%%%%%%%%%%%%%%%%%%%%%%%%%%%%%%%%%%%%%%%%%%%%%%%%%%%%%%%%

\end{document}